\documentclass[twocolumn,showpacs,preprintnumbers,prc]{revtex4}
\usepackage{amsmath}
\usepackage{amssymb}
\usepackage{graphicx}
\usepackage{dcolumn}
\usepackage{float}
\usepackage{grffile}
\usepackage{multirow}
\usepackage{hyperref}\hypersetup{colorlinks=true,linkcolor=blue,citecolor=blue,urlcolor=blue,bookmarksnumbered=false,bookmarksopen=false}
\usepackage{mathptmx}
\DeclareMathAlphabet{\mathcal}{OMS}{cmsy}{m}{n}

\newcommand{\eq}[1]{(\ref{#1})}
\newcommand{\pol}[1]{\varsigma^{(#1)}}
\newcommand{\bra}[1]{\langle #1 |}
\newcommand{\ket}[1]{| #1 \rangle}
\newcommand{\bm}[1]{\boldsymbol{#1}}
\newcommand{\M}{\mathcal{M}}
\newcommand{\B}{\mathcal{B}}
\newcommand{\T}{\mathcal{T}}
\newcommand{\R}{\mathcal{R}}
\newcommand{\thcm}{\theta_\text{c.m.}}
\newcolumntype{C}[1]{>{\centering\let\newline\\\arraybackslash\hspace{0pt}}m{#1}}

\renewcommand{\d}{\mathrm{d}}
\renewcommand{\Re}{\text{Re}}
\renewcommand{\Im}{\text{Im}}

\allowdisplaybreaks
\begin{document}

\title{The incompleteness of complete pseudoscalar-meson photoproduction}

\author{Tom Vrancx}
\email{Tom.Vrancx@UGent.be}
\author{Jan Ryckebusch}
\email{Jan.Ryckebusch@UGent.be}
\author{Tom Van Cuyck}
\author{Pieter Vancraeyveld}

\affiliation{Department of Physics and Astronomy,\\
 Ghent University, Proeftuinstraat 86, B-9000 Gent, Belgium}
\date{\today}

\begin{abstract}
\begin{description}
\item[Background:]
A complete set is a minimum set of observables which allows one to determine the underlying reaction amplitudes unambiguously. Pseudoscalar-meson photoproduction from the nucleon is characterized by four such amplitudes and complete sets involve single- and double-polarization observables.

\item[Purpose:]
Identify complete sets of observables, and study how measurements with finite error bars impact their potential to determine the reaction amplitudes unambiguously.

\item[Method:]
The authors provide arguments to employ the transversity representation in order to determine the amplitudes in pseudoscalar-meson photoproduction. It is studied whether the amplitudes in the transversity basis for the $\gamma p \to K^+\Lambda$ reaction can be estimated without ambiguity. To this end, data from the GRAAL collaboration and simulations from a realistic model are analyzed.

\item[Results:]
It is illustrated that the moduli of normalized transversity amplitudes can be determined from precise single-polarization data. Starting from simulations with achievable experimental resolution, it is quite likely to obtain imaginary solutions for the relative phases of the amplitudes. Also the real solutions face a discrete phase ambiguity which makes it impossible to obtain a statistically significant solution for the relative phases at realistic experimental conditions.

\item[Conclusions:]
Single polarization observables are effective in determining the moduli of the amplitudes in a transversity basis. Determining the relative phases of the amplitudes from double-polarization observables is far less evident. The availability of a complete set of observables does not allow one to unambiguously determine the reaction amplitudes with statistical significance.
\end{description}
\end{abstract}



\maketitle 

\section{Introduction}
\label{sec:intro}

Pseudoscalar-meson photoproduction from the nucleon continues to be an invaluable source of information about the operation of quantum chromodynamics (QCD) in the nonperturbative regime \cite{CLAS2012a, CLAS2012b, LEGS-GRAAL2012, MAMI2012, ELSA2012}. Various classes of models, like constituent-quark approaches, have been developed to understand the structure and dynamics of hadrons in the low-energy regime of QCD. Experimental determination of the reaction amplitudes represents the most stringent test of those models and may open up a new chapter in the understanding of the energy eigenvalues and decay properties of hadron states.

Quantum mechanics dictates that measurable quantities can be expressed as bilinear combinations of complex amplitudes. Pseudoscalar meson photoproduction involves only two kinematical degrees of freedom, for example the energy of the incident photon and the scattering angle of the pseudoscalar meson. Since the (real) photon, the target, and the recoiling baryon  have two spin degrees of freedom, eight possible amplitudes can be constructed. Due to angular momentum conservation, and depending on the adopted representation, half of these amplitudes either vanish identically or can be expressed in terms of the other four amplitudes. This leaves one with a set of four independent amplitudes. These amplitudes are complex functions (of the two kinematical variables) and therefore eight real functions are to be distinguished: four \emph{moduli} and four \emph{phases}. As quantum states are determined up to a constant phase factor, only the relative phases of the amplitudes can be extracted. This means that 
pseudoscalar photoproduction can be quantified in terms of seven real-valued functions of the two kinematical variables. Equivalently, at fixed kinematics seven real values suffice to determine all observables.

A set containing a minimum number of observables from which, at fixed kinematics, the seven real values can be determined unanimously is referred to as a \emph{complete set}. In a seminal paper dating back to 1975, Barker, Donnachie, and Storrow argued that a complete set requires nine observables of a specific type \cite{Barker:1975bp}. In 1996, this was contested by Keaton and Workman \cite{Keaton1996} and by Wen-Tai Chiang and Tabakin \cite{Chiang:1996em}. The latter proved that eight well-chosen observables suffice to unambiguously determine the four moduli and three independent relative phases.

In theory, a complete set of eight observables suffices to retrieve the generating seven variables. The need for an additional observable (a set of eight equations involving only seven variables) to uniquely determine the relative phases, is a reflection of the fact that the latter are linked to the observables through nonlinear equations. A set of seven well-chosen observables generally yields multiple solutions for the phases. Therefore, an extra observable is required to mark the correct solution. In reality, however, observables have a finite precision and deviate from the exact values. This compromises the solution of the phase ambiguity \cite{Ireland2010}. It then remains a question whether a set of eight observables is still sufficient to reach a situation of complete knowledge about the amplitudes.

Thanks to recent technological advances in producing high-quality polarized beams and in developing polarized nucleon targets \cite{Hoblit2009}, it becomes possible to measure a sufficiently large amount of single- and double-polarization observables in pion and kaon photoproduction. As a result, a status of complete quantum mechanical information of pseudoscalar meson photoproduction comes within reach. The self-analyzing character of the $\Lambda$ is an enormous asset for achieving truly complete measurements for $\gamma p \to K ^{+} \Lambda$ and experimental efforts are under way \cite{CLAS2012b}. For example, the CLAS collaboration at Jefferson Lab has many $\gamma p \to K ^{+} \Lambda$ polarization data in the pipeline.

It should be stressed that in the quality and quantity of the experimental results, there is some kind of hierarchy. Indeed, double-polarization data is most often outnumbered by single-polarization results, which in their turn are outnumbered by the size of the differential cross section database. In view of this, the transversity representation of the amplitudes is a very promising one. Indeed, in this basis the single-polarization observables are linked to the squared moduli of the amplitudes by means of linear equations. Accordingly, the transversity basis occupies a central position in this work. 

The outline of the remainder of this paper is as follows. In Sec.\ \ref{sec:observables}, the transversity amplitudes are introduced and all possible observables for pseudoscalar meson photoproduction are expressed in this basis. In Sec.\ \ref{sec:formalism}, Wen-Tai Chiang and Tabakin's formalism for solving complete sets is briefly reviewed. This formalism is solid in the exact case. When experimental uncertainty is involved, however, a consistency issue for a unique determination of the phases arises. A method to resolve this inconsistency is proposed. As a test of this method, in Sec.\ \ref{sec:results} it is applied to simulations from a realistic model for the $\gamma p \to K^+ \Lambda$ reaction. The conclusions are summarized in Sec.\ \ref{sec:conclusion}. In the Appendix, both the helicity and the CGLN expansions of the observables are considered, and a connection to the transversity basis is established.

\section{Observables for pseudoscalar meson production}
\label{sec:observables}

\subsection{Reaction amplitudes}
\label{subsec:reacamplitudes}
The convention is adopted that the $xz$--plane coincides with the reaction plane and that the positive $z$--axis is along the direction of the photon's three-momentum. The two independent kinematic variables that will be considered throughout this work are the total energy of the reaction, or \emph{invariant mass}, $W$ and the \emph{scattering angle} of the pseudoscalar meson in the center-of-mass frame $\thcm$. As mentioned in Sec.\ \ref{sec:intro}, pseudoscalar-meson photoproduction can be quantified by four complex reaction amplitudes $\M_i(W,\cos\thcm)$ $(i = 1,2,3,4)$. The unpolarized differential cross section, for example, is given by
\begin{equation}
\frac{\d\sigma(W,\cos\thcm)}{\d\Omega} = \varrho \sum_{i=1}^4 \bigl|\M_i(W,\cos\thcm)\bigr|^2,\label{eq:angcrossection}
\end{equation}
where $\varrho$ is a kinematic factor. There are various equivalent representations for the $ \M_i $, all of which have a distinct kinematic factor $\varrho$. The representation is determined by the choices made with regard to the quantization axis of the involved particles with a non-vanishing spin: the nucleon target, the incoming photon, and the recoiling baryon.

The Dirac spinors for a particle with rest mass $m$, four-momentum $p^\mu = (E,\vec{p})$, and a spin vector with polar angles $(\theta,\phi)$ are defined as
\begin{align}
u_\pm(p^\mu,\theta,\phi) = \frac{1}{\sqrt{2m(E + m)}}
\begin{pmatrix}
(E + m) I_2\\
\vec{\sigma} \cdot \vec{p} 
\end{pmatrix}
s_\pm(\theta,\phi), \label{eq:spinor}
\end{align}
following the Bj\o{}rken-Drell convention \cite{BjorkenDrellBook}. Here, $I_2$ represents the $2\times2$ identity matrix and $\vec{\sigma} = (\sigma_x,\sigma_y,\sigma_z)$ is the Pauli vector. Further, the Pauli spinors $s_+(\theta,\phi)$ (``spin up'') and $s_-(\theta,\phi)$ (``spin
down'') are given by
\begin{align}
s_+(\theta,\phi) &=
\begin{pmatrix}
\cos\frac{\theta}{2}\\
e^{i\phi}\sin\frac{\theta}{2}
\end{pmatrix},
\nonumber\\
s_-(\theta,\phi) &=
\begin{pmatrix}
-e^{-i\phi}\sin\frac{\theta}{2}\\
\cos\frac{\theta}{2}
\end{pmatrix}.
\label{eq:rotationmatrices}
\end{align}
The following shorthand notations are introduced
\begin{align}
\ket{\pm}_x &= u_\pm(p^\mu,\tfrac{\pi}{2},0), \nonumber\\
\ket{\pm}_y &= u_\pm(p^\mu,\tfrac{\pi}{2},\tfrac{\pi}{2}), \nonumber\\
\ket{\pm}_z &= u_\pm(p^\mu,0,0).\label{eq:xyzkets}
\end{align}
Using Eq.~\eq{eq:spinor} it can readily be verified that the $ \ket{\pm}_x$ and $ \ket{\pm}_z$ can be expressed in terms of the 
$ \ket{\pm}_y$ as follows
\begin{align}
\ket{\pm}_x &= \frac{1 \mp i}{2}\Bigl(\ket{-}_y \pm \ket{+}_y\Bigr), \nonumber\\
\ket{\pm}_z &= \frac{1}{\sqrt{2}}\Bigl(\ket{\pm}_y - i \ket{\mp}_y\Bigr).\label{eq:rotate}
\end{align}
The $\M_i(W,\cos\thcm)$ are the amplitudes of the operator $\epsilon^\mu_\lambda J_{\mu}$ for a fixed polarization of the initial and final state particles. The $\epsilon^\mu_\lambda$ is the photon's polarization four-vector and $J^{\mu}$ the transition current operator. The photon beam polarizations denoted by ``$\B = x$'' ($+\vec{e}_x$ direction) and ``$\B = y$'' ($+\vec{e}_y$ direction) correspond with
\begin{align}
\epsilon^\mu_x & = (0,1,0,0), \nonumber \\
\epsilon^\mu_y & = (0,0,1,0),
\label{eq:photolinear}
\end{align}
and give rise to the following currents
\begin{align}
J_{x} & = \epsilon^\mu_x J_\mu, \nonumber\\
J_{y} & = \epsilon^\mu_y J_\mu.
\end{align}
The photon beam polarizations denoted by ``$\B = \pm \pi/4$'' (oblique polarization, $\epsilon^\mu_{\pm \pi/4} = \tfrac{1}{\sqrt{2}}(0,1,\pm 1,0)$) and ``$\B = \pm$'' (circular polarization, $\epsilon^\mu_\pm = \tfrac{1}{\sqrt{2}}(0,1,\pm i,0)$) correspond with
\begin{align}
J_{\pm \pi/4} &= -\frac{1}{\sqrt{2}}\left(J_x \pm J_y\right),\label{eq:Jpi4}\\
\qquad J_{\pm} &= -\frac{1}{\sqrt{2}}\left(J_x \pm iJ_y\right)\label{eq:Jpm}.
\end{align}
%
%
%
\subsection{Observables in the transversity basis}
\label{subsec:transversity}
The so-called \emph{transversity amplitudes} $b_i$ express the $\M_i$ in terms of the spinors $\ket{\pm}_y$ (quantization axis perpendicular to the reaction plane) and the linear photon polarizations $J_x$ and $J_y$, i.e.\
\begin{align}
b_1 &= {}_y\bra{+} J_y \ket{+}_y,\nonumber\\
b_2 &= {}_y\bra{-} J_y \ket{-}_y,\nonumber\\
b_3 &= {}_y\bra{+} J_x \ket{-}_y,\nonumber\\
b_4 &= {}_y\bra{-} J_x \ket{+}_y.
\label{eq:transversityamplitudes}
\end{align}
In these definitions, the bra (ket) refers to the recoil (target). The differential cross section for a given beam $\B$, target $\T$, and recoil polarization $\R$ is denoted as
\begin{align}
\varsigma^{(\B,\T,\R)} = \frac{\d \sigma}{\d \Omega}^{(\B,\T,\R)}.\label{eq:cs_BTR}
\end{align}
An unpolarized state is denoted by ``$0$''. For example, $\T = 0$ denotes an unpolarized target and in computing the cross section \eq{eq:cs_BTR} for $\T = 0$, an averaging over both target polarizations is implicitly assumed.

An asymmetry $A$ can generally be expressed as
\begin{align}
A = \frac{\pol{\B_1,\T_1,\R_1} - \pol{\B_2,\T_2,\R_2}}{\pol{\B_1,\T_1,\R_1} + \pol{\B_2,\T_2,\R_2}}. \label{eq:asymmetry_definition}
\end{align}
A \emph{single} asymmetry comprises one polarized and two unpolarized states. Hence, there are three possible single-polarization observables, namely the beam asymmetry $\Sigma$ ($\B \neq 0$), the target asymmetry $T$ ($\T \neq 0$), and the recoil asymmetry $P$ ($\R \neq 0$). The explicit definitions of these three single asymmetries can be found in Table~\ref{tab:transversity_representation}. A \emph{double} asymmetry, involves two polarized and one unpolarized state. There are three types of double asymmetries: the target-recoil asymmetries ($\B_1 = \B_2 = 0$), the beam-recoil asymmetries ($\T_1 = \T_2 = 0$), and the beam-target asymmetries ($\R_1 = \R_2 = 0$). The definitions for the various double asymmetries are contained in Table~\ref{tab:transversity_representation}.

The aim of the current section is to represent the single and double asymmetries in the transversity basis. As representative examples, the transversity expansions of the single asymmetry $\Sigma$ and of the double asymmetry $C_x$ are derived. The beam asymmetry $\Sigma$ is defined as
\begin{align}
\Sigma = \frac{\pol{y,0,0} - \pol{x,0,0}}{\pol{y,0,0} + \pol{x,0,0}}.
\end{align}
Using 
\begin{align}
  \pol{y,0,0} &= \frac{1}{2} \Bigl( \pol{y,+y,+y} + \pol{y,+y,-y} 
\nonumber \\
&\quad\, + \pol{y,-y,+y} + \pol{y,-y,-y} \Bigr), \nonumber \\
  &= \frac{\varrho}{2} \Bigl(\bigl|\underbrace{{}_y\bra{+} J_y \ket{+}_y}_{=\,b_1}\bigr|^2 + \bigl|\underbrace{{}_y\bra{-} J_y \ket{+}_y}_{=\,0}\bigr|^2
\nonumber \\
&\quad\, + \bigl|\underbrace{{}_y\bra{+} J_y \ket{-}_y}_{=\,0}\bigr|^2 + \bigl|\underbrace{{}_y\bra{-} J_y \ket{-}_y}_{=\,b_2}\bigr|^2\Bigr), \nonumber \\
  &= \frac{\varrho}{2} \left(|b_1|^2 + |b_2|^2\right) , \\
\intertext{and}
\pol{x,0,0} &= \frac{1}{2} \Bigl(\pol{x,+y,+y} + \pol{x,+y,-y} 
\nonumber \\
&\quad\, + \pol{x,-y,+y} + \pol{x,-y,-y}\Bigr),\nonumber\\
&= \frac{\varrho}{2} \Bigl(\bigl|\underbrace{{}_y\bra{+} J_x \ket{+}_y}_{=\,0}\bigr|^2 + \bigl|\underbrace{{}_y\bra{-} J_x \ket{+}_y}_{=\,b_4}\bigr|^2 
\nonumber \\ 
&\quad\, + \bigl|\underbrace{{}_y\bra{+} J_x \ket{-}_y}_{=\,b_3}\bigr|^2 + \bigl|\underbrace{{}_y\bra{-} J_x \ket{-}_y}_{=\,0}\bigr|^2\Bigr),\nonumber\\
&= \frac{\varrho}{2} \left(|b_3|^2 + |b_4|^2\right),
\end{align}
one obtains
\begin{align}
\Sigma = \frac{|b_1|^2 + |b_2|^2 - |b_3|^2 - |b_4|^2}{|b_1|^2 + |b_2|^2 + |b_3|^2 + |b_4|^2}.
\end{align}
At this point the \emph{normalized} transversity amplitudes $a_i$ 
\begin{align}
a_i = \frac{b_i}{\sqrt{|b_1|^2 + |b_2|^2 + |b_3|^2 + |b_4|^2}} ,
\label{eq:normalizedtransversityamplitudes}
\end{align}
are introduced. 
Upon using the notation $r_i = |a_i|$, the normalization condition of the $a_i$ reads
\begin{align}
r_1^2 + r_2^2 + r_3^2 + r_4^2 = 1,\label{eq:normalization}
\end{align}
which means that there are only three independent moduli, or six real values to be determined. For the $b_i$, this is respectively four and seven. The $|b_i|$ can be obtained from the $r_i$, combined with the magnitude of the unpolarized differential cross section. Using the $r_i$, the beam asymmetry $\Sigma$ can be expressed as
\begin{align}
\Sigma &= r_1^2 + r_2^2 - r_3^2 - r_4^2.
\end{align}
Similar calculations as those for $\Sigma$ yield the expressions for $T$ and $P$, contained in Table~\ref{tab:transversity_representation}.
%
\begin{table*}[]
\caption{The expressions for the single and double asymmetries in the normalized transversity basis. Both the expressions obtained in this work and those found in the literature \cite{Adelseck1990} are listed. The convention for the beam-target asymmetry $E$ is adopted from Ref.\ \cite{Sandorfi:2011nv} instead of Ref.\ \cite{Adelseck1990}.}
  \centering
  \vspace{10pt}
  \renewcommand{\arraystretch}{1.3}
\begin{tabular}{c|c|cc}
\hline
\noalign{\smallskip}\hline
& \multirow{2}{*}{$\bm{(\B_1,\T_1,\R_1) \quad (\B_2,\T_2,\R_2)}$} & \multicolumn{2}{c}{\textbf{Transversity representation}} \\
& & This work & Literature \cite{Adelseck1990}\\
\hline
$\Sigma$  & $ (y,0,0)\quad (x,0,0)$ & \multicolumn{2}{c}{$r_1^2 + r_2^2 - r_3^2 - r_4^2$} \\
$ T $ & $ (0,+y,0) \quad (0,-y,0)$ & \multicolumn{2}{c}{$r_1^2 - r_2^2 - r_3^2 + r_4^2$} \\
$ P $ & $ (0,0,+y) \quad (0,0,-y)$ & \multicolumn{2}{c}{$r_1^2 - r_2^2 + r_3^2 - r_4^2$} \\
\hline 
$C_x$ & $(+,0,+x) \quad (+,0,-x)$ & \phantom{$-$}$-2\Im(a_1a_4^* + a_2a_3^*)$\phantom{$-$} & \phantom{$-$}$-2\Im(a_1a_4^* - a_2a_3^*)$\phantom{$-$}\\
$C_z$ & $(+,0,+z) \quad (+,0,-z)$ & $+2\Re(a_1a_4^* - a_2a_3^*)$ & $+2\Re(a_1a_4^* + a_2a_3^*)$\\
$O_x$ & $(+\frac{\pi}{4},0,+x) \quad (+\frac{\pi}{4},0,-x)$ & $+2\Re(a_1a_4^* + a_2a_3^*)$ & $+2\Re(a_1a_4^* - a_2a_3^*)$\\
$O_z$ & $(+\frac{\pi}{4},0,+z) \quad (+\frac{\pi}{4},0,-z)$ & $+2\Im(a_1a_4^* - a_2a_3^*)$ & $+2\Im(a_1a_4^* + a_2a_3^*)$\\
\hline
$E$ & $(+,-z,0) \quad (+,+z,0)$ & $+2\Re(a_1a_3^* - a_2a_4^*)$ & $-2\Re(a_1a_3^* + a_2a_4^*)$ \\
$F$ & $(+,+x,0) \quad (+,-x,0)$ & $-2\Im(a_1a_3^* + a_2a_4^*)$ & $+2\Im(a_1a_3^* - a_2a_4^*)$ \\
$G$ & $(+\frac{\pi}{4},+z,0) \quad (+\frac{\pi}{4},-z,0)$ & $-2\Im(a_1a_3^* - a_2a_4^*)$ & $+2\Im(a_1a_3^* + a_2a_4^*)$ \\
$H$ & $(+\frac{\pi}{4},+x,0) \quad (+\frac{\pi}{4},-x,0)$ & $+2\Re(a_1a_3^* + a_2a_4^*)$ & $-2\Re(a_1a_3^* - a_2a_4^*)$ \\
\hline
$T_x$ & $(0,+x,+x) \quad (0,+x,-x)$ & $+2\Re(a_1a_2^* + a_3a_4^*)$ & $+2\Re(a_1a_2^* - a_3a_4^*)$ \\
$T_z$ & $(0,+x,+z) \quad (0,+x,-z)$ & $+2\Im(a_1a_2^* + a_3a_4^*)$ & $+2\Im(a_1a_2^* - a_3a_4^*)$ \\
$L_x$ & $(0,+z,+x) \quad (0,+z,-x)$ & $-2\Im(a_1a_2^* - a_3a_4^*)$ & $-2\Im(a_1a_2^* + a_3a_4^*)$ \\
$L_z$ & $(0,+z,+z) \quad (0,+z,-z)$ & $+2\Re(a_1a_2^* - a_3a_4^*)$ & $+2\Re(a_1a_2^* + a_3a_4^*)$ \\
\hline
\noalign{\smallskip}\hline
\end{tabular}
\label{tab:transversity_representation}
\end{table*}
%

Next the beam-recoil asymmetry $C_x$ 
\begin{align}
C_x &= \frac{\pol{+,0,+x} - \pol{+,0,-x}}{\pol{+,0,+x} + \pol{+,0,-x}},
\label{eq:defofcx}
\end{align}
is considered.
The first term in the numerator can be expressed as
\begin{align}
\pol{+,0,+x} &= \frac{1}{2}\left(\pol{+,+y,+x} + \pol{+,-y,+x}\right),\nonumber\\
&= \frac{\varrho}{2}\Bigl(\bigl|{}_x\bra{+} J_+ \ket{+}_y\bigr|^2 + \bigl|{}_x\bra{+} J_+ \ket{-}_y\bigr|^2\Bigr).
\end{align}
Using Eq.~\eq{eq:Jpm}, one obtains
\begin{align}
\pol{+,0,+x} &= \frac{\varrho}{4}\Bigl(\bigl|{}_x\bra{+} J_x \ket{+}_y + i\,{}_x\bra{+} J_y \ket{+}_y \bigr|^2 \nonumber \\
&\quad\, + \bigl|{}_x\bra{+} J_x \ket{-}_y + i\,{}_x\bra{+} J_y \ket{-}_y\bigr|^2\Bigr).
\end{align}
Then, Eq.~\eq{eq:rotate} is employed to transform the $_{x}\bra{+}$ spinors into $_{y}\bra{\pm}$ spinors, i.e.\
\begin{align}
\pol{+,0,+x} &= \frac{\varrho}{8}\Bigl(\bigl|\underbrace{{}_y\bra{+} J_x \ket{+}_y}_{=\,0} + \underbrace{{}_y\bra{-} J_x \ket{+}_y}_{=\,b_4} 
\nonumber \\
&\quad\,+ \, i\underbrace{{}_y\bra{+} J_y \ket{+}_y}_{=\,b_1} +\, i\underbrace{{}_y\bra{-} J_y \ket{+}_y}_{=\,0} \bigr|^2 \nonumber\\
  &\quad\,+ \bigl|\underbrace{{}_y\bra{+} J_x \ket{-}_y}_{=\,b_3} + \underbrace{{}_y\bra{-} J_x \ket{-}_y}_{=\,0} 
\nonumber \\
&\quad\, +\,i\underbrace{{}_y\bra{+} J_y \ket{-}_y}_{=\,0} +\,i\underbrace{{}_y\bra{-} J_y \ket{-}_y}_{=\,b_2}\bigr|^2\Bigr),\nonumber\\
  &= \frac{\varrho}{8}\left(|ib_1 + b_4|^2 + |ib_2 + b_3|^2\right).
\label{eq:part1}
\end{align}
An analogous calculation for $\pol{+,0,-x}$ yields
\begin{align}
\pol{+,0,-x} = \frac{\varrho}{8}\left(|ib_1 - b_4|^2 + |ib_2 - b_3|^2\right).
\label{eq:part2}
\end{align}
Inserting expressions \eq{eq:part1} and \eq{eq:part2} in the definition
\eq{eq:defofcx} leads to
\begin{align}
C_x &= \frac{i(b_1b_4^* - b_1^*b_4 + b_2b_3^* - b_2^*b_3)}{|b_1|^2 + |b_2|^2 + |b_3|^2 + |b_4|^2},\nonumber\\
&= -2\Im(a_1a_4^* + a_2a_3^*).\label{eq:C_x-transversity}
\end{align}
Analogous derivations for the remaining beam-recoil ($C_z, O_x, O_x$), the beam-target ($E, F, G, H$), and the tar\-get-recoil asymmetries ($T_x, T_z, L_x, L_z$) yield the expressions listed in Table~\ref{tab:transversity_representation}.

As it is unfrequently used in the field, the transversity representation for the observables is not contained in the extensive recent review by Sandorfi \textit{et al}.~\cite{Sandorfi:2011nv}. However, when it comes to extracting the invariant amplitudes from complete measurements, it will turn out that the transversity representation is highly beneficial. This will the subject of Sec.~\ref{subsec:merits_transversity}. The transversity expressions for the double asymmetries obtained in this work, are not consistent with those listed in literature \cite{Barker:1975bp,Adelseck1990,Ireland2010}.

By inspecting Table~\ref{tab:transversity_representation} it is clear that the substitution $a_3 \to -a_3$ makes the expressions derived in this work consistent with those contained in Ref.~\cite{Adelseck1990}. To the knowledge of the authors, the expressions for the asymmetries in the transversity basis were first published by Barker, Donnachie, and Storrow in 1975 \cite{Barker:1975bp}. Slightly different expressions are contained in a 1990 paper by Adelseck and Saghai \cite{Adelseck1990}. However, explicit derivations are not contained in either of both publications. In Appendices \ref{app:helicity_representation} and \ref{app:CGLN_representation}, the asymmetries are expanded in the helicity and the CGLN basis, respectively. The helicity representation is presented in Table~\ref{tab:helicity_representation} and is consistent with the one obtained by Fasano \textit{et al}.~\cite{Fasano:1992es}. On the other hand, the CGLN expansion was found to coincide with the one calculated by Sandorfi \textit{
et al}.~\cite{Sandorfi:2011nv}. Interestingly enough, both the helicity and CGLN representations were derived from the transversity expressions obtained in this work. Starting from the transversity expressions listed in the last column of Table~\ref{tab:transversity_representation} the helicity and CGLN expansions available in the literature are not retrieved.

\section{Inferring the transversity amplitudes from the asymmetries}
\label{sec:formalism}

\subsection{The moduli}
\label{subsec:moduli}

In Sec.~\ref{subsec:transversity} the notation $r_i = |a_i|$ was introduced. Hence, the (normalized) transversity amplitudes can be expressed as
\begin{align}
a_j = r_je^{i\alpha_j}.
\end{align}
Here, $\alpha_i$ represents the phase of the transversity amplitude $a_i$. From the expressions for $\Sigma$, $T$ and $P$ in Table~\ref{tab:transversity_representation} and from the normalization condition \eq{eq:normalization}, one can readily solve for the moduli of the amplitudes in terms of the single asymmetries, to obtain
\begin{align}
r_1 &= \frac{1}{2}\sqrt{1 + \Sigma + T + P},\nonumber\\
r_2 &= \frac{1}{2}\sqrt{1 + \Sigma - T - P},\nonumber\\
r_3 &= \frac{1}{2}\sqrt{1 - \Sigma - T + P},\nonumber\\
r_4 &= \frac{1}{2}\sqrt{1 - \Sigma + T - P}.
\label{eq:moduli_expressions}
\end{align}
Hence, a measurement of $\Sigma$, $T$ and $P$ at fixed kinematics $(W, \cos\thcm)$ yields the moduli of the transversity amplitudes $r_i$.

In Sec.\ \ref{subsec:transversity}, it was mentioned that only three of the four moduli are independent. Any of the four combinations of three independent moduli can be chosen without violating generality. As will be explained in Sec.\ \ref{subsubsec:analysis}, however, it is not beneficial to eliminate one of the moduli.

\subsection{The relative phases}
\label{subsec:phases}
\subsubsection{Independent and nonindependent relative phases}
\label{subsubsec:independent_phases}
From the double asymmetries one can extract the relative phases $\alpha_{ij} = \alpha_i - \alpha_j$, given that the moduli $r_i$ are known. There are six possible combinations for the relative phases, namely $\alpha_{12}$, $\alpha_{13}$, $\alpha_{14}$, $\alpha_{23}$, $\alpha_{24}$, and $\alpha_{34}$, all of which can be extracted from the double asymmetries (see Table~\ref{tab:transversity_representation}). However, only three of these are independent variables; the remaining three can be expressed as linear combinations of the three independent relative phases. For example, with $\alpha_4$ as the reference phase, $ \left( \alpha_{14}, \alpha_{24}, \alpha_{34} \right)$ are the independent phases and $ \left( \alpha_{12}, \alpha_{13}, \alpha_{23} \right)$ are the nonindependent ones
\begin{align}
\alpha_{12} &= \alpha_{14} - \alpha_{24}, \nonumber\\
\alpha_{13} &= \alpha_{14} - \alpha_{34}, \nonumber\\
\alpha_{23} &= \alpha_{24} - \alpha_{34}.
\end{align}
Alternatively, one can express the phases relative to $\alpha_1$, $\alpha_2$, or $\alpha_3$. From now on, $\alpha_4$ will serve as the reference phase and the independent phases are denoted as $\delta_i = \alpha_i - \alpha_4$ $(i = 1,2,3)$. The nonindependent phases are labeled as $\Delta_{ij} = \delta_i - \delta_j$ $(i \neq j)$. Table~\ref{tab:reference_phases} shows the correspondence between the independent and the nonindependent phases for the four distinct reference phases.

\subsubsection{Complete sets}
\label{subsubsec:complete_sets}
In Ref.\ \cite{Chiang:1996em}, Wen-Tai Chiang and Tabakin proved that a specific set of four double asymmetries is sufficient to extract two of the independent phases, $\delta_i$ and $\delta_j$, and two of the nonindependent phases, $\Delta_{ik}$ and $\Delta_{jk}$ ($i \neq j \neq k$). The remaining independent phase $\delta_k$ can be constructed as $\delta_k = \delta_i - \Delta_{ik}$ or $\delta_k = \delta_j - \Delta_{jk}$. Hence, a \emph{complete set}, consisting of three single and four double asymmetries, determines the four moduli and the three independent relative phases of the transversity amplitudes $a_i$.

The collection of complete sets, which is listed in tables III--VIII of Ref.\ \cite{Chiang:1996em}, can be divided into two categories. An example of a complete set of the first kind is $\{C_x, O_x, E, F\}$ (the single asymmetries $\Sigma$, $T$, and $P$ are implicitly assumed to be included as they are a vital part of any complete set in the transversity basis). From Table~\ref{tab:transversity_representation} one obtains
\begin{align}
r_1r_4\sin\delta_1 + r_2r_3\sin\Delta_{23} &= -\frac{C_x}{2}, \nonumber\\
r_1r_4\cos\delta_1 + r_2r_3\cos\Delta_{23} &= \frac{O_x}{2},\label{eq:Cx-Ox}
\end{align}
and
\begin{align}
r_1r_3\cos\Delta_{13} - r_2r_4\cos\delta_2 &= \frac{E}{2}, \nonumber\\
r_1r_3\sin\Delta_{13} + r_2r_4\sin\delta_2 &= -\frac{F}{2}.\label{eq:E-F}
\end{align}
Since Eq.~\eq{eq:Cx-Ox} contains both a sine and a cosine of the unknowns, there are generally two solutions for $\{\delta_1, \Delta_{23}\}$. The same reasoning applies to $E$ and $F$, and hence there are four possible solutions for $\{\delta_1, \delta_2, \Delta_{13}, \Delta_{23}\}$. Yet, only one of these solutions will satisfy the (trivial) relation
\begin{align}
\delta_1 + \Delta_{23} - \delta_2 - \Delta_{13} = 0,\label{eq:phases_constraint}
\end{align}
and that specific solution is the actual solution.

If, for example, one replaces $F$ with $H$ in Eq.~\eq{eq:E-F}, a complete set of the second kind is obtained. Table~\ref{tab:transversity_representation} yields
\begin{align}
r_1r_3\cos\Delta_{13} + r_2r_4\cos\delta_2 &= \frac{H}{2}.\label{eq:H}
\end{align}
Since $\{E,H\}$ contain only the cosine of the unknowns, there are four possible solutions for $\{\delta_2, \Delta_{13}\}$. Hence, an eightfold ambiguity for $\{\delta_1, \delta_2, \Delta_{13}, \Delta_{23}\}$ emerges. Only one of the solutions, however, obeys the constraint \eq{eq:phases_constraint}. Similarly, the complete set $\{C_x, O_x, F, G\}$ is subject to an eightfold phase ambiguity.

\begin{table}[t!]
\caption{Correspondence between the independent ($\delta_i$) and nonindependent ($\Delta_{ij}$) phases for the reference phase $\alpha_4$ and the independent ($\delta_i^{\;\alpha_k}$) and nonindependent ($\Delta_{ij} = \delta_i^{\;\alpha_k} - \delta_j^{\;\alpha_k} = \alpha_i - \alpha_j$) phases for the reference phase $\alpha_k$ ($k = 1,2,3$).}\label{tab:reference_phases}
\centering
\renewcommand{\arraystretch}{1.3}
\begin{tabular}{c|cccccc}
\hline
\noalign{\smallskip}\hline
$\bm{\alpha_4}$ & $\bm{\delta_1}$ & $\bm{\delta_2}$ & $\bm{\delta_3}$ & $\bm{\Delta_{12}}$ & $\bm{\Delta_{13}}$ & $\bm{\Delta_{23}}$ \\
\hline
$\alpha_1$ & $-\delta_4^{\;\alpha_1}$ & $\Delta_{24}$ & $\Delta_{34}$ & $-\delta_2^{\;\alpha_1}$ & $-\delta_3^{\;\alpha_1}$ & $\Delta_{23}$ \\
$\alpha_2$ & $\Delta_{14}$ & $-\delta_4^{\;\alpha_2}$ & $\Delta_{34}$ & $\delta_1^{\;\alpha_2}$ & $\Delta_{13}$ & $-\delta_3^{\;\alpha_2}$ \\
$\alpha_3$ & $\Delta_{14}$ & $\Delta_{24}$ & $-\delta_4^{\;\alpha_3}$ & $\Delta_{12}$ & $\delta_1^{\;\alpha_3}$ & $\delta_2^{\;\alpha_3}$ \\
\hline
\noalign{\smallskip}\hline
\end{tabular}
\end{table}

\subsection{Introducing experimental error}
\label{subsec:experimental_error}
As indicated in Sec.\ \ref{subsubsec:complete_sets}, a complete set provides access to two independent and two nonindependent phases, i.e.\ the set $\{\delta_i, \delta_j, \Delta_{ik}, \Delta_{jk}\}$ $(i\neq j\neq k)$. There are two ways to calculate the third independent phase $\delta_k$: \ $\delta_k = \delta_i - \Delta_{ik}$ or $\delta_k = \delta_j - \Delta_{jk}$. For infinite experimental resolution both expressions for $\delta_k$ are equivalent. When experimental error is introduced, the estimated values $\{\widehat{\delta}_i, \widehat{\delta}_j, \widehat{\Delta}_{ik}, \widehat{\Delta}_{jk}\}$ deviate from their actual values $\{\delta_i, \delta_j, \Delta_{ik}, \Delta_{jk}\}$ (estimators are marked with a `` $\widehat{\phantom{a}}$ '') and the two expressions for $\widehat{\delta}_k$ yield a different result in general. Since there is no prior preference, an equal weight can be assigned to both 
estimators 
\begin{align}
\widehat{\delta}_k = \frac{1}{2}\bigl(\widehat{\delta}_i - \widehat{\Delta}_{ik}\bigr) + \frac{1}{2}\bigl(\widehat{\delta}_j - \widehat{\Delta}_{jk}\bigr),
\end{align}
thereby resolving the $\widehat{\delta}_k$ ambiguity. However, this is not the only problem that emerges from introducing experimental error. Consider again the complete set $\{C_x, O_x, E, F\}$ of Eqs.~\eq{eq:Cx-Ox}--\eq{eq:E-F}. For this set of observables the $\widehat{\delta}_i$ read
\begin{align}
\begin{cases}
\widehat{\delta}_1, \vspace{5pt}\\
\widehat{\delta}_2, \vspace{5pt}\\
\widehat{\delta}_3 = \frac{1}{2}\bigl(\widehat{\delta}_1 - \widehat{\Delta}_{13}\bigr) + \frac{1}{2}\bigl(\widehat{\delta}_2 - \widehat{\Delta}_{23}\bigr).
\end{cases}
\label{eq:independent_phases_alpha4}
\end{align}
Suppose that, for example, $\alpha_1$ was assigned as the reference phase, instead of $\alpha_4$. Then, $\{C_x, O_x, E, F\}$ provides access to $\{\widehat{\delta}_3^{\;\alpha_1}, \widehat{\delta}_4^{\;\alpha_1}, \widehat{\Delta}_{23}, \widehat{\Delta}_{24}\}$, as can be inferred from Table~\ref{tab:reference_phases}. The corresponding independent set of phases reads
\begin{align}
\begin{cases}
\widehat{\delta}_2^{\;\alpha_1} = \frac{1}{2}\bigl(\widehat{\delta}_3^{\;\alpha_1} + \widehat{\Delta}_{23}\bigr) + \frac{1}{2}\bigl(\widehat{\delta}_4^{\;\alpha_1} + \widehat{\Delta}_{24}\bigr),\vspace{5pt}\\
\widehat{\delta}_3^{\;\alpha_1}, \vspace{5pt}\\
\widehat{\delta}_4^{\;\alpha_1}.
\end{cases}
\label{eq:independent_phases_alpha1}
\end{align}
From \eq{eq:independent_phases_alpha1} one can estimate the $\delta_i$ through $\{\widehat{\delta}_1^{\;\prime} = - \widehat{\delta}_4^{\;\alpha_1},\, \widehat{\delta}_2^{\;\prime} = \widehat{\delta}_2^{\;\alpha_1} - \widehat{\delta}_4^{\;\alpha_1},\, \widehat{\delta}_3^{\;\prime} =  \widehat{\delta}_3^{\;\alpha_1} - \widehat{\delta}_4^{\;\alpha_1}\}$ and Table~\ref{tab:reference_phases}. This yields
\begin{align}
\begin{cases}
\widehat{\delta}_1^{\;\prime} = \widehat{\delta}_1, \vspace{5pt}\\
\widehat{\delta}_2^{\;\prime} = \frac{1}{2}\widehat{\delta}_2 + \frac{1}{2}\bigl(\widehat{\delta}_1 + \widehat{\Delta}_{23} - \widehat{\Delta}_{13}\bigr), \vspace{5pt}\\
\widehat{\delta}_3^{\;\prime} = \widehat{\delta}_1 - \widehat{\Delta}_{13}. 
\end{cases}
\label{eq:independent_phases_alpha4_1}
\end{align}
The set \eq{eq:independent_phases_alpha4_1} is not consistent with the set \eq{eq:independent_phases_alpha4}. Indeed, the estimates for the independent phases (there are four possible sets) depends on the choice of reference phase. 

\begin{figure*}[!t]
\begin{center}
\footnotesize
\begingroup
  \makeatletter
  \providecommand\color[2][]{%
    \GenericError{(gnuplot) \space\space\space\@spaces}{%
      Package color not loaded in conjunction with
      terminal option `colourtext'%
    }{See the gnuplot documentation for explanation.%
    }{Either use 'blacktext' in gnuplot or load the package
      color.sty in LaTeX.}%
    \renewcommand\color[2][]{}%
  }%
  \providecommand\includegraphics[2][]{%
    \GenericError{(gnuplot) \space\space\space\@spaces}{%
      Package graphicx or graphics not loaded%
    }{See the gnuplot documentation for explanation.%
    }{The gnuplot epslatex terminal needs graphicx.sty or graphics.sty.}%
    \renewcommand\includegraphics[2][]{}%
  }%
  \providecommand\rotatebox[2]{#2}%
  \@ifundefined{ifGPcolor}{%
    \newif\ifGPcolor
    \GPcolortrue
  }{}%
  \@ifundefined{ifGPblacktext}{%
    \newif\ifGPblacktext
    \GPblacktexttrue
  }{}%
  \let\gplgaddtomacro\g@addto@macro
  \gdef\gplbacktext{}%
  \gdef\gplfronttext{}%
  \makeatother
  \ifGPblacktext
    \def\colorrgb#1{}%
    \def\colorgray#1{}%
  \else
    \ifGPcolor
      \def\colorrgb#1{\color[rgb]{#1}}%
      \def\colorgray#1{\color[gray]{#1}}%
      \expandafter\def\csname LTw\endcsname{\color{white}}%
      \expandafter\def\csname LTb\endcsname{\color{black}}%
      \expandafter\def\csname LTa\endcsname{\color{black}}%
      \expandafter\def\csname LT0\endcsname{\color[rgb]{1,0,0}}%
      \expandafter\def\csname LT1\endcsname{\color[rgb]{0,1,0}}%
      \expandafter\def\csname LT2\endcsname{\color[rgb]{0,0,1}}%
      \expandafter\def\csname LT3\endcsname{\color[rgb]{1,0,1}}%
      \expandafter\def\csname LT4\endcsname{\color[rgb]{0,1,1}}%
      \expandafter\def\csname LT5\endcsname{\color[rgb]{1,1,0}}%
      \expandafter\def\csname LT6\endcsname{\color[rgb]{0,0,0}}%
      \expandafter\def\csname LT7\endcsname{\color[rgb]{1,0.3,0}}%
      \expandafter\def\csname LT8\endcsname{\color[rgb]{0.5,0.5,0.5}}%
    \else
      \def\colorrgb#1{\color{black}}%
      \def\colorgray#1{\color[gray]{#1}}%
      \expandafter\def\csname LTw\endcsname{\color{white}}%
      \expandafter\def\csname LTb\endcsname{\color{black}}%
      \expandafter\def\csname LTa\endcsname{\color{black}}%
      \expandafter\def\csname LT0\endcsname{\color{black}}%
      \expandafter\def\csname LT1\endcsname{\color{black}}%
      \expandafter\def\csname LT2\endcsname{\color{black}}%
      \expandafter\def\csname LT3\endcsname{\color{black}}%
      \expandafter\def\csname LT4\endcsname{\color{black}}%
      \expandafter\def\csname LT5\endcsname{\color{black}}%
      \expandafter\def\csname LT6\endcsname{\color{black}}%
      \expandafter\def\csname LT7\endcsname{\color{black}}%
      \expandafter\def\csname LT8\endcsname{\color{black}}%
    \fi
  \fi
  \setlength{\unitlength}{0.0500bp}%
  \begin{picture}(4626.00,3148.00)%
    \gplgaddtomacro\gplbacktext{%
      \csname LTb\endcsname%
      \put(2274,2881){\makebox(0,0){\normalsize $\phantom{\eta}r_1\phantom{\eta}$}}%
    }%
    \gplgaddtomacro\gplfronttext{%
      \csname LTb\endcsname%
      \put(937,211){\makebox(0,0){\strut{}}}%
      \put(1314,211){\makebox(0,0){\strut{}}}%
      \put(1692,211){\makebox(0,0){\strut{}}}%
      \put(2069,211){\makebox(0,0){\strut{}}}%
      \put(2446,211){\makebox(0,0){\strut{}}}%
      \put(2823,211){\makebox(0,0){\strut{}}}%
      \put(3200,211){\makebox(0,0){\strut{}}}%
      \put(3578,211){\makebox(0,0){\strut{}}}%
      \put(3955,211){\makebox(0,0){\strut{}}}%
      \put(438,438){\makebox(0,0)[r]{\strut{}-1}}%
      \put(438,983){\makebox(0,0)[r]{\strut{}-0.5}}%
      \put(438,1528){\makebox(0,0)[r]{\strut{} 0}}%
      \put(438,2073){\makebox(0,0)[r]{\strut{} 0.5}}%
      \put(438,2618){\makebox(0,0)[r]{\strut{} 1}}%
      \put(-96,1528){\rotatebox{-270}{\makebox(0,0){\normalsize $\cos\thcm$}}}%
    }%
    \gplbacktext
    \put(490,353){\includegraphics[scale=0.66]{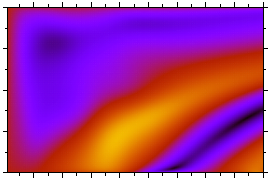}}%
    \gplfronttext
  \end{picture}%
\endgroup
\hspace{-41pt}\vspace{-15pt}
\begingroup
  \makeatletter
  \providecommand\color[2][]{%
    \GenericError{(gnuplot) \space\space\space\@spaces}{%
      Package color not loaded in conjunction with
      terminal option `colourtext'%
    }{See the gnuplot documentation for explanation.%
    }{Either use 'blacktext' in gnuplot or load the package
      color.sty in LaTeX.}%
    \renewcommand\color[2][]{}%
  }%
  \providecommand\includegraphics[2][]{%
    \GenericError{(gnuplot) \space\space\space\@spaces}{%
      Package graphicx or graphics not loaded%
    }{See the gnuplot documentation for explanation.%
    }{The gnuplot epslatex terminal needs graphicx.sty or graphics.sty.}%
    \renewcommand\includegraphics[2][]{}%
  }%
  \providecommand\rotatebox[2]{#2}%
  \@ifundefined{ifGPcolor}{%
    \newif\ifGPcolor
    \GPcolortrue
  }{}%
  \@ifundefined{ifGPblacktext}{%
    \newif\ifGPblacktext
    \GPblacktexttrue
  }{}%
  \let\gplgaddtomacro\g@addto@macro
  \gdef\gplbacktext{}%
  \gdef\gplfronttext{}%
  \makeatother
  \ifGPblacktext
    \def\colorrgb#1{}%
    \def\colorgray#1{}%
  \else
    \ifGPcolor
      \def\colorrgb#1{\color[rgb]{#1}}%
      \def\colorgray#1{\color[gray]{#1}}%
      \expandafter\def\csname LTw\endcsname{\color{white}}%
      \expandafter\def\csname LTb\endcsname{\color{black}}%
      \expandafter\def\csname LTa\endcsname{\color{black}}%
      \expandafter\def\csname LT0\endcsname{\color[rgb]{1,0,0}}%
      \expandafter\def\csname LT1\endcsname{\color[rgb]{0,1,0}}%
      \expandafter\def\csname LT2\endcsname{\color[rgb]{0,0,1}}%
      \expandafter\def\csname LT3\endcsname{\color[rgb]{1,0,1}}%
      \expandafter\def\csname LT4\endcsname{\color[rgb]{0,1,1}}%
      \expandafter\def\csname LT5\endcsname{\color[rgb]{1,1,0}}%
      \expandafter\def\csname LT6\endcsname{\color[rgb]{0,0,0}}%
      \expandafter\def\csname LT7\endcsname{\color[rgb]{1,0.3,0}}%
      \expandafter\def\csname LT8\endcsname{\color[rgb]{0.5,0.5,0.5}}%
    \else
      \def\colorrgb#1{\color{black}}%
      \def\colorgray#1{\color[gray]{#1}}%
      \expandafter\def\csname LTw\endcsname{\color{white}}%
      \expandafter\def\csname LTb\endcsname{\color{black}}%
      \expandafter\def\csname LTa\endcsname{\color{black}}%
      \expandafter\def\csname LT0\endcsname{\color{black}}%
      \expandafter\def\csname LT1\endcsname{\color{black}}%
      \expandafter\def\csname LT2\endcsname{\color{black}}%
      \expandafter\def\csname LT3\endcsname{\color{black}}%
      \expandafter\def\csname LT4\endcsname{\color{black}}%
      \expandafter\def\csname LT5\endcsname{\color{black}}%
      \expandafter\def\csname LT6\endcsname{\color{black}}%
      \expandafter\def\csname LT7\endcsname{\color{black}}%
      \expandafter\def\csname LT8\endcsname{\color{black}}%
    \fi
  \fi
  \setlength{\unitlength}{0.0500bp}%
  \begin{picture}(4626.00,3148.00)%
    \gplgaddtomacro\gplbacktext{%
      \csname LTb\endcsname%
      \put(2274,2881){\makebox(0,0){\normalsize $\phantom{\eta}r_2\phantom{\eta}$}}%
    }%
    \gplgaddtomacro\gplfronttext{%
      \csname LTb\endcsname%
      \put(937,211){\makebox(0,0){\strut{}}}%
      \put(1314,211){\makebox(0,0){\strut{}}}%
      \put(1692,211){\makebox(0,0){\strut{}}}%
      \put(2069,211){\makebox(0,0){\strut{}}}%
      \put(2446,211){\makebox(0,0){\strut{}}}%
      \put(2823,211){\makebox(0,0){\strut{}}}%
      \put(3200,211){\makebox(0,0){\strut{}}}%
      \put(3578,211){\makebox(0,0){\strut{}}}%
      \put(3955,211){\makebox(0,0){\strut{}}}%
      \put(438,438){\makebox(0,0)[r]{\strut{}}}%
      \put(438,983){\makebox(0,0)[r]{\strut{}}}%
      \put(438,1528){\makebox(0,0)[r]{\strut{}}}%
      \put(438,2073){\makebox(0,0)[r]{\strut{}}}%
      \put(438,2618){\makebox(0,0)[r]{\strut{}}}%
      \put(4272,437){\makebox(0,0)[l]{\strut{} 0}}%
      \put(4272,873){\makebox(0,0)[l]{\strut{} 0.2}}%
      \put(4272,1309){\makebox(0,0)[l]{\strut{} 0.4}}%
      \put(4272,1745){\makebox(0,0)[l]{\strut{} 0.6}}%
      \put(4272,2181){\makebox(0,0)[l]{\strut{} 0.8}}%
      \put(4272,2618){\makebox(0,0)[l]{\strut{} 1}}%
    }%
    \gplbacktext
    \put(500,353){\includegraphics[scale=0.66]{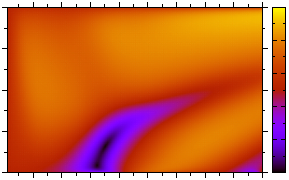}}%
    \gplfronttext
  \end{picture}%
\endgroup

\begingroup
  \makeatletter
  \providecommand\color[2][]{%
    \GenericError{(gnuplot) \space\space\space\@spaces}{%
      Package color not loaded in conjunction with
      terminal option `colourtext'%
    }{See the gnuplot documentation for explanation.%
    }{Either use 'blacktext' in gnuplot or load the package
      color.sty in LaTeX.}%
    \renewcommand\color[2][]{}%
  }%
  \providecommand\includegraphics[2][]{%
    \GenericError{(gnuplot) \space\space\space\@spaces}{%
      Package graphicx or graphics not loaded%
    }{See the gnuplot documentation for explanation.%
    }{The gnuplot epslatex terminal needs graphicx.sty or graphics.sty.}%
    \renewcommand\includegraphics[2][]{}%
  }%
  \providecommand\rotatebox[2]{#2}%
  \@ifundefined{ifGPcolor}{%
    \newif\ifGPcolor
    \GPcolortrue
  }{}%
  \@ifundefined{ifGPblacktext}{%
    \newif\ifGPblacktext
    \GPblacktexttrue
  }{}%
  \let\gplgaddtomacro\g@addto@macro
  \gdef\gplbacktext{}%
  \gdef\gplfronttext{}%
  \makeatother
  \ifGPblacktext
    \def\colorrgb#1{}%
    \def\colorgray#1{}%
  \else
    \ifGPcolor
      \def\colorrgb#1{\color[rgb]{#1}}%
      \def\colorgray#1{\color[gray]{#1}}%
      \expandafter\def\csname LTw\endcsname{\color{white}}%
      \expandafter\def\csname LTb\endcsname{\color{black}}%
      \expandafter\def\csname LTa\endcsname{\color{black}}%
      \expandafter\def\csname LT0\endcsname{\color[rgb]{1,0,0}}%
      \expandafter\def\csname LT1\endcsname{\color[rgb]{0,1,0}}%
      \expandafter\def\csname LT2\endcsname{\color[rgb]{0,0,1}}%
      \expandafter\def\csname LT3\endcsname{\color[rgb]{1,0,1}}%
      \expandafter\def\csname LT4\endcsname{\color[rgb]{0,1,1}}%
      \expandafter\def\csname LT5\endcsname{\color[rgb]{1,1,0}}%
      \expandafter\def\csname LT6\endcsname{\color[rgb]{0,0,0}}%
      \expandafter\def\csname LT7\endcsname{\color[rgb]{1,0.3,0}}%
      \expandafter\def\csname LT8\endcsname{\color[rgb]{0.5,0.5,0.5}}%
    \else
      \def\colorrgb#1{\color{black}}%
      \def\colorgray#1{\color[gray]{#1}}%
      \expandafter\def\csname LTw\endcsname{\color{white}}%
      \expandafter\def\csname LTb\endcsname{\color{black}}%
      \expandafter\def\csname LTa\endcsname{\color{black}}%
      \expandafter\def\csname LT0\endcsname{\color{black}}%
      \expandafter\def\csname LT1\endcsname{\color{black}}%
      \expandafter\def\csname LT2\endcsname{\color{black}}%
      \expandafter\def\csname LT3\endcsname{\color{black}}%
      \expandafter\def\csname LT4\endcsname{\color{black}}%
      \expandafter\def\csname LT5\endcsname{\color{black}}%
      \expandafter\def\csname LT6\endcsname{\color{black}}%
      \expandafter\def\csname LT7\endcsname{\color{black}}%
      \expandafter\def\csname LT8\endcsname{\color{black}}%
    \fi
  \fi
  \setlength{\unitlength}{0.0500bp}%
  \begin{picture}(4626.00,3148.00)%
    \gplgaddtomacro\gplbacktext{%
      \csname LTb\endcsname%
      \put(2274,2881){\makebox(0,0){\normalsize $\phantom{\eta}r_3\phantom{\eta}$}}%
    }%
    \gplgaddtomacro\gplfronttext{%
      \csname LTb\endcsname%
      \put(937,211){\makebox(0,0){\strut{}1.7}}%
      \put(1314,211){\makebox(0,0){\strut{}1.8}}%
      \put(1692,211){\makebox(0,0){\strut{}1.9}}%
      \put(2069,211){\makebox(0,0){\strut{}2.0}}%
      \put(2446,211){\makebox(0,0){\strut{}2.1}}%
      \put(2823,211){\makebox(0,0){\strut{}2.2}}%
      \put(3200,211){\makebox(0,0){\strut{}2.3}}%
      \put(3578,211){\makebox(0,0){\strut{}2.4}}%
      \put(3955,211){\makebox(0,0){\strut{}2.5}}%
      \put(2275,-159){\makebox(0,0){\normalsize $W$ (GeV)}}%
      \put(438,438){\makebox(0,0)[r]{\strut{}-1}}%
      \put(438,983){\makebox(0,0)[r]{\strut{}-0.5}}%
      \put(438,1528){\makebox(0,0)[r]{\strut{} 0}}%
      \put(438,2073){\makebox(0,0)[r]{\strut{} 0.5}}%
      \put(438,2618){\makebox(0,0)[r]{\strut{} 1}}%
      \put(-96,1528){\rotatebox{-270}{\makebox(0,0){\normalsize $\cos\thcm$}}}%
    }%
    \gplbacktext
    \put(490,353){\includegraphics[scale=0.66]{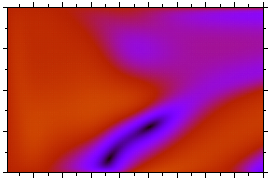}}%
    \gplfronttext
  \end{picture}%
\endgroup
\hspace{-41pt}
\begingroup
  \makeatletter
  \providecommand\color[2][]{%
    \GenericError{(gnuplot) \space\space\space\@spaces}{%
      Package color not loaded in conjunction with
      terminal option `colourtext'%
    }{See the gnuplot documentation for explanation.%
    }{Either use 'blacktext' in gnuplot or load the package
      color.sty in LaTeX.}%
    \renewcommand\color[2][]{}%
  }%
  \providecommand\includegraphics[2][]{%
    \GenericError{(gnuplot) \space\space\space\@spaces}{%
      Package graphicx or graphics not loaded%
    }{See the gnuplot documentation for explanation.%
    }{The gnuplot epslatex terminal needs graphicx.sty or graphics.sty.}%
    \renewcommand\includegraphics[2][]{}%
  }%
  \providecommand\rotatebox[2]{#2}%
  \@ifundefined{ifGPcolor}{%
    \newif\ifGPcolor
    \GPcolortrue
  }{}%
  \@ifundefined{ifGPblacktext}{%
    \newif\ifGPblacktext
    \GPblacktexttrue
  }{}%
  \let\gplgaddtomacro\g@addto@macro
  \gdef\gplbacktext{}%
  \gdef\gplfronttext{}%
  \makeatother
  \ifGPblacktext
    \def\colorrgb#1{}%
    \def\colorgray#1{}%
  \else
    \ifGPcolor
      \def\colorrgb#1{\color[rgb]{#1}}%
      \def\colorgray#1{\color[gray]{#1}}%
      \expandafter\def\csname LTw\endcsname{\color{white}}%
      \expandafter\def\csname LTb\endcsname{\color{black}}%
      \expandafter\def\csname LTa\endcsname{\color{black}}%
      \expandafter\def\csname LT0\endcsname{\color[rgb]{1,0,0}}%
      \expandafter\def\csname LT1\endcsname{\color[rgb]{0,1,0}}%
      \expandafter\def\csname LT2\endcsname{\color[rgb]{0,0,1}}%
      \expandafter\def\csname LT3\endcsname{\color[rgb]{1,0,1}}%
      \expandafter\def\csname LT4\endcsname{\color[rgb]{0,1,1}}%
      \expandafter\def\csname LT5\endcsname{\color[rgb]{1,1,0}}%
      \expandafter\def\csname LT6\endcsname{\color[rgb]{0,0,0}}%
      \expandafter\def\csname LT7\endcsname{\color[rgb]{1,0.3,0}}%
      \expandafter\def\csname LT8\endcsname{\color[rgb]{0.5,0.5,0.5}}%
    \else
      \def\colorrgb#1{\color{black}}%
      \def\colorgray#1{\color[gray]{#1}}%
      \expandafter\def\csname LTw\endcsname{\color{white}}%
      \expandafter\def\csname LTb\endcsname{\color{black}}%
      \expandafter\def\csname LTa\endcsname{\color{black}}%
      \expandafter\def\csname LT0\endcsname{\color{black}}%
      \expandafter\def\csname LT1\endcsname{\color{black}}%
      \expandafter\def\csname LT2\endcsname{\color{black}}%
      \expandafter\def\csname LT3\endcsname{\color{black}}%
      \expandafter\def\csname LT4\endcsname{\color{black}}%
      \expandafter\def\csname LT5\endcsname{\color{black}}%
      \expandafter\def\csname LT6\endcsname{\color{black}}%
      \expandafter\def\csname LT7\endcsname{\color{black}}%
      \expandafter\def\csname LT8\endcsname{\color{black}}%
    \fi
  \fi
  \setlength{\unitlength}{0.0500bp}%
  \begin{picture}(4626.00,3148.00)%
    \gplgaddtomacro\gplbacktext{%
      \csname LTb\endcsname%
      \put(2274,2881){\makebox(0,0){\normalsize $\phantom{\eta}r_4\phantom{\eta}$}}%
    }%
    \gplgaddtomacro\gplfronttext{%
      \csname LTb\endcsname%
      \put(937,211){\makebox(0,0){\strut{}1.7}}%
      \put(1314,211){\makebox(0,0){\strut{}1.8}}%
      \put(1692,211){\makebox(0,0){\strut{}1.9}}%
      \put(2069,211){\makebox(0,0){\strut{}2.0}}%
      \put(2446,211){\makebox(0,0){\strut{}2.1}}%
      \put(2823,211){\makebox(0,0){\strut{}2.2}}%
      \put(3200,211){\makebox(0,0){\strut{}2.3}}%
      \put(3578,211){\makebox(0,0){\strut{}2.4}}%
      \put(3955,211){\makebox(0,0){\strut{}2.5}}%
      \put(2275,-159){\makebox(0,0){\normalsize $W$ (GeV)}}%
      \put(438,438){\makebox(0,0)[r]{\strut{}}}%
      \put(438,983){\makebox(0,0)[r]{\strut{}}}%
      \put(438,1528){\makebox(0,0)[r]{\strut{}}}%
      \put(438,2073){\makebox(0,0)[r]{\strut{}}}%
      \put(438,2618){\makebox(0,0)[r]{\strut{}}}%
      \put(4272,437){\makebox(0,0)[l]{\strut{} 0}}%
      \put(4272,873){\makebox(0,0)[l]{\strut{} 0.2}}%
      \put(4272,1309){\makebox(0,0)[l]{\strut{} 0.4}}%
      \put(4272,1745){\makebox(0,0)[l]{\strut{} 0.6}}%
      \put(4272,2181){\makebox(0,0)[l]{\strut{} 0.8}}%
      \put(4272,2618){\makebox(0,0)[l]{\strut{} 1}}%
    }%
    \gplbacktext
    \put(500,353){\includegraphics[scale=0.66]{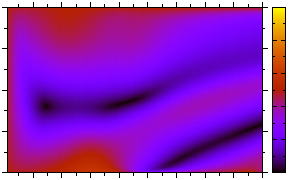}}%
    \gplfronttext
  \end{picture}%
\endgroup
\end{center}
\caption{(Color online) The energy and angular dependence of the moduli $r_i$ of the normalized transversity amplitudes for the $\gamma p \to  K^{+} \Lambda$ reaction. The calculations are performed with the RPR-2011 model ($\Delta W \approx 2.38$ MeV, $\Delta \cos \thcm \approx 8.33\times10^{-3}$).}
\label{fig:modulilow}
\end{figure*}

\begin{figure}[!t]
\begin{center}
\footnotesize
\begingroup
  \makeatletter
  \providecommand\color[2][]{%
    \GenericError{(gnuplot) \space\space\space\@spaces}{%
      Package color not loaded in conjunction with
      terminal option `colourtext'%
    }{See the gnuplot documentation for explanation.%
    }{Either use 'blacktext' in gnuplot or load the package
      color.sty in LaTeX.}%
    \renewcommand\color[2][]{}%
  }%
  \providecommand\includegraphics[2][]{%
    \GenericError{(gnuplot) \space\space\space\@spaces}{%
      Package graphicx or graphics not loaded%
    }{See the gnuplot documentation for explanation.%
    }{The gnuplot epslatex terminal needs graphicx.sty or graphics.sty.}%
    \renewcommand\includegraphics[2][]{}%
  }%
  \providecommand\rotatebox[2]{#2}%
  \@ifundefined{ifGPcolor}{%
    \newif\ifGPcolor
    \GPcolortrue
  }{}%
  \@ifundefined{ifGPblacktext}{%
    \newif\ifGPblacktext
    \GPblacktexttrue
  }{}%
  \let\gplgaddtomacro\g@addto@macro
  \gdef\gplbacktext{}%
  \gdef\gplfronttext{}%
  \makeatother
  \ifGPblacktext
    \def\colorrgb#1{}%
    \def\colorgray#1{}%
  \else
    \ifGPcolor
      \def\colorrgb#1{\color[rgb]{#1}}%
      \def\colorgray#1{\color[gray]{#1}}%
      \expandafter\def\csname LTw\endcsname{\color{white}}%
      \expandafter\def\csname LTb\endcsname{\color{black}}%
      \expandafter\def\csname LTa\endcsname{\color{black}}%
      \expandafter\def\csname LT0\endcsname{\color[rgb]{1,0,0}}%
      \expandafter\def\csname LT1\endcsname{\color[rgb]{0,1,0}}%
      \expandafter\def\csname LT2\endcsname{\color[rgb]{0,0,1}}%
      \expandafter\def\csname LT3\endcsname{\color[rgb]{1,0,1}}%
      \expandafter\def\csname LT4\endcsname{\color[rgb]{0,1,1}}%
      \expandafter\def\csname LT5\endcsname{\color[rgb]{1,1,0}}%
      \expandafter\def\csname LT6\endcsname{\color[rgb]{0,0,0}}%
      \expandafter\def\csname LT7\endcsname{\color[rgb]{1,0.3,0}}%
      \expandafter\def\csname LT8\endcsname{\color[rgb]{0.5,0.5,0.5}}%
    \else
      \def\colorrgb#1{\color{black}}%
      \def\colorgray#1{\color[gray]{#1}}%
      \expandafter\def\csname LTw\endcsname{\color{white}}%
      \expandafter\def\csname LTb\endcsname{\color{black}}%
      \expandafter\def\csname LTa\endcsname{\color{black}}%
      \expandafter\def\csname LT0\endcsname{\color{black}}%
      \expandafter\def\csname LT1\endcsname{\color{black}}%
      \expandafter\def\csname LT2\endcsname{\color{black}}%
      \expandafter\def\csname LT3\endcsname{\color{black}}%
      \expandafter\def\csname LT4\endcsname{\color{black}}%
      \expandafter\def\csname LT5\endcsname{\color{black}}%
      \expandafter\def\csname LT6\endcsname{\color{black}}%
      \expandafter\def\csname LT7\endcsname{\color{black}}%
      \expandafter\def\csname LT8\endcsname{\color{black}}%
    \fi
  \fi
  \setlength{\unitlength}{0.0500bp}%
  \begin{picture}(4626.00,3148.00)%
    \gplgaddtomacro\gplbacktext{%
      \csname LTb\endcsname%
      \put(2274,2883){\makebox(0,0){\normalsize $\phantom{\eta}\delta_1\phantom{\eta}$}}%
    }%
    \gplgaddtomacro\gplfronttext{%
      \csname LTb\endcsname%
      \put(937,209){\makebox(0,0){\strut{}}}%
      \put(1314,209){\makebox(0,0){\strut{}}}%
      \put(1692,209){\makebox(0,0){\strut{}}}%
      \put(2069,209){\makebox(0,0){\strut{}}}%
      \put(2446,209){\makebox(0,0){\strut{}}}%
      \put(2823,209){\makebox(0,0){\strut{}}}%
      \put(3200,209){\makebox(0,0){\strut{}}}%
      \put(3578,209){\makebox(0,0){\strut{}}}%
      \put(3955,209){\makebox(0,0){\strut{}}}%
      \put(438,436){\makebox(0,0)[r]{\strut{}-1}}%
      \put(438,977){\makebox(0,0)[r]{\strut{}-0.5}}%
      \put(438,1518){\makebox(0,0)[r]{\strut{} 0}}%
      \put(438,2059){\makebox(0,0)[r]{\strut{} 0.5}}%
      \put(438,2600){\makebox(0,0)[r]{\strut{} 1}}%
      \put(-96,1518){\rotatebox{-270}{\makebox(0,0){\normalsize $\cos\thcm$}}}%
      \put(4343,435){\makebox(0,0)[l]{\strut{}$0$}}%
      \put(4343,976){\makebox(0,0)[l]{\strut{}$\dfrac{\pi}{2}$}}%
      \put(4343,1517){\makebox(0,0)[l]{\strut{}$\pi$}}%
      \put(4343,2058){\makebox(0,0)[l]{\strut{}$\dfrac{3\pi}{2}$}}%
      \put(4343,2600){\makebox(0,0)[l]{\strut{}$2\pi$}}%
      \put(4343,435){\makebox(0,0)[l]{\strut{}}}%
      \put(4343,976){\makebox(0,0)[l]{\strut{}}}%
      \put(4343,1517){\makebox(0,0)[l]{\strut{}}}%
      \put(4343,2058){\makebox(0,0)[l]{\strut{}}}%
      \put(4343,2600){\makebox(0,0)[l]{\strut{}}}%
    }%
    \gplbacktext
    \put(500,353){\includegraphics[scale=0.66]{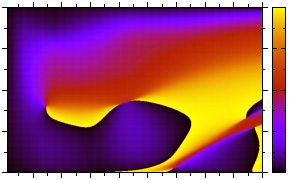}}%
    \gplfronttext
  \end{picture}%
\endgroup\vspace{-15pt}
\begingroup
  \makeatletter
  \providecommand\color[2][]{%
    \GenericError{(gnuplot) \space\space\space\@spaces}{%
      Package color not loaded in conjunction with
      terminal option `colourtext'%
    }{See the gnuplot documentation for explanation.%
    }{Either use 'blacktext' in gnuplot or load the package
      color.sty in LaTeX.}%
    \renewcommand\color[2][]{}%
  }%
  \providecommand\includegraphics[2][]{%
    \GenericError{(gnuplot) \space\space\space\@spaces}{%
      Package graphicx or graphics not loaded%
    }{See the gnuplot documentation for explanation.%
    }{The gnuplot epslatex terminal needs graphicx.sty or graphics.sty.}%
    \renewcommand\includegraphics[2][]{}%
  }%
  \providecommand\rotatebox[2]{#2}%
  \@ifundefined{ifGPcolor}{%
    \newif\ifGPcolor
    \GPcolortrue
  }{}%
  \@ifundefined{ifGPblacktext}{%
    \newif\ifGPblacktext
    \GPblacktexttrue
  }{}%
  \let\gplgaddtomacro\g@addto@macro
  \gdef\gplbacktext{}%
  \gdef\gplfronttext{}%
  \makeatother
  \ifGPblacktext
    \def\colorrgb#1{}%
    \def\colorgray#1{}%
  \else
    \ifGPcolor
      \def\colorrgb#1{\color[rgb]{#1}}%
      \def\colorgray#1{\color[gray]{#1}}%
      \expandafter\def\csname LTw\endcsname{\color{white}}%
      \expandafter\def\csname LTb\endcsname{\color{black}}%
      \expandafter\def\csname LTa\endcsname{\color{black}}%
      \expandafter\def\csname LT0\endcsname{\color[rgb]{1,0,0}}%
      \expandafter\def\csname LT1\endcsname{\color[rgb]{0,1,0}}%
      \expandafter\def\csname LT2\endcsname{\color[rgb]{0,0,1}}%
      \expandafter\def\csname LT3\endcsname{\color[rgb]{1,0,1}}%
      \expandafter\def\csname LT4\endcsname{\color[rgb]{0,1,1}}%
      \expandafter\def\csname LT5\endcsname{\color[rgb]{1,1,0}}%
      \expandafter\def\csname LT6\endcsname{\color[rgb]{0,0,0}}%
      \expandafter\def\csname LT7\endcsname{\color[rgb]{1,0.3,0}}%
      \expandafter\def\csname LT8\endcsname{\color[rgb]{0.5,0.5,0.5}}%
    \else
      \def\colorrgb#1{\color{black}}%
      \def\colorgray#1{\color[gray]{#1}}%
      \expandafter\def\csname LTw\endcsname{\color{white}}%
      \expandafter\def\csname LTb\endcsname{\color{black}}%
      \expandafter\def\csname LTa\endcsname{\color{black}}%
      \expandafter\def\csname LT0\endcsname{\color{black}}%
      \expandafter\def\csname LT1\endcsname{\color{black}}%
      \expandafter\def\csname LT2\endcsname{\color{black}}%
      \expandafter\def\csname LT3\endcsname{\color{black}}%
      \expandafter\def\csname LT4\endcsname{\color{black}}%
      \expandafter\def\csname LT5\endcsname{\color{black}}%
      \expandafter\def\csname LT6\endcsname{\color{black}}%
      \expandafter\def\csname LT7\endcsname{\color{black}}%
      \expandafter\def\csname LT8\endcsname{\color{black}}%
    \fi
  \fi
  \setlength{\unitlength}{0.0500bp}%
  \begin{picture}(4626.00,3148.00)%
    \gplgaddtomacro\gplbacktext{%
      \csname LTb\endcsname%
      \put(2274,2883){\makebox(0,0){\normalsize $\phantom{\eta}\delta_2\phantom{\eta}$}}%
    }%
    \gplgaddtomacro\gplfronttext{%
      \csname LTb\endcsname%
      \put(937,209){\makebox(0,0){\strut{}}}%
      \put(1314,209){\makebox(0,0){\strut{}}}%
      \put(1692,209){\makebox(0,0){\strut{}}}%
      \put(2069,209){\makebox(0,0){\strut{}}}%
      \put(2446,209){\makebox(0,0){\strut{}}}%
      \put(2823,209){\makebox(0,0){\strut{}}}%
      \put(3200,209){\makebox(0,0){\strut{}}}%
      \put(3578,209){\makebox(0,0){\strut{}}}%
      \put(3955,209){\makebox(0,0){\strut{}}}%
      \put(438,436){\makebox(0,0)[r]{\strut{}-1}}%
      \put(438,977){\makebox(0,0)[r]{\strut{}-0.5}}%
      \put(438,1518){\makebox(0,0)[r]{\strut{} 0}}%
      \put(438,2059){\makebox(0,0)[r]{\strut{} 0.5}}%
      \put(438,2600){\makebox(0,0)[r]{\strut{} 1}}%
      \put(-96,1518){\rotatebox{-270}{\makebox(0,0){\normalsize $\cos\thcm$}}}%
      \put(4343,435){\makebox(0,0)[l]{\strut{}$0$}}%
      \put(4343,976){\makebox(0,0)[l]{\strut{}$\dfrac{\pi}{2}$}}%
      \put(4343,1517){\makebox(0,0)[l]{\strut{}$\pi$}}%
      \put(4343,2058){\makebox(0,0)[l]{\strut{}$\dfrac{3\pi}{2}$}}%
      \put(4343,2600){\makebox(0,0)[l]{\strut{}$2\pi$}}%
      \put(4343,435){\makebox(0,0)[l]{\strut{}}}%
      \put(4343,976){\makebox(0,0)[l]{\strut{}}}%
      \put(4343,1517){\makebox(0,0)[l]{\strut{}}}%
      \put(4343,2058){\makebox(0,0)[l]{\strut{}}}%
      \put(4343,2600){\makebox(0,0)[l]{\strut{}}}%
    }%
    \gplbacktext
    \put(500,353){\includegraphics[scale=0.66]{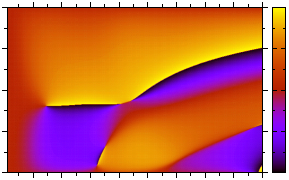}}%
    \gplfronttext
  \end{picture}%
\endgroup\vspace{-15pt}
\begingroup
  \makeatletter
  \providecommand\color[2][]{%
    \GenericError{(gnuplot) \space\space\space\@spaces}{%
      Package color not loaded in conjunction with
      terminal option `colourtext'%
    }{See the gnuplot documentation for explanation.%
    }{Either use 'blacktext' in gnuplot or load the package
      color.sty in LaTeX.}%
    \renewcommand\color[2][]{}%
  }%
  \providecommand\includegraphics[2][]{%
    \GenericError{(gnuplot) \space\space\space\@spaces}{%
      Package graphicx or graphics not loaded%
    }{See the gnuplot documentation for explanation.%
    }{The gnuplot epslatex terminal needs graphicx.sty or graphics.sty.}%
    \renewcommand\includegraphics[2][]{}%
  }%
  \providecommand\rotatebox[2]{#2}%
  \@ifundefined{ifGPcolor}{%
    \newif\ifGPcolor
    \GPcolortrue
  }{}%
  \@ifundefined{ifGPblacktext}{%
    \newif\ifGPblacktext
    \GPblacktexttrue
  }{}%
  \let\gplgaddtomacro\g@addto@macro
  \gdef\gplbacktext{}%
  \gdef\gplfronttext{}%
  \makeatother
  \ifGPblacktext
    \def\colorrgb#1{}%
    \def\colorgray#1{}%
  \else
    \ifGPcolor
      \def\colorrgb#1{\color[rgb]{#1}}%
      \def\colorgray#1{\color[gray]{#1}}%
      \expandafter\def\csname LTw\endcsname{\color{white}}%
      \expandafter\def\csname LTb\endcsname{\color{black}}%
      \expandafter\def\csname LTa\endcsname{\color{black}}%
      \expandafter\def\csname LT0\endcsname{\color[rgb]{1,0,0}}%
      \expandafter\def\csname LT1\endcsname{\color[rgb]{0,1,0}}%
      \expandafter\def\csname LT2\endcsname{\color[rgb]{0,0,1}}%
      \expandafter\def\csname LT3\endcsname{\color[rgb]{1,0,1}}%
      \expandafter\def\csname LT4\endcsname{\color[rgb]{0,1,1}}%
      \expandafter\def\csname LT5\endcsname{\color[rgb]{1,1,0}}%
      \expandafter\def\csname LT6\endcsname{\color[rgb]{0,0,0}}%
      \expandafter\def\csname LT7\endcsname{\color[rgb]{1,0.3,0}}%
      \expandafter\def\csname LT8\endcsname{\color[rgb]{0.5,0.5,0.5}}%
    \else
      \def\colorrgb#1{\color{black}}%
      \def\colorgray#1{\color[gray]{#1}}%
      \expandafter\def\csname LTw\endcsname{\color{white}}%
      \expandafter\def\csname LTb\endcsname{\color{black}}%
      \expandafter\def\csname LTa\endcsname{\color{black}}%
      \expandafter\def\csname LT0\endcsname{\color{black}}%
      \expandafter\def\csname LT1\endcsname{\color{black}}%
      \expandafter\def\csname LT2\endcsname{\color{black}}%
      \expandafter\def\csname LT3\endcsname{\color{black}}%
      \expandafter\def\csname LT4\endcsname{\color{black}}%
      \expandafter\def\csname LT5\endcsname{\color{black}}%
      \expandafter\def\csname LT6\endcsname{\color{black}}%
      \expandafter\def\csname LT7\endcsname{\color{black}}%
      \expandafter\def\csname LT8\endcsname{\color{black}}%
    \fi
  \fi
  \setlength{\unitlength}{0.0500bp}%
  \begin{picture}(4626.00,3148.00)%
    \gplgaddtomacro\gplbacktext{%
      \csname LTb\endcsname%
      \put(2274,2883){\makebox(0,0){\normalsize $\phantom{\eta}\delta_3\phantom{\eta}$}}%
    }%
    \gplgaddtomacro\gplfronttext{%
      \csname LTb\endcsname%
      \put(937,209){\makebox(0,0){\strut{}1.7}}%
      \put(1314,209){\makebox(0,0){\strut{}1.8}}%
      \put(1692,209){\makebox(0,0){\strut{}1.9}}%
      \put(2069,209){\makebox(0,0){\strut{}2.0}}%
      \put(2446,209){\makebox(0,0){\strut{}2.1}}%
      \put(2823,209){\makebox(0,0){\strut{}2.2}}%
      \put(3200,209){\makebox(0,0){\strut{}2.3}}%
      \put(3578,209){\makebox(0,0){\strut{}2.4}}%
      \put(3955,209){\makebox(0,0){\strut{}2.5}}%
      \put(2275,-161){\makebox(0,0){\normalsize $W$ (GeV)}}%
      \put(438,436){\makebox(0,0)[r]{\strut{}-1}}%
      \put(438,977){\makebox(0,0)[r]{\strut{}-0.5}}%
      \put(438,1518){\makebox(0,0)[r]{\strut{} 0}}%
      \put(438,2059){\makebox(0,0)[r]{\strut{} 0.5}}%
      \put(438,2600){\makebox(0,0)[r]{\strut{} 1}}%
      \put(-96,1518){\rotatebox{-270}{\makebox(0,0){\normalsize $\cos\thcm$}}}%
      \put(4343,435){\makebox(0,0)[l]{\strut{}$0$}}%
      \put(4343,976){\makebox(0,0)[l]{\strut{}$\dfrac{\pi}{2}$}}%
      \put(4343,1517){\makebox(0,0)[l]{\strut{}$\pi$}}%
      \put(4343,2058){\makebox(0,0)[l]{\strut{}$\dfrac{3\pi}{2}$}}%
      \put(4343,2600){\makebox(0,0)[l]{\strut{}$2\pi$}}%
      \put(4343,435){\makebox(0,0)[l]{\strut{}}}%
      \put(4343,976){\makebox(0,0)[l]{\strut{}}}%
      \put(4343,1517){\makebox(0,0)[l]{\strut{}}}%
      \put(4343,2058){\makebox(0,0)[l]{\strut{}}}%
      \put(4343,2600){\makebox(0,0)[l]{\strut{}}}%
    }%
    \gplbacktext
    \put(500,353){\includegraphics[scale=0.66]{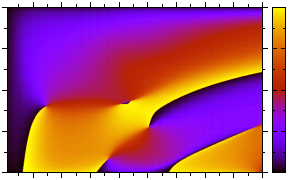}}%
    \gplfronttext
  \end{picture}%
\endgroup
\end{center}
\caption{(Color online) The energy and angular dependence of the independent phases $\delta_i$ of the normalized transversity amplitudes of the $\gamma p \to K^{+} \Lambda$ reaction. The calculations are performed with the RPR-2011 model ($\Delta W \approx 1.43$ MeV, $\Delta \cos \thcm = 5\times10^{-3}$).}
\label{fig:phaseslow}
\end{figure}

One would like to have a consistent set of estimators for the independent phases $\delta_i^{\;\alpha_j}$ ($i \neq j$) and $\delta_i^{\;\alpha_4}\equiv \delta_i$ ($i = 1,2,3$). The notation $\widetilde{\delta}_i^{\;\alpha_j}$ is adopted for the consistent estimators. The aforementioned reference-phase ambiguity can be resolved by imposing that
\begin{align}
\widetilde{\delta}_i^{\;\alpha_j} = \widetilde{\delta}_i^{\;\alpha_k} - \widetilde{\delta}_j^{\;\alpha_k}. \label{eq:consistency_property}
\end{align}
Since this requirement is linear in the independent phases, the consistent estimators are also linear in $\{\widehat{\delta}_i^{\;\alpha_l}, \widehat{\delta}_j^{\;\alpha_l}, \widehat{\Delta}_{ik}, \widehat{\Delta}_{jk}\}$ $\left( i \neq j \neq k \neq l \right)$. The most general expression for a consistent set of estimated phases, reads 
\begin{align}
\begin{cases}
\widetilde{\delta}_i^{\;\alpha_l} = c_1 \widehat{\delta}_i^{\;\alpha_l} + (1 - c_1)\bigl(\widehat{\delta}_j^{\;\alpha_l} + \widehat{\Delta}_{ik} - \widehat{\Delta}_{jk}\bigr),\vspace{5pt}\\
\widetilde{\delta}_j^{\;\alpha_l} = c_1 \widehat{\delta}_j^{\;\alpha_l} + (1 - c_1)\bigl(\widehat{\delta}_i^{\;\alpha_l} + \widehat{\Delta}_{jk} - \widehat{\Delta}_{ik}\bigr),\vspace{5pt}\\
\widetilde{\delta}_k^{\;\alpha_l} = c_2\bigl(\widehat{\delta}_i^{\;\alpha_l} - \widehat{\Delta}_{ik}\bigr) + (1 - c_2)\bigl(\widehat{\delta}_j^{\;\alpha_l} - \widehat{\Delta}_{jk}\bigr).
\end{cases}
\label{eq:consistent_set_coeff}
\end{align}
Here, three constraints have been imposed. Firstly, the $\widetilde{\delta}_{\{i,j,k\}}^{\;\alpha_l}$ should yield the exact set $\delta_{\{i,j,k\}}^{\;\alpha_l}$ for vanishing experimental error bars. Secondly, the expressions for $\widetilde{\delta}_i^{\;\alpha_l}$ and $\widetilde{\delta}_j^{\;\alpha_l}$ should be equal for $i \leftrightarrow j$. This leaves one with two unknowns $c_1$ and $c_2$. Thirdly, $c_1$ and $c_2$ cannot depend on $\alpha_l$, which is implied by Eq.~\eq{eq:consistency_property}. By applying Eq.~\eq{eq:consistency_property} on expressions \eq{eq:consistent_set_coeff} and by using Table~\ref{tab:reference_phases}, one readily finds that $c_1 = \frac{3}{4}$ and $c_2 = \frac{1}{2}$. Therefore, the consistent set of estimators for the phases, reads
\begin{align}
\begin{cases}
\widetilde{\delta}_i^{\;\alpha_l} = \frac{3}{4}\widehat{\delta}_i^{\;\alpha_l} + \frac{1}{4}\bigl(\widehat{\delta}_j^{\;\alpha_l} + \widehat{\Delta}_{ik} - \widehat{\Delta}_{jk}\bigr),\vspace{5pt}\\
\widetilde{\delta}_j^{\;\alpha_l} = \frac{3}{4}\widehat{\delta}_j^{\;\alpha_l} + \frac{1}{4}\bigl(\widehat{\delta}_i^{\;\alpha_l} + \widehat{\Delta}_{jk} - \widehat{\Delta}_{ik}\bigr),\vspace{5pt}\\
\widetilde{\delta}_k^{\;\alpha_l} = \frac{1}{2}\bigl(\widehat{\delta}_i^{\;\alpha_l} - \widehat{\Delta}_{ik}\bigr) + \frac{1}{2}\bigl(\widehat{\delta}_j^{\;\alpha_l} - \widehat{\Delta}_{jk}\bigr).
\end{cases}
\label{eq:consistent_set}
\end{align}
The estimates $\widetilde{\delta}_i$'s ($i = 1,2,3$) for the independent phases are now insensitive to the choices made with regard to the reference phase.

\section{Results}
\label{sec:results}

\subsection{The angular and energy dependence of the transversity amplitudes}
\label{subsec:Eandtheta}

A complete measurement comprises a minimal set of asymmetries from which the accessible parameters of the normalized transversity amplitudes $a_{i}$ can be estimated. These parameters include three independent moduli $r_i$ and three independent relative phases $\delta _i$. In Sec.\ \ref{sec:formalism}, it was shown how the moduli can be obtained from the single asymmetries and how a set of three independent phases can be estimated in a consistent way from a complete measurement which also involves double asymmetries. 

Figures \ref{fig:modulilow} and \ref{fig:phaseslow} show the ($W$, $\cos \thcm$) dependence of the $r_i$ and the $\delta_i$ for the $\gamma p \to K ^{+} \Lambda$ reaction as predicted by a realistic model, namely the latest version of the Regge-plus-Resonance (RPR) model, i.e.\ RPR-2011 \cite{tamaraprc,lesleyprl, lesleyprc}. This model has a Reggeized $t$-channel background and includes a total of 8 $s$-channel resonances, namely $S_{11}(1535)$,  $S_{11}(1650)$, $F_{15}(1680)$, $P_{13}(1720)$, $P_{11}(1900)$, $P_{13}(1900)$, $D_{13}(1900)$, and $F_{15}(2000)$. The most apparent feature of the RPR-2011 predictions for the moduli and phases are the strong variations with energy $W$ at backward scattering angles, due to $s$-channel resonances. The smooth energy dependence of the moduli and phases at very forward scattering angles reflects the important $t$-channel background contributions of $\gamma p \to K ^{+} \Lambda$. At forward kaon angles, where most of the strength resides, the RPR-2011 
model predicts a dominant role for $r_{2}$.

\begin{figure*}[!t]
\begin{center}
\footnotesize
\hspace{23pt}
\begingroup
  \makeatletter
  \providecommand\color[2][]{%
    \GenericError{(gnuplot) \space\space\space\@spaces}{%
      Package color not loaded in conjunction with
      terminal option `colourtext'%
    }{See the gnuplot documentation for explanation.%
    }{Either use 'blacktext' in gnuplot or load the package
      color.sty in LaTeX.}%
    \renewcommand\color[2][]{}%
  }%
  \providecommand\includegraphics[2][]{%
    \GenericError{(gnuplot) \space\space\space\@spaces}{%
      Package graphicx or graphics not loaded%
    }{See the gnuplot documentation for explanation.%
    }{The gnuplot epslatex terminal needs graphicx.sty or graphics.sty.}%
    \renewcommand\includegraphics[2][]{}%
  }%
  \providecommand\rotatebox[2]{#2}%
  \@ifundefined{ifGPcolor}{%
    \newif\ifGPcolor
    \GPcolortrue
  }{}%
  \@ifundefined{ifGPblacktext}{%
    \newif\ifGPblacktext
    \GPblacktexttrue
  }{}%
  \let\gplgaddtomacro\g@addto@macro
  \gdef\gplbacktext{}%
  \gdef\gplfronttext{}%
  \makeatother
  \ifGPblacktext
    \def\colorrgb#1{}%
    \def\colorgray#1{}%
  \else
    \ifGPcolor
      \def\colorrgb#1{\color[rgb]{#1}}%
      \def\colorgray#1{\color[gray]{#1}}%
      \expandafter\def\csname LTw\endcsname{\color{white}}%
      \expandafter\def\csname LTb\endcsname{\color{black}}%
      \expandafter\def\csname LTa\endcsname{\color{black}}%
      \expandafter\def\csname LT0\endcsname{\color[rgb]{1,0,0}}%
      \expandafter\def\csname LT1\endcsname{\color[rgb]{0,1,0}}%
      \expandafter\def\csname LT2\endcsname{\color[rgb]{0,0,1}}%
      \expandafter\def\csname LT3\endcsname{\color[rgb]{1,0,1}}%
      \expandafter\def\csname LT4\endcsname{\color[rgb]{0,1,1}}%
      \expandafter\def\csname LT5\endcsname{\color[rgb]{1,1,0}}%
      \expandafter\def\csname LT6\endcsname{\color[rgb]{0,0,0}}%
      \expandafter\def\csname LT7\endcsname{\color[rgb]{1,0.3,0}}%
      \expandafter\def\csname LT8\endcsname{\color[rgb]{0.5,0.5,0.5}}%
    \else
      \def\colorrgb#1{\color{black}}%
      \def\colorgray#1{\color[gray]{#1}}%
      \expandafter\def\csname LTw\endcsname{\color{white}}%
      \expandafter\def\csname LTb\endcsname{\color{black}}%
      \expandafter\def\csname LTa\endcsname{\color{black}}%
      \expandafter\def\csname LT0\endcsname{\color{black}}%
      \expandafter\def\csname LT1\endcsname{\color{black}}%
      \expandafter\def\csname LT2\endcsname{\color{black}}%
      \expandafter\def\csname LT3\endcsname{\color{black}}%
      \expandafter\def\csname LT4\endcsname{\color{black}}%
      \expandafter\def\csname LT5\endcsname{\color{black}}%
      \expandafter\def\csname LT6\endcsname{\color{black}}%
      \expandafter\def\csname LT7\endcsname{\color{black}}%
      \expandafter\def\csname LT8\endcsname{\color{black}}%
    \fi
  \fi
  \setlength{\unitlength}{0.0500bp}%
  \begin{picture}(2744.00,1346.00)%
    \gplgaddtomacro\gplbacktext{%
      \csname LTb\endcsname%
      \put(-60,182){\makebox(0,0)[r]{\strut{} 0}}%
      \put(-60,425){\makebox(0,0)[r]{\strut{} 0.2}}%
      \put(-60,668){\makebox(0,0)[r]{\strut{} 0.4}}%
      \put(-60,912){\makebox(0,0)[r]{\strut{} 0.6}}%
      \put(-60,1155){\makebox(0,0)[r]{\strut{} 0.8}}%
      \put(239,-120){\makebox(0,0){\strut{}}}%
      \put(692,-120){\makebox(0,0){\strut{}}}%
      \put(1145,-120){\makebox(0,0){\strut{}}}%
      \put(1598,-120){\makebox(0,0){\strut{}}}%
      \put(2051,-120){\makebox(0,0){\strut{}}}%
      \put(2504,-120){\makebox(0,0){\strut{}}}%
      \put(-550,607){\rotatebox{-270}{\makebox(0,0){\normalsize $r_1$}}}%
      \put(1371,1315){\makebox(0,0){\normalsize $0.808 \lesssim \cos\thcm \lesssim 0.861 $}}%
    }%
    \gplgaddtomacro\gplfronttext{%
      \csname LTb\endcsname%
      \put(2291,985){\makebox(0,0)[r]{\strut{}RPR-2011\;}}%
      \csname LTb\endcsname%
      \put(2291,749){\makebox(0,0)[r]{\strut{}GRAAL\;}}%
    }%
    \gplbacktext
    \put(0,0){\includegraphics{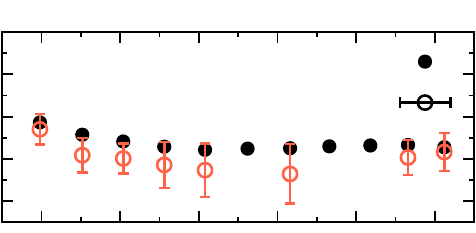}}%
    \gplfronttext
  \end{picture}%
\endgroup\hspace{20pt}
\begingroup
  \makeatletter
  \providecommand\color[2][]{%
    \GenericError{(gnuplot) \space\space\space\@spaces}{%
      Package color not loaded in conjunction with
      terminal option `colourtext'%
    }{See the gnuplot documentation for explanation.%
    }{Either use 'blacktext' in gnuplot or load the package
      color.sty in LaTeX.}%
    \renewcommand\color[2][]{}%
  }%
  \providecommand\includegraphics[2][]{%
    \GenericError{(gnuplot) \space\space\space\@spaces}{%
      Package graphicx or graphics not loaded%
    }{See the gnuplot documentation for explanation.%
    }{The gnuplot epslatex terminal needs graphicx.sty or graphics.sty.}%
    \renewcommand\includegraphics[2][]{}%
  }%
  \providecommand\rotatebox[2]{#2}%
  \@ifundefined{ifGPcolor}{%
    \newif\ifGPcolor
    \GPcolortrue
  }{}%
  \@ifundefined{ifGPblacktext}{%
    \newif\ifGPblacktext
    \GPblacktexttrue
  }{}%
  \let\gplgaddtomacro\g@addto@macro
  \gdef\gplbacktext{}%
  \gdef\gplfronttext{}%
  \makeatother
  \ifGPblacktext
    \def\colorrgb#1{}%
    \def\colorgray#1{}%
  \else
    \ifGPcolor
      \def\colorrgb#1{\color[rgb]{#1}}%
      \def\colorgray#1{\color[gray]{#1}}%
      \expandafter\def\csname LTw\endcsname{\color{white}}%
      \expandafter\def\csname LTb\endcsname{\color{black}}%
      \expandafter\def\csname LTa\endcsname{\color{black}}%
      \expandafter\def\csname LT0\endcsname{\color[rgb]{1,0,0}}%
      \expandafter\def\csname LT1\endcsname{\color[rgb]{0,1,0}}%
      \expandafter\def\csname LT2\endcsname{\color[rgb]{0,0,1}}%
      \expandafter\def\csname LT3\endcsname{\color[rgb]{1,0,1}}%
      \expandafter\def\csname LT4\endcsname{\color[rgb]{0,1,1}}%
      \expandafter\def\csname LT5\endcsname{\color[rgb]{1,1,0}}%
      \expandafter\def\csname LT6\endcsname{\color[rgb]{0,0,0}}%
      \expandafter\def\csname LT7\endcsname{\color[rgb]{1,0.3,0}}%
      \expandafter\def\csname LT8\endcsname{\color[rgb]{0.5,0.5,0.5}}%
    \else
      \def\colorrgb#1{\color{black}}%
      \def\colorgray#1{\color[gray]{#1}}%
      \expandafter\def\csname LTw\endcsname{\color{white}}%
      \expandafter\def\csname LTb\endcsname{\color{black}}%
      \expandafter\def\csname LTa\endcsname{\color{black}}%
      \expandafter\def\csname LT0\endcsname{\color{black}}%
      \expandafter\def\csname LT1\endcsname{\color{black}}%
      \expandafter\def\csname LT2\endcsname{\color{black}}%
      \expandafter\def\csname LT3\endcsname{\color{black}}%
      \expandafter\def\csname LT4\endcsname{\color{black}}%
      \expandafter\def\csname LT5\endcsname{\color{black}}%
      \expandafter\def\csname LT6\endcsname{\color{black}}%
      \expandafter\def\csname LT7\endcsname{\color{black}}%
      \expandafter\def\csname LT8\endcsname{\color{black}}%
    \fi
  \fi
  \setlength{\unitlength}{0.0500bp}%
  \begin{picture}(2744.00,1346.00)%
    \gplgaddtomacro\gplbacktext{%
      \csname LTb\endcsname%
      \put(-60,243){\makebox(0,0)[r]{\strut{} 0}}%
      \put(-60,608){\makebox(0,0)[r]{\strut{} 0.2}}%
      \put(-60,973){\makebox(0,0)[r]{\strut{} 0.4}}%
      \put(239,-120){\makebox(0,0){\strut{}}}%
      \put(692,-120){\makebox(0,0){\strut{}}}%
      \put(1145,-120){\makebox(0,0){\strut{}}}%
      \put(1598,-120){\makebox(0,0){\strut{}}}%
      \put(2051,-120){\makebox(0,0){\strut{}}}%
      \put(2504,-120){\makebox(0,0){\strut{}}}%
      \put(1371,1315){\makebox(0,0){\normalsize $0.145 \lesssim \cos\thcm \lesssim 0.173 $}}%
    }%
    \gplgaddtomacro\gplfronttext{%
    }%
    \gplbacktext
    \put(0,0){\includegraphics{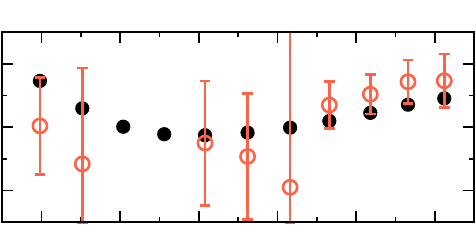}}%
    \gplfronttext
  \end{picture}%
\endgroup\hspace{20pt}
\begingroup
  \makeatletter
  \providecommand\color[2][]{%
    \GenericError{(gnuplot) \space\space\space\@spaces}{%
      Package color not loaded in conjunction with
      terminal option `colourtext'%
    }{See the gnuplot documentation for explanation.%
    }{Either use 'blacktext' in gnuplot or load the package
      color.sty in LaTeX.}%
    \renewcommand\color[2][]{}%
  }%
  \providecommand\includegraphics[2][]{%
    \GenericError{(gnuplot) \space\space\space\@spaces}{%
      Package graphicx or graphics not loaded%
    }{See the gnuplot documentation for explanation.%
    }{The gnuplot epslatex terminal needs graphicx.sty or graphics.sty.}%
    \renewcommand\includegraphics[2][]{}%
  }%
  \providecommand\rotatebox[2]{#2}%
  \@ifundefined{ifGPcolor}{%
    \newif\ifGPcolor
    \GPcolortrue
  }{}%
  \@ifundefined{ifGPblacktext}{%
    \newif\ifGPblacktext
    \GPblacktexttrue
  }{}%
  \let\gplgaddtomacro\g@addto@macro
  \gdef\gplbacktext{}%
  \gdef\gplfronttext{}%
  \makeatother
  \ifGPblacktext
    \def\colorrgb#1{}%
    \def\colorgray#1{}%
  \else
    \ifGPcolor
      \def\colorrgb#1{\color[rgb]{#1}}%
      \def\colorgray#1{\color[gray]{#1}}%
      \expandafter\def\csname LTw\endcsname{\color{white}}%
      \expandafter\def\csname LTb\endcsname{\color{black}}%
      \expandafter\def\csname LTa\endcsname{\color{black}}%
      \expandafter\def\csname LT0\endcsname{\color[rgb]{1,0,0}}%
      \expandafter\def\csname LT1\endcsname{\color[rgb]{0,1,0}}%
      \expandafter\def\csname LT2\endcsname{\color[rgb]{0,0,1}}%
      \expandafter\def\csname LT3\endcsname{\color[rgb]{1,0,1}}%
      \expandafter\def\csname LT4\endcsname{\color[rgb]{0,1,1}}%
      \expandafter\def\csname LT5\endcsname{\color[rgb]{1,1,0}}%
      \expandafter\def\csname LT6\endcsname{\color[rgb]{0,0,0}}%
      \expandafter\def\csname LT7\endcsname{\color[rgb]{1,0.3,0}}%
      \expandafter\def\csname LT8\endcsname{\color[rgb]{0.5,0.5,0.5}}%
    \else
      \def\colorrgb#1{\color{black}}%
      \def\colorgray#1{\color[gray]{#1}}%
      \expandafter\def\csname LTw\endcsname{\color{white}}%
      \expandafter\def\csname LTb\endcsname{\color{black}}%
      \expandafter\def\csname LTa\endcsname{\color{black}}%
      \expandafter\def\csname LT0\endcsname{\color{black}}%
      \expandafter\def\csname LT1\endcsname{\color{black}}%
      \expandafter\def\csname LT2\endcsname{\color{black}}%
      \expandafter\def\csname LT3\endcsname{\color{black}}%
      \expandafter\def\csname LT4\endcsname{\color{black}}%
      \expandafter\def\csname LT5\endcsname{\color{black}}%
      \expandafter\def\csname LT6\endcsname{\color{black}}%
      \expandafter\def\csname LT7\endcsname{\color{black}}%
      \expandafter\def\csname LT8\endcsname{\color{black}}%
    \fi
  \fi
  \setlength{\unitlength}{0.0500bp}%
  \begin{picture}(2744.00,1346.00)%
    \gplgaddtomacro\gplbacktext{%
      \csname LTb\endcsname%
      \put(-60,60){\makebox(0,0)[r]{\strut{} 0.2}}%
      \put(-60,373){\makebox(0,0)[r]{\strut{} 0.4}}%
      \put(-60,686){\makebox(0,0)[r]{\strut{} 0.6}}%
      \put(-60,999){\makebox(0,0)[r]{\strut{} 0.8}}%
      \put(239,-120){\makebox(0,0){\strut{}}}%
      \put(692,-120){\makebox(0,0){\strut{}}}%
      \put(1145,-120){\makebox(0,0){\strut{}}}%
      \put(1598,-120){\makebox(0,0){\strut{}}}%
      \put(2051,-120){\makebox(0,0){\strut{}}}%
      \put(2504,-120){\makebox(0,0){\strut{}}}%
      \put(1371,1315){\makebox(0,0){\normalsize $-0.806 \lesssim \cos\thcm \lesssim -0.750 $}}%
    }%
    \gplgaddtomacro\gplfronttext{%
    }%
    \gplbacktext
    \put(0,0){\includegraphics{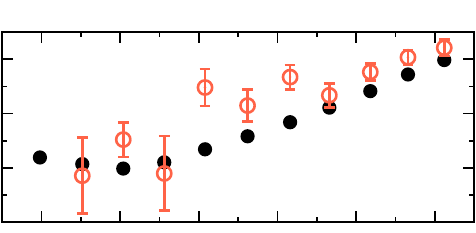}}%
    \gplfronttext
  \end{picture}%
\endgroup
\\ $ $ \\
\hspace{23pt}
\begingroup
  \makeatletter
  \providecommand\color[2][]{%
    \GenericError{(gnuplot) \space\space\space\@spaces}{%
      Package color not loaded in conjunction with
      terminal option `colourtext'%
    }{See the gnuplot documentation for explanation.%
    }{Either use 'blacktext' in gnuplot or load the package
      color.sty in LaTeX.}%
    \renewcommand\color[2][]{}%
  }%
  \providecommand\includegraphics[2][]{%
    \GenericError{(gnuplot) \space\space\space\@spaces}{%
      Package graphicx or graphics not loaded%
    }{See the gnuplot documentation for explanation.%
    }{The gnuplot epslatex terminal needs graphicx.sty or graphics.sty.}%
    \renewcommand\includegraphics[2][]{}%
  }%
  \providecommand\rotatebox[2]{#2}%
  \@ifundefined{ifGPcolor}{%
    \newif\ifGPcolor
    \GPcolortrue
  }{}%
  \@ifundefined{ifGPblacktext}{%
    \newif\ifGPblacktext
    \GPblacktexttrue
  }{}%
  \let\gplgaddtomacro\g@addto@macro
  \gdef\gplbacktext{}%
  \gdef\gplfronttext{}%
  \makeatother
  \ifGPblacktext
    \def\colorrgb#1{}%
    \def\colorgray#1{}%
  \else
    \ifGPcolor
      \def\colorrgb#1{\color[rgb]{#1}}%
      \def\colorgray#1{\color[gray]{#1}}%
      \expandafter\def\csname LTw\endcsname{\color{white}}%
      \expandafter\def\csname LTb\endcsname{\color{black}}%
      \expandafter\def\csname LTa\endcsname{\color{black}}%
      \expandafter\def\csname LT0\endcsname{\color[rgb]{1,0,0}}%
      \expandafter\def\csname LT1\endcsname{\color[rgb]{0,1,0}}%
      \expandafter\def\csname LT2\endcsname{\color[rgb]{0,0,1}}%
      \expandafter\def\csname LT3\endcsname{\color[rgb]{1,0,1}}%
      \expandafter\def\csname LT4\endcsname{\color[rgb]{0,1,1}}%
      \expandafter\def\csname LT5\endcsname{\color[rgb]{1,1,0}}%
      \expandafter\def\csname LT6\endcsname{\color[rgb]{0,0,0}}%
      \expandafter\def\csname LT7\endcsname{\color[rgb]{1,0.3,0}}%
      \expandafter\def\csname LT8\endcsname{\color[rgb]{0.5,0.5,0.5}}%
    \else
      \def\colorrgb#1{\color{black}}%
      \def\colorgray#1{\color[gray]{#1}}%
      \expandafter\def\csname LTw\endcsname{\color{white}}%
      \expandafter\def\csname LTb\endcsname{\color{black}}%
      \expandafter\def\csname LTa\endcsname{\color{black}}%
      \expandafter\def\csname LT0\endcsname{\color{black}}%
      \expandafter\def\csname LT1\endcsname{\color{black}}%
      \expandafter\def\csname LT2\endcsname{\color{black}}%
      \expandafter\def\csname LT3\endcsname{\color{black}}%
      \expandafter\def\csname LT4\endcsname{\color{black}}%
      \expandafter\def\csname LT5\endcsname{\color{black}}%
      \expandafter\def\csname LT6\endcsname{\color{black}}%
      \expandafter\def\csname LT7\endcsname{\color{black}}%
      \expandafter\def\csname LT8\endcsname{\color{black}}%
    \fi
  \fi
  \setlength{\unitlength}{0.0500bp}%
  \begin{picture}(2744.00,1180.00)%
    \gplgaddtomacro\gplbacktext{%
      \csname LTb\endcsname%
      \put(-60,60){\makebox(0,0)[r]{\strut{} 0.6}}%
      \put(-60,425){\makebox(0,0)[r]{\strut{} 0.7}}%
      \put(-60,790){\makebox(0,0)[r]{\strut{} 0.8}}%
      \put(-60,1155){\makebox(0,0)[r]{\strut{} 0.9}}%
      \put(239,-120){\makebox(0,0){\strut{}}}%
      \put(692,-120){\makebox(0,0){\strut{}}}%
      \put(1145,-120){\makebox(0,0){\strut{}}}%
      \put(1598,-120){\makebox(0,0){\strut{}}}%
      \put(2051,-120){\makebox(0,0){\strut{}}}%
      \put(2504,-120){\makebox(0,0){\strut{}}}%
      \put(-550,607){\rotatebox{-270}{\makebox(0,0){\normalsize $r_2$}}}%
    }%
    \gplgaddtomacro\gplfronttext{%
    }%
    \gplbacktext
    \put(0,0){\includegraphics{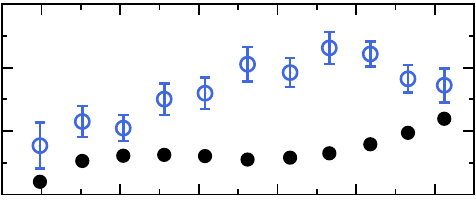}}%
    \gplfronttext
  \end{picture}%
\endgroup\hspace{20pt}
\begingroup
  \makeatletter
  \providecommand\color[2][]{%
    \GenericError{(gnuplot) \space\space\space\@spaces}{%
      Package color not loaded in conjunction with
      terminal option `colourtext'%
    }{See the gnuplot documentation for explanation.%
    }{Either use 'blacktext' in gnuplot or load the package
      color.sty in LaTeX.}%
    \renewcommand\color[2][]{}%
  }%
  \providecommand\includegraphics[2][]{%
    \GenericError{(gnuplot) \space\space\space\@spaces}{%
      Package graphicx or graphics not loaded%
    }{See the gnuplot documentation for explanation.%
    }{The gnuplot epslatex terminal needs graphicx.sty or graphics.sty.}%
    \renewcommand\includegraphics[2][]{}%
  }%
  \providecommand\rotatebox[2]{#2}%
  \@ifundefined{ifGPcolor}{%
    \newif\ifGPcolor
    \GPcolortrue
  }{}%
  \@ifundefined{ifGPblacktext}{%
    \newif\ifGPblacktext
    \GPblacktexttrue
  }{}%
  \let\gplgaddtomacro\g@addto@macro
  \gdef\gplbacktext{}%
  \gdef\gplfronttext{}%
  \makeatother
  \ifGPblacktext
    \def\colorrgb#1{}%
    \def\colorgray#1{}%
  \else
    \ifGPcolor
      \def\colorrgb#1{\color[rgb]{#1}}%
      \def\colorgray#1{\color[gray]{#1}}%
      \expandafter\def\csname LTw\endcsname{\color{white}}%
      \expandafter\def\csname LTb\endcsname{\color{black}}%
      \expandafter\def\csname LTa\endcsname{\color{black}}%
      \expandafter\def\csname LT0\endcsname{\color[rgb]{1,0,0}}%
      \expandafter\def\csname LT1\endcsname{\color[rgb]{0,1,0}}%
      \expandafter\def\csname LT2\endcsname{\color[rgb]{0,0,1}}%
      \expandafter\def\csname LT3\endcsname{\color[rgb]{1,0,1}}%
      \expandafter\def\csname LT4\endcsname{\color[rgb]{0,1,1}}%
      \expandafter\def\csname LT5\endcsname{\color[rgb]{1,1,0}}%
      \expandafter\def\csname LT6\endcsname{\color[rgb]{0,0,0}}%
      \expandafter\def\csname LT7\endcsname{\color[rgb]{1,0.3,0}}%
      \expandafter\def\csname LT8\endcsname{\color[rgb]{0.5,0.5,0.5}}%
    \else
      \def\colorrgb#1{\color{black}}%
      \def\colorgray#1{\color[gray]{#1}}%
      \expandafter\def\csname LTw\endcsname{\color{white}}%
      \expandafter\def\csname LTb\endcsname{\color{black}}%
      \expandafter\def\csname LTa\endcsname{\color{black}}%
      \expandafter\def\csname LT0\endcsname{\color{black}}%
      \expandafter\def\csname LT1\endcsname{\color{black}}%
      \expandafter\def\csname LT2\endcsname{\color{black}}%
      \expandafter\def\csname LT3\endcsname{\color{black}}%
      \expandafter\def\csname LT4\endcsname{\color{black}}%
      \expandafter\def\csname LT5\endcsname{\color{black}}%
      \expandafter\def\csname LT6\endcsname{\color{black}}%
      \expandafter\def\csname LT7\endcsname{\color{black}}%
      \expandafter\def\csname LT8\endcsname{\color{black}}%
    \fi
  \fi
  \setlength{\unitlength}{0.0500bp}%
  \begin{picture}(2744.00,1180.00)%
    \gplgaddtomacro\gplbacktext{%
      \csname LTb\endcsname%
      \put(-60,60){\makebox(0,0)[r]{\strut{} 0.6}}%
      \put(-60,425){\makebox(0,0)[r]{\strut{} 0.7}}%
      \put(-60,790){\makebox(0,0)[r]{\strut{} 0.8}}%
      \put(-60,1155){\makebox(0,0)[r]{\strut{} 0.9}}%
      \put(239,-120){\makebox(0,0){\strut{}}}%
      \put(692,-120){\makebox(0,0){\strut{}}}%
      \put(1145,-120){\makebox(0,0){\strut{}}}%
      \put(1598,-120){\makebox(0,0){\strut{}}}%
      \put(2051,-120){\makebox(0,0){\strut{}}}%
      \put(2504,-120){\makebox(0,0){\strut{}}}%
    }%
    \gplgaddtomacro\gplfronttext{%
    }%
    \gplbacktext
    \put(0,0){\includegraphics{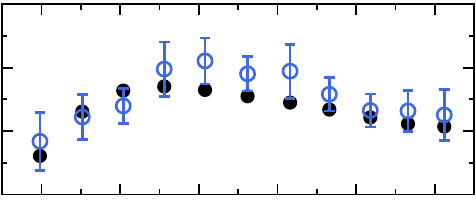}}%
    \gplfronttext
  \end{picture}%
\endgroup\hspace{20pt}
\begingroup
  \makeatletter
  \providecommand\color[2][]{%
    \GenericError{(gnuplot) \space\space\space\@spaces}{%
      Package color not loaded in conjunction with
      terminal option `colourtext'%
    }{See the gnuplot documentation for explanation.%
    }{Either use 'blacktext' in gnuplot or load the package
      color.sty in LaTeX.}%
    \renewcommand\color[2][]{}%
  }%
  \providecommand\includegraphics[2][]{%
    \GenericError{(gnuplot) \space\space\space\@spaces}{%
      Package graphicx or graphics not loaded%
    }{See the gnuplot documentation for explanation.%
    }{The gnuplot epslatex terminal needs graphicx.sty or graphics.sty.}%
    \renewcommand\includegraphics[2][]{}%
  }%
  \providecommand\rotatebox[2]{#2}%
  \@ifundefined{ifGPcolor}{%
    \newif\ifGPcolor
    \GPcolortrue
  }{}%
  \@ifundefined{ifGPblacktext}{%
    \newif\ifGPblacktext
    \GPblacktexttrue
  }{}%
  \let\gplgaddtomacro\g@addto@macro
  \gdef\gplbacktext{}%
  \gdef\gplfronttext{}%
  \makeatother
  \ifGPblacktext
    \def\colorrgb#1{}%
    \def\colorgray#1{}%
  \else
    \ifGPcolor
      \def\colorrgb#1{\color[rgb]{#1}}%
      \def\colorgray#1{\color[gray]{#1}}%
      \expandafter\def\csname LTw\endcsname{\color{white}}%
      \expandafter\def\csname LTb\endcsname{\color{black}}%
      \expandafter\def\csname LTa\endcsname{\color{black}}%
      \expandafter\def\csname LT0\endcsname{\color[rgb]{1,0,0}}%
      \expandafter\def\csname LT1\endcsname{\color[rgb]{0,1,0}}%
      \expandafter\def\csname LT2\endcsname{\color[rgb]{0,0,1}}%
      \expandafter\def\csname LT3\endcsname{\color[rgb]{1,0,1}}%
      \expandafter\def\csname LT4\endcsname{\color[rgb]{0,1,1}}%
      \expandafter\def\csname LT5\endcsname{\color[rgb]{1,1,0}}%
      \expandafter\def\csname LT6\endcsname{\color[rgb]{0,0,0}}%
      \expandafter\def\csname LT7\endcsname{\color[rgb]{1,0.3,0}}%
      \expandafter\def\csname LT8\endcsname{\color[rgb]{0.5,0.5,0.5}}%
    \else
      \def\colorrgb#1{\color{black}}%
      \def\colorgray#1{\color[gray]{#1}}%
      \expandafter\def\csname LTw\endcsname{\color{white}}%
      \expandafter\def\csname LTb\endcsname{\color{black}}%
      \expandafter\def\csname LTa\endcsname{\color{black}}%
      \expandafter\def\csname LT0\endcsname{\color{black}}%
      \expandafter\def\csname LT1\endcsname{\color{black}}%
      \expandafter\def\csname LT2\endcsname{\color{black}}%
      \expandafter\def\csname LT3\endcsname{\color{black}}%
      \expandafter\def\csname LT4\endcsname{\color{black}}%
      \expandafter\def\csname LT5\endcsname{\color{black}}%
      \expandafter\def\csname LT6\endcsname{\color{black}}%
      \expandafter\def\csname LT7\endcsname{\color{black}}%
      \expandafter\def\csname LT8\endcsname{\color{black}}%
    \fi
  \fi
  \setlength{\unitlength}{0.0500bp}%
  \begin{picture}(2744.00,1180.00)%
    \gplgaddtomacro\gplbacktext{%
      \csname LTb\endcsname%
      \put(-60,60){\makebox(0,0)[r]{\strut{} 0}}%
      \put(-60,334){\makebox(0,0)[r]{\strut{} 0.2}}%
      \put(-60,608){\makebox(0,0)[r]{\strut{} 0.4}}%
      \put(-60,881){\makebox(0,0)[r]{\strut{} 0.6}}%
      \put(-60,1155){\makebox(0,0)[r]{\strut{} 0.8}}%
      \put(239,-120){\makebox(0,0){\strut{}}}%
      \put(692,-120){\makebox(0,0){\strut{}}}%
      \put(1145,-120){\makebox(0,0){\strut{}}}%
      \put(1598,-120){\makebox(0,0){\strut{}}}%
      \put(2051,-120){\makebox(0,0){\strut{}}}%
      \put(2504,-120){\makebox(0,0){\strut{}}}%
    }%
    \gplgaddtomacro\gplfronttext{%
    }%
    \gplbacktext
    \put(0,0){\includegraphics{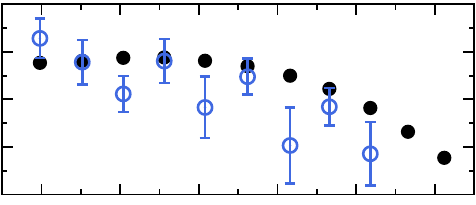}}%
    \gplfronttext
  \end{picture}%
\endgroup
\\ $ $ \\
\hspace{23pt}
\begingroup
  \makeatletter
  \providecommand\color[2][]{%
    \GenericError{(gnuplot) \space\space\space\@spaces}{%
      Package color not loaded in conjunction with
      terminal option `colourtext'%
    }{See the gnuplot documentation for explanation.%
    }{Either use 'blacktext' in gnuplot or load the package
      color.sty in LaTeX.}%
    \renewcommand\color[2][]{}%
  }%
  \providecommand\includegraphics[2][]{%
    \GenericError{(gnuplot) \space\space\space\@spaces}{%
      Package graphicx or graphics not loaded%
    }{See the gnuplot documentation for explanation.%
    }{The gnuplot epslatex terminal needs graphicx.sty or graphics.sty.}%
    \renewcommand\includegraphics[2][]{}%
  }%
  \providecommand\rotatebox[2]{#2}%
  \@ifundefined{ifGPcolor}{%
    \newif\ifGPcolor
    \GPcolortrue
  }{}%
  \@ifundefined{ifGPblacktext}{%
    \newif\ifGPblacktext
    \GPblacktexttrue
  }{}%
  \let\gplgaddtomacro\g@addto@macro
  \gdef\gplbacktext{}%
  \gdef\gplfronttext{}%
  \makeatother
  \ifGPblacktext
    \def\colorrgb#1{}%
    \def\colorgray#1{}%
  \else
    \ifGPcolor
      \def\colorrgb#1{\color[rgb]{#1}}%
      \def\colorgray#1{\color[gray]{#1}}%
      \expandafter\def\csname LTw\endcsname{\color{white}}%
      \expandafter\def\csname LTb\endcsname{\color{black}}%
      \expandafter\def\csname LTa\endcsname{\color{black}}%
      \expandafter\def\csname LT0\endcsname{\color[rgb]{1,0,0}}%
      \expandafter\def\csname LT1\endcsname{\color[rgb]{0,1,0}}%
      \expandafter\def\csname LT2\endcsname{\color[rgb]{0,0,1}}%
      \expandafter\def\csname LT3\endcsname{\color[rgb]{1,0,1}}%
      \expandafter\def\csname LT4\endcsname{\color[rgb]{0,1,1}}%
      \expandafter\def\csname LT5\endcsname{\color[rgb]{1,1,0}}%
      \expandafter\def\csname LT6\endcsname{\color[rgb]{0,0,0}}%
      \expandafter\def\csname LT7\endcsname{\color[rgb]{1,0.3,0}}%
      \expandafter\def\csname LT8\endcsname{\color[rgb]{0.5,0.5,0.5}}%
    \else
      \def\colorrgb#1{\color{black}}%
      \def\colorgray#1{\color[gray]{#1}}%
      \expandafter\def\csname LTw\endcsname{\color{white}}%
      \expandafter\def\csname LTb\endcsname{\color{black}}%
      \expandafter\def\csname LTa\endcsname{\color{black}}%
      \expandafter\def\csname LT0\endcsname{\color{black}}%
      \expandafter\def\csname LT1\endcsname{\color{black}}%
      \expandafter\def\csname LT2\endcsname{\color{black}}%
      \expandafter\def\csname LT3\endcsname{\color{black}}%
      \expandafter\def\csname LT4\endcsname{\color{black}}%
      \expandafter\def\csname LT5\endcsname{\color{black}}%
      \expandafter\def\csname LT6\endcsname{\color{black}}%
      \expandafter\def\csname LT7\endcsname{\color{black}}%
      \expandafter\def\csname LT8\endcsname{\color{black}}%
    \fi
  \fi
  \setlength{\unitlength}{0.0500bp}%
  \begin{picture}(2744.00,1180.00)%
    \gplgaddtomacro\gplbacktext{%
      \csname LTb\endcsname%
      \put(-60,60){\makebox(0,0)[r]{\strut{} 0.4}}%
      \put(-60,498){\makebox(0,0)[r]{\strut{} 0.5}}%
      \put(-60,936){\makebox(0,0)[r]{\strut{} 0.6}}%
      \put(239,-120){\makebox(0,0){\strut{}}}%
      \put(692,-120){\makebox(0,0){\strut{}}}%
      \put(1145,-120){\makebox(0,0){\strut{}}}%
      \put(1598,-120){\makebox(0,0){\strut{}}}%
      \put(2051,-120){\makebox(0,0){\strut{}}}%
      \put(2504,-120){\makebox(0,0){\strut{}}}%
      \put(-550,607){\rotatebox{-270}{\makebox(0,0){\normalsize $r_3$}}}%
    }%
    \gplgaddtomacro\gplfronttext{%
    }%
    \gplbacktext
    \put(0,0){\includegraphics{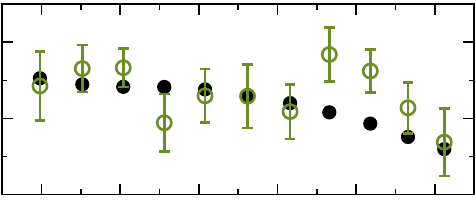}}%
    \gplfronttext
  \end{picture}%
\endgroup\hspace{20pt}
\begingroup
  \makeatletter
  \providecommand\color[2][]{%
    \GenericError{(gnuplot) \space\space\space\@spaces}{%
      Package color not loaded in conjunction with
      terminal option `colourtext'%
    }{See the gnuplot documentation for explanation.%
    }{Either use 'blacktext' in gnuplot or load the package
      color.sty in LaTeX.}%
    \renewcommand\color[2][]{}%
  }%
  \providecommand\includegraphics[2][]{%
    \GenericError{(gnuplot) \space\space\space\@spaces}{%
      Package graphicx or graphics not loaded%
    }{See the gnuplot documentation for explanation.%
    }{The gnuplot epslatex terminal needs graphicx.sty or graphics.sty.}%
    \renewcommand\includegraphics[2][]{}%
  }%
  \providecommand\rotatebox[2]{#2}%
  \@ifundefined{ifGPcolor}{%
    \newif\ifGPcolor
    \GPcolortrue
  }{}%
  \@ifundefined{ifGPblacktext}{%
    \newif\ifGPblacktext
    \GPblacktexttrue
  }{}%
  \let\gplgaddtomacro\g@addto@macro
  \gdef\gplbacktext{}%
  \gdef\gplfronttext{}%
  \makeatother
  \ifGPblacktext
    \def\colorrgb#1{}%
    \def\colorgray#1{}%
  \else
    \ifGPcolor
      \def\colorrgb#1{\color[rgb]{#1}}%
      \def\colorgray#1{\color[gray]{#1}}%
      \expandafter\def\csname LTw\endcsname{\color{white}}%
      \expandafter\def\csname LTb\endcsname{\color{black}}%
      \expandafter\def\csname LTa\endcsname{\color{black}}%
      \expandafter\def\csname LT0\endcsname{\color[rgb]{1,0,0}}%
      \expandafter\def\csname LT1\endcsname{\color[rgb]{0,1,0}}%
      \expandafter\def\csname LT2\endcsname{\color[rgb]{0,0,1}}%
      \expandafter\def\csname LT3\endcsname{\color[rgb]{1,0,1}}%
      \expandafter\def\csname LT4\endcsname{\color[rgb]{0,1,1}}%
      \expandafter\def\csname LT5\endcsname{\color[rgb]{1,1,0}}%
      \expandafter\def\csname LT6\endcsname{\color[rgb]{0,0,0}}%
      \expandafter\def\csname LT7\endcsname{\color[rgb]{1,0.3,0}}%
      \expandafter\def\csname LT8\endcsname{\color[rgb]{0.5,0.5,0.5}}%
    \else
      \def\colorrgb#1{\color{black}}%
      \def\colorgray#1{\color[gray]{#1}}%
      \expandafter\def\csname LTw\endcsname{\color{white}}%
      \expandafter\def\csname LTb\endcsname{\color{black}}%
      \expandafter\def\csname LTa\endcsname{\color{black}}%
      \expandafter\def\csname LT0\endcsname{\color{black}}%
      \expandafter\def\csname LT1\endcsname{\color{black}}%
      \expandafter\def\csname LT2\endcsname{\color{black}}%
      \expandafter\def\csname LT3\endcsname{\color{black}}%
      \expandafter\def\csname LT4\endcsname{\color{black}}%
      \expandafter\def\csname LT5\endcsname{\color{black}}%
      \expandafter\def\csname LT6\endcsname{\color{black}}%
      \expandafter\def\csname LT7\endcsname{\color{black}}%
      \expandafter\def\csname LT8\endcsname{\color{black}}%
    \fi
  \fi
  \setlength{\unitlength}{0.0500bp}%
  \begin{picture}(2744.00,1180.00)%
    \gplgaddtomacro\gplbacktext{%
      \csname LTb\endcsname%
      \put(-60,60){\makebox(0,0)[r]{\strut{} 0.4}}%
      \put(-60,373){\makebox(0,0)[r]{\strut{} 0.5}}%
      \put(-60,686){\makebox(0,0)[r]{\strut{} 0.6}}%
      \put(-60,999){\makebox(0,0)[r]{\strut{} 0.7}}%
      \put(239,-120){\makebox(0,0){\strut{}}}%
      \put(692,-120){\makebox(0,0){\strut{}}}%
      \put(1145,-120){\makebox(0,0){\strut{}}}%
      \put(1598,-120){\makebox(0,0){\strut{}}}%
      \put(2051,-120){\makebox(0,0){\strut{}}}%
      \put(2504,-120){\makebox(0,0){\strut{}}}%
    }%
    \gplgaddtomacro\gplfronttext{%
    }%
    \gplbacktext
    \put(0,0){\includegraphics{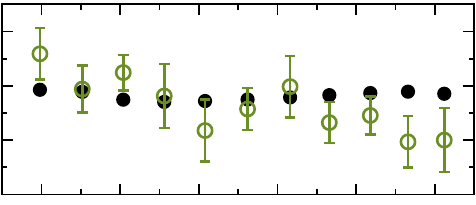}}%
    \gplfronttext
  \end{picture}%
\endgroup\hspace{20pt}
\begingroup
  \makeatletter
  \providecommand\color[2][]{%
    \GenericError{(gnuplot) \space\space\space\@spaces}{%
      Package color not loaded in conjunction with
      terminal option `colourtext'%
    }{See the gnuplot documentation for explanation.%
    }{Either use 'blacktext' in gnuplot or load the package
      color.sty in LaTeX.}%
    \renewcommand\color[2][]{}%
  }%
  \providecommand\includegraphics[2][]{%
    \GenericError{(gnuplot) \space\space\space\@spaces}{%
      Package graphicx or graphics not loaded%
    }{See the gnuplot documentation for explanation.%
    }{The gnuplot epslatex terminal needs graphicx.sty or graphics.sty.}%
    \renewcommand\includegraphics[2][]{}%
  }%
  \providecommand\rotatebox[2]{#2}%
  \@ifundefined{ifGPcolor}{%
    \newif\ifGPcolor
    \GPcolortrue
  }{}%
  \@ifundefined{ifGPblacktext}{%
    \newif\ifGPblacktext
    \GPblacktexttrue
  }{}%
  \let\gplgaddtomacro\g@addto@macro
  \gdef\gplbacktext{}%
  \gdef\gplfronttext{}%
  \makeatother
  \ifGPblacktext
    \def\colorrgb#1{}%
    \def\colorgray#1{}%
  \else
    \ifGPcolor
      \def\colorrgb#1{\color[rgb]{#1}}%
      \def\colorgray#1{\color[gray]{#1}}%
      \expandafter\def\csname LTw\endcsname{\color{white}}%
      \expandafter\def\csname LTb\endcsname{\color{black}}%
      \expandafter\def\csname LTa\endcsname{\color{black}}%
      \expandafter\def\csname LT0\endcsname{\color[rgb]{1,0,0}}%
      \expandafter\def\csname LT1\endcsname{\color[rgb]{0,1,0}}%
      \expandafter\def\csname LT2\endcsname{\color[rgb]{0,0,1}}%
      \expandafter\def\csname LT3\endcsname{\color[rgb]{1,0,1}}%
      \expandafter\def\csname LT4\endcsname{\color[rgb]{0,1,1}}%
      \expandafter\def\csname LT5\endcsname{\color[rgb]{1,1,0}}%
      \expandafter\def\csname LT6\endcsname{\color[rgb]{0,0,0}}%
      \expandafter\def\csname LT7\endcsname{\color[rgb]{1,0.3,0}}%
      \expandafter\def\csname LT8\endcsname{\color[rgb]{0.5,0.5,0.5}}%
    \else
      \def\colorrgb#1{\color{black}}%
      \def\colorgray#1{\color[gray]{#1}}%
      \expandafter\def\csname LTw\endcsname{\color{white}}%
      \expandafter\def\csname LTb\endcsname{\color{black}}%
      \expandafter\def\csname LTa\endcsname{\color{black}}%
      \expandafter\def\csname LT0\endcsname{\color{black}}%
      \expandafter\def\csname LT1\endcsname{\color{black}}%
      \expandafter\def\csname LT2\endcsname{\color{black}}%
      \expandafter\def\csname LT3\endcsname{\color{black}}%
      \expandafter\def\csname LT4\endcsname{\color{black}}%
      \expandafter\def\csname LT5\endcsname{\color{black}}%
      \expandafter\def\csname LT6\endcsname{\color{black}}%
      \expandafter\def\csname LT7\endcsname{\color{black}}%
      \expandafter\def\csname LT8\endcsname{\color{black}}%
    \fi
  \fi
  \setlength{\unitlength}{0.0500bp}%
  \begin{picture}(2744.00,1180.00)%
    \gplgaddtomacro\gplbacktext{%
      \csname LTb\endcsname%
      \put(-60,197){\makebox(0,0)[r]{\strut{} 0.2}}%
      \put(-60,471){\makebox(0,0)[r]{\strut{} 0.4}}%
      \put(-60,744){\makebox(0,0)[r]{\strut{} 0.6}}%
      \put(-60,1018){\makebox(0,0)[r]{\strut{} 0.8}}%
      \put(239,-120){\makebox(0,0){\strut{}}}%
      \put(692,-120){\makebox(0,0){\strut{}}}%
      \put(1145,-120){\makebox(0,0){\strut{}}}%
      \put(1598,-120){\makebox(0,0){\strut{}}}%
      \put(2051,-120){\makebox(0,0){\strut{}}}%
      \put(2504,-120){\makebox(0,0){\strut{}}}%
    }%
    \gplgaddtomacro\gplfronttext{%
    }%
    \gplbacktext
    \put(0,0){\includegraphics{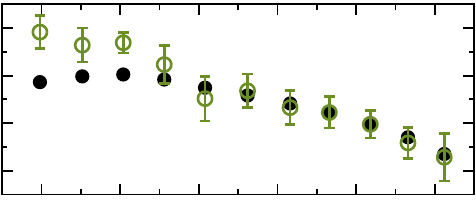}}%
    \gplfronttext
  \end{picture}%
\endgroup
\\ $ $ \\
\hspace{23pt}
\begingroup
  \makeatletter
  \providecommand\color[2][]{%
    \GenericError{(gnuplot) \space\space\space\@spaces}{%
      Package color not loaded in conjunction with
      terminal option `colourtext'%
    }{See the gnuplot documentation for explanation.%
    }{Either use 'blacktext' in gnuplot or load the package
      color.sty in LaTeX.}%
    \renewcommand\color[2][]{}%
  }%
  \providecommand\includegraphics[2][]{%
    \GenericError{(gnuplot) \space\space\space\@spaces}{%
      Package graphicx or graphics not loaded%
    }{See the gnuplot documentation for explanation.%
    }{The gnuplot epslatex terminal needs graphicx.sty or graphics.sty.}%
    \renewcommand\includegraphics[2][]{}%
  }%
  \providecommand\rotatebox[2]{#2}%
  \@ifundefined{ifGPcolor}{%
    \newif\ifGPcolor
    \GPcolortrue
  }{}%
  \@ifundefined{ifGPblacktext}{%
    \newif\ifGPblacktext
    \GPblacktexttrue
  }{}%
  \let\gplgaddtomacro\g@addto@macro
  \gdef\gplbacktext{}%
  \gdef\gplfronttext{}%
  \makeatother
  \ifGPblacktext
    \def\colorrgb#1{}%
    \def\colorgray#1{}%
  \else
    \ifGPcolor
      \def\colorrgb#1{\color[rgb]{#1}}%
      \def\colorgray#1{\color[gray]{#1}}%
      \expandafter\def\csname LTw\endcsname{\color{white}}%
      \expandafter\def\csname LTb\endcsname{\color{black}}%
      \expandafter\def\csname LTa\endcsname{\color{black}}%
      \expandafter\def\csname LT0\endcsname{\color[rgb]{1,0,0}}%
      \expandafter\def\csname LT1\endcsname{\color[rgb]{0,1,0}}%
      \expandafter\def\csname LT2\endcsname{\color[rgb]{0,0,1}}%
      \expandafter\def\csname LT3\endcsname{\color[rgb]{1,0,1}}%
      \expandafter\def\csname LT4\endcsname{\color[rgb]{0,1,1}}%
      \expandafter\def\csname LT5\endcsname{\color[rgb]{1,1,0}}%
      \expandafter\def\csname LT6\endcsname{\color[rgb]{0,0,0}}%
      \expandafter\def\csname LT7\endcsname{\color[rgb]{1,0.3,0}}%
      \expandafter\def\csname LT8\endcsname{\color[rgb]{0.5,0.5,0.5}}%
    \else
      \def\colorrgb#1{\color{black}}%
      \def\colorgray#1{\color[gray]{#1}}%
      \expandafter\def\csname LTw\endcsname{\color{white}}%
      \expandafter\def\csname LTb\endcsname{\color{black}}%
      \expandafter\def\csname LTa\endcsname{\color{black}}%
      \expandafter\def\csname LT0\endcsname{\color{black}}%
      \expandafter\def\csname LT1\endcsname{\color{black}}%
      \expandafter\def\csname LT2\endcsname{\color{black}}%
      \expandafter\def\csname LT3\endcsname{\color{black}}%
      \expandafter\def\csname LT4\endcsname{\color{black}}%
      \expandafter\def\csname LT5\endcsname{\color{black}}%
      \expandafter\def\csname LT6\endcsname{\color{black}}%
      \expandafter\def\csname LT7\endcsname{\color{black}}%
      \expandafter\def\csname LT8\endcsname{\color{black}}%
    \fi
  \fi
  \setlength{\unitlength}{0.0500bp}%
  \begin{picture}(2744.00,1180.00)%
    \gplgaddtomacro\gplbacktext{%
      \csname LTb\endcsname%
      \put(-60,60){\makebox(0,0)[r]{\strut{} 0}}%
      \put(-60,425){\makebox(0,0)[r]{\strut{} 0.2}}%
      \put(-60,790){\makebox(0,0)[r]{\strut{} 0.4}}%
      \put(-60,1155){\makebox(0,0)[r]{\strut{} 0.6}}%
      \put(239,-120){\makebox(0,0){\strut{}1.65}}%
      \put(692,-120){\makebox(0,0){\strut{}1.70}}%
      \put(1145,-120){\makebox(0,0){\strut{}1.75}}%
      \put(1598,-120){\makebox(0,0){\strut{}1.80}}%
      \put(2051,-120){\makebox(0,0){\strut{}1.85}}%
      \put(2504,-120){\makebox(0,0){\strut{}1.90}}%
      \put(-550,607){\rotatebox{-270}{\makebox(0,0){\normalsize $r_4$}}}%
      \put(1371,-490){\makebox(0,0){\normalsize $W$ (GeV)}}%
    }%
    \gplgaddtomacro\gplfronttext{%
    }%
    \gplbacktext
    \put(0,0){\includegraphics{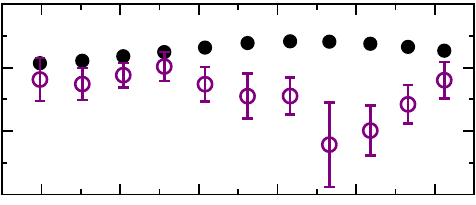}}%
    \gplfronttext
  \end{picture}%
\endgroup\hspace{20pt}
\begingroup
  \makeatletter
  \providecommand\color[2][]{%
    \GenericError{(gnuplot) \space\space\space\@spaces}{%
      Package color not loaded in conjunction with
      terminal option `colourtext'%
    }{See the gnuplot documentation for explanation.%
    }{Either use 'blacktext' in gnuplot or load the package
      color.sty in LaTeX.}%
    \renewcommand\color[2][]{}%
  }%
  \providecommand\includegraphics[2][]{%
    \GenericError{(gnuplot) \space\space\space\@spaces}{%
      Package graphicx or graphics not loaded%
    }{See the gnuplot documentation for explanation.%
    }{The gnuplot epslatex terminal needs graphicx.sty or graphics.sty.}%
    \renewcommand\includegraphics[2][]{}%
  }%
  \providecommand\rotatebox[2]{#2}%
  \@ifundefined{ifGPcolor}{%
    \newif\ifGPcolor
    \GPcolortrue
  }{}%
  \@ifundefined{ifGPblacktext}{%
    \newif\ifGPblacktext
    \GPblacktexttrue
  }{}%
  \let\gplgaddtomacro\g@addto@macro
  \gdef\gplbacktext{}%
  \gdef\gplfronttext{}%
  \makeatother
  \ifGPblacktext
    \def\colorrgb#1{}%
    \def\colorgray#1{}%
  \else
    \ifGPcolor
      \def\colorrgb#1{\color[rgb]{#1}}%
      \def\colorgray#1{\color[gray]{#1}}%
      \expandafter\def\csname LTw\endcsname{\color{white}}%
      \expandafter\def\csname LTb\endcsname{\color{black}}%
      \expandafter\def\csname LTa\endcsname{\color{black}}%
      \expandafter\def\csname LT0\endcsname{\color[rgb]{1,0,0}}%
      \expandafter\def\csname LT1\endcsname{\color[rgb]{0,1,0}}%
      \expandafter\def\csname LT2\endcsname{\color[rgb]{0,0,1}}%
      \expandafter\def\csname LT3\endcsname{\color[rgb]{1,0,1}}%
      \expandafter\def\csname LT4\endcsname{\color[rgb]{0,1,1}}%
      \expandafter\def\csname LT5\endcsname{\color[rgb]{1,1,0}}%
      \expandafter\def\csname LT6\endcsname{\color[rgb]{0,0,0}}%
      \expandafter\def\csname LT7\endcsname{\color[rgb]{1,0.3,0}}%
      \expandafter\def\csname LT8\endcsname{\color[rgb]{0.5,0.5,0.5}}%
    \else
      \def\colorrgb#1{\color{black}}%
      \def\colorgray#1{\color[gray]{#1}}%
      \expandafter\def\csname LTw\endcsname{\color{white}}%
      \expandafter\def\csname LTb\endcsname{\color{black}}%
      \expandafter\def\csname LTa\endcsname{\color{black}}%
      \expandafter\def\csname LT0\endcsname{\color{black}}%
      \expandafter\def\csname LT1\endcsname{\color{black}}%
      \expandafter\def\csname LT2\endcsname{\color{black}}%
      \expandafter\def\csname LT3\endcsname{\color{black}}%
      \expandafter\def\csname LT4\endcsname{\color{black}}%
      \expandafter\def\csname LT5\endcsname{\color{black}}%
      \expandafter\def\csname LT6\endcsname{\color{black}}%
      \expandafter\def\csname LT7\endcsname{\color{black}}%
      \expandafter\def\csname LT8\endcsname{\color{black}}%
    \fi
  \fi
  \setlength{\unitlength}{0.0500bp}%
  \begin{picture}(2744.00,1180.00)%
    \gplgaddtomacro\gplbacktext{%
      \csname LTb\endcsname%
      \put(-60,182){\makebox(0,0)[r]{\strut{} 0.1}}%
      \put(-60,425){\makebox(0,0)[r]{\strut{} 0.2}}%
      \put(-60,668){\makebox(0,0)[r]{\strut{} 0.3}}%
      \put(-60,912){\makebox(0,0)[r]{\strut{} 0.4}}%
      \put(-60,1155){\makebox(0,0)[r]{\strut{} 0.5}}%
      \put(239,-120){\makebox(0,0){\strut{}1.65}}%
      \put(692,-120){\makebox(0,0){\strut{}1.70}}%
      \put(1145,-120){\makebox(0,0){\strut{}1.75}}%
      \put(1598,-120){\makebox(0,0){\strut{}1.80}}%
      \put(2051,-120){\makebox(0,0){\strut{}1.85}}%
      \put(2504,-120){\makebox(0,0){\strut{}1.90}}%
      \put(1371,-490){\makebox(0,0){\normalsize $W$ (GeV)}}%
    }%
    \gplgaddtomacro\gplfronttext{%
    }%
    \gplbacktext
    \put(0,0){\includegraphics{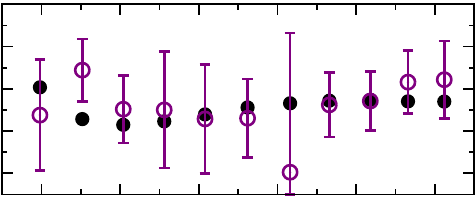}}%
    \gplfronttext
  \end{picture}%
\endgroup\hspace{20pt}
\begingroup
  \makeatletter
  \providecommand\color[2][]{%
    \GenericError{(gnuplot) \space\space\space\@spaces}{%
      Package color not loaded in conjunction with
      terminal option `colourtext'%
    }{See the gnuplot documentation for explanation.%
    }{Either use 'blacktext' in gnuplot or load the package
      color.sty in LaTeX.}%
    \renewcommand\color[2][]{}%
  }%
  \providecommand\includegraphics[2][]{%
    \GenericError{(gnuplot) \space\space\space\@spaces}{%
      Package graphicx or graphics not loaded%
    }{See the gnuplot documentation for explanation.%
    }{The gnuplot epslatex terminal needs graphicx.sty or graphics.sty.}%
    \renewcommand\includegraphics[2][]{}%
  }%
  \providecommand\rotatebox[2]{#2}%
  \@ifundefined{ifGPcolor}{%
    \newif\ifGPcolor
    \GPcolortrue
  }{}%
  \@ifundefined{ifGPblacktext}{%
    \newif\ifGPblacktext
    \GPblacktexttrue
  }{}%
  \let\gplgaddtomacro\g@addto@macro
  \gdef\gplbacktext{}%
  \gdef\gplfronttext{}%
  \makeatother
  \ifGPblacktext
    \def\colorrgb#1{}%
    \def\colorgray#1{}%
  \else
    \ifGPcolor
      \def\colorrgb#1{\color[rgb]{#1}}%
      \def\colorgray#1{\color[gray]{#1}}%
      \expandafter\def\csname LTw\endcsname{\color{white}}%
      \expandafter\def\csname LTb\endcsname{\color{black}}%
      \expandafter\def\csname LTa\endcsname{\color{black}}%
      \expandafter\def\csname LT0\endcsname{\color[rgb]{1,0,0}}%
      \expandafter\def\csname LT1\endcsname{\color[rgb]{0,1,0}}%
      \expandafter\def\csname LT2\endcsname{\color[rgb]{0,0,1}}%
      \expandafter\def\csname LT3\endcsname{\color[rgb]{1,0,1}}%
      \expandafter\def\csname LT4\endcsname{\color[rgb]{0,1,1}}%
      \expandafter\def\csname LT5\endcsname{\color[rgb]{1,1,0}}%
      \expandafter\def\csname LT6\endcsname{\color[rgb]{0,0,0}}%
      \expandafter\def\csname LT7\endcsname{\color[rgb]{1,0.3,0}}%
      \expandafter\def\csname LT8\endcsname{\color[rgb]{0.5,0.5,0.5}}%
    \else
      \def\colorrgb#1{\color{black}}%
      \def\colorgray#1{\color[gray]{#1}}%
      \expandafter\def\csname LTw\endcsname{\color{white}}%
      \expandafter\def\csname LTb\endcsname{\color{black}}%
      \expandafter\def\csname LTa\endcsname{\color{black}}%
      \expandafter\def\csname LT0\endcsname{\color{black}}%
      \expandafter\def\csname LT1\endcsname{\color{black}}%
      \expandafter\def\csname LT2\endcsname{\color{black}}%
      \expandafter\def\csname LT3\endcsname{\color{black}}%
      \expandafter\def\csname LT4\endcsname{\color{black}}%
      \expandafter\def\csname LT5\endcsname{\color{black}}%
      \expandafter\def\csname LT6\endcsname{\color{black}}%
      \expandafter\def\csname LT7\endcsname{\color{black}}%
      \expandafter\def\csname LT8\endcsname{\color{black}}%
    \fi
  \fi
  \setlength{\unitlength}{0.0500bp}%
  \begin{picture}(2744.00,1180.00)%
    \gplgaddtomacro\gplbacktext{%
      \csname LTb\endcsname%
      \put(-60,279){\makebox(0,0)[r]{\strut{} 0.2}}%
      \put(-60,717){\makebox(0,0)[r]{\strut{} 0.4}}%
      \put(-60,1155){\makebox(0,0)[r]{\strut{} 0.6}}%
      \put(239,-120){\makebox(0,0){\strut{}1.65}}%
      \put(692,-120){\makebox(0,0){\strut{}1.70}}%
      \put(1145,-120){\makebox(0,0){\strut{}1.75}}%
      \put(1598,-120){\makebox(0,0){\strut{}1.80}}%
      \put(2051,-120){\makebox(0,0){\strut{}1.85}}%
      \put(2504,-120){\makebox(0,0){\strut{}1.90}}%
      \put(1371,-490){\makebox(0,0){\normalsize $W$ (GeV)}}%
    }%
    \gplgaddtomacro\gplfronttext{%
    }%
    \gplbacktext
    \put(0,0){\includegraphics{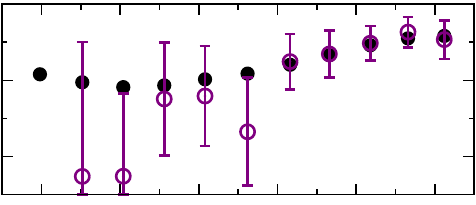}}%
    \gplfronttext
  \end{picture}%
\endgroup
\\ $ $ \\ $ $ \\
\end{center}
\caption{(Color online) The energy dependence of the moduli $r_{i}$ of the normalized transversity amplitudes for the $\gamma p \to K^{+} \Lambda$ reaction. The data are extracted from the GRAAL results for the single-polarization observables reported in Refs.\ \cite{Lleres2007,Lleres2009}. The dots are the bin-centered RPR-2011 predictions.}
\label{fig:moduli_GRAAL_RPR}
\end{figure*}

\subsection{The merits of the transversity basis}
\label{subsec:merits_transversity}
In Sec.~\ref{sec:formalism}, it was shown how the transversity amplitudes can be inferred from a complete set of measured observables. The first and essential step comprised the determination of the moduli from the three single asymmetries. Indeed, as became clear, the extraction of the $\delta_i$ requires prior knowledge about the moduli $r_i$. Luckily enough, single asymmetries are more easily obtained experimentally than double asymmetries. Consequently, the single asymmetries database generally have better statistics compared to double polarization observables. The published database for e.g.\ $\gamma p \to K ^+ \Lambda$ includes 2260 single asymmetries ($\Sigma$: 178, $T$: 69, $P$: 2013), in contrast to only 456 double asymmetries ($C_x$: 162, $C_z$: 162, $O_{x'}$: 66, $O_{z'}$: 66) \cite{lesleyprc}. The beam-recoil asymmetries $\{O_{x'}, O_{z'}\}$ are related to $\{O_x, O_z\}$ through Eq.~\eq{eq:primed_doubles}. In the transversity representation the parameters that are most easily extracted 
from the data, namely the moduli, are related to that class of asymmetries that are more readily measured, namely the single asymmetries. This is not the case for the helicity basis, where the moduli are related to the set of double asymmetries $\{C_{z'},E,L_{z'}\}$. It is important to stress that determining invariant amplitudes at some kinematical point requires knowledge of a complete set at the same kinematics. To date, however, there is not a single pseudoscalar photoproduction reaction for which a complete dataset has been published.

For the $\gamma p \to K ^+ \Lambda$ reaction, the GRAAL collaboration has measured $\{\Sigma, T, P\}$ at 66 kinematical points \cite{Lleres2007, Lleres2009}. These data cover $1.65 \lesssim W \lesssim 1.91$~GeV (bins of $\Delta W \approx 50$ MeV) and $-0.81 \lesssim \cos\thcm \lesssim 0.86$ (bins of $\Delta \cos\thcm \approx 0.3$), and can be used to estimate the moduli $r_i$ through Eqs.\ \eq{eq:moduli_expressions}. Figure \ref{fig:moduli_GRAAL_RPR} shows these extracted moduli at three $\cos\thcm$ intervals (33 of the 66 kinematical points for which data is available) along with the corresponding predictions from RPR-2011. It is observed that at a few kinematical points, some of the $\widehat{r}_i$ cannot be extracted from the data. This occurs whenever the measured set $\{\Sigma, T, P\}$ deviates too strongly from the `exact' set so that one or more arguments of the square roots \eq{eq:moduli_expressions} become negative. For the GRAAL data, only 17 of the $66 \times 4 = 264$ $\widehat{r}_i$ 
values are imaginary ($r_1$: 12, $r_2$: 3, $r_4$: 2). Overall, the RPR-2011 model offers a fair description of the energy and angular dependence of the extracted $\widehat{r}_i$ except for the $\widehat{r}_2$ and $\widehat{r}_4$ at $0.808 \lesssim \cos\thcm \lesssim 0.861$.

Obviously, the GRAAL bin width of $\Delta W \approx 50$ MeV suffices to map the energy dependence of the $r_i$. The most extensive $\gamma p \to K^{+} \Lambda$ data set to date is due to the CLAS collaboration at Jefferson Lab \cite{cracken2010}. The data cover $1.62 \le W \le 2.84$~GeV (bins of $\Delta W=10$~MeV) and $ -0.85 \le \cos \thcm \le 0.95$ (bins of $\Delta \cos \thcm=0.1$) and include data for $\frac {\d \sigma} { \d \Omega}$ and the recoil polarization $P$.

In addition to $\{\Sigma, T, P\}$, GRAAL provides data for $\{O_{x'}, O_{z'}\}$ at the same kinematical points. Since a complete set of asymmetries is required to extract the $\delta_i$, these beam-recoil data cannot be employed in order to gain additional information about the transversity amplitudes.

As is seen from Table \ref{tab:helicity_representation}, the GRAAL data for $\{\Sigma, T, P, O_{x'}, O_{z'}\}$ cannot be used to extract the moduli of the helicity amplitudes. The latter would require experimental data for $\{C_{z'}, E, L_{z'}\}$ at the same kinematics, which is not available to this day. Clearly, given the current status of the experimental $\gamma p \to K ^+ \Lambda$ program, the transversity representation offers the best perspectives to learn about the reaction amplitudes.

The extracted information about the energy and angular dependence of the amplitudes $r_i(W,\cos\thcm)$ of Fig.~\ref{fig:moduli_GRAAL_RPR} is complementary to what could be obtained about the partial waves in a so-called truncated partial wave analysis (TPWA) \cite{Tiator:2012ah,Tiator:2011tu,Workman:2011hi}. In TPWA one aims at extracting information about the energy dependence of the partial waves from the experimental data. The partial waves serve as expansion  parameters for the angular dependence of the observables. The amount of  partial waves which can be included (usually denoted by $l_{\max}$) in the fits, depends on the quality and quantity of the measured $\cos\thcm$ dependence of the observables. It is anticipated that in a reaction channel like $\gamma p \to K^+ \Lambda$ with substantial background contributions, $l_{\max}$ cannot be truncated to small values. In the TPWA approach the underlying dynamics is parameterized in terms of functions which depend on $W$, whereas in an amplitude analysis 
as presented here, one attempts to map the full $(W,\cos\thcm)$ dependence. In the forthcoming section, it will be shown that the availability of complete sets of eight observables with realistic error bars at given $(W,\cos\thcm)$ does not guarantee that one can retrieve the underlying transversity amplitudes. In a recent analysis of $\gamma N\to \pi N$ simulated data \cite{Workman:2011hi} it was shown that the availability of the $\cos\thcm$ dependence of six observables at given $W$ yields unique partial-wave solutions.

\subsection{Extracting the transversity amplitudes from RPR-2011 simulations}
\label{subsec:amplitude_extraction}
In this section the completeness of complete sets is investigated. The analysis is performed for the representative complete set $\{C_x,O_x,E,F\}$ of the first kind. The goal is to determine the moduli $\widehat{r}_i$ and the consistent independent phases $\widetilde{\delta}_i$ of the transversity amplitudes from simulated observables with finite experimental resolution generated with the RPR-2011 model. It is investigated to what extent the retrieved amplitudes comply with the input amplitudes from the simulations.

\subsubsection{Strategy}
\label{subsubsec:strategy}
A measured asymmetry is simulated by generating a fixed number of events from a Gaussian distribution. The mean of this distribution is the RPR-2011 prediction for a certain value of $W$ and $\cos\thcm$. The standard deviation is a specific experimental resolution $\sigma_\text{exp}$. The mean and standard deviation of the generated events determine the value and the error of the simulated data point. If the error of the simulated data point is smaller than the specified $\sigma_\text{exp}$, the former is rejected.

From a simulated set $\{C_x,O_x,$\-$E,F\}$, one can obtain the $\widehat{r}_i$ and the $\widetilde{\delta}_i$ by means of  Eqs.~\eq{eq:moduli_expressions} and \eq{eq:consistent_set}. In order to estimate the error on the $\widehat{r}_i$ and the $\widetilde{\delta}_i$, standard error propagation is applied. Since seven asymmetries are required to estimate the six parameters of the normalized transversity amplitudes, the correlation between the asymmetries has to be taken into account upon estimating errors. The squared error of a certain function $f$ of the asymmetries is calculated as
\begin{multline}
\sigma^2(f) = \sum_{i}	\left(\frac{\partial f}{\partial A_i}\right)^2\sigma^2(A_i) \\+
\sum_{\substack{i, j\\ i \neq j}}\frac{\partial f}{\partial A_i}\frac{\partial f}{\partial A_j}\sigma(A_i,A_j),
\label{eq:error_propagation}
\end{multline}
with $A_i,A_j \in \{\Sigma,T,P,C_x,O_x,E,F\}$ and $\sigma(A_i,A_j)$ the covariance between $A_i$ and $A_j$.

\subsubsection{Analysis}
\label{subsubsec:analysis}
As indicated in Sec.\ \ref{subsec:merits_transversity}, the extraction of the $\widehat{r}_i$ from a measured set of single asymmetries through Eqs.\ \eq{eq:moduli_expressions} is rather straightforward. In some isolated situations at least one of the retrieved $\widehat{r}_i$ became imaginary. The moduli estimates, however, are required to extract the relevant phases from a complete set, as is seen from Eqs.\ \eq{eq:Cx-Ox}, \eq{eq:E-F}, and \eq{eq:H}. Consequently, an imaginary estimate for one of the moduli results in imaginary estimates for the phases, no matter the achieved precision of the double-asymmetry observables. Through the normalization condition \eq{eq:normalization} only three of the four moduli are independent. One imaginary modulus estimate, be it an independent or a nonindependent one, is sufficient to jeopardize the phase analysis. For this reason, the four moduli are treated on equal grounds and a selection of a set of three independent moduli has been avoided all along.

\begin{table}[!t]
\caption{The model's values and two simulated datasets for the complete set $\{C_x,O_x,E,F\}$ at $W = 1700$ MeV and $\cos\thcm = -0.5$ with $\sigma_\text{exp} = 0.05$.}\label{tab:compared_measurements}
\centering
\renewcommand{\arraystretch}{1.3}
\begin{tabular}{c|C{45pt}|C{85pt}|C{85pt}}
\hline
\noalign{\smallskip}\hline
& {\bf Model} & {\bf Simulation A} & {\bf Simulation B}\\
\hline
$\Sigma$ & $\phantom{-}0.0489$ & $\phantom{-}0.0543 \pm 0.0519$ & $\phantom{-}0.0458 \pm 0.0533$ \\
$T$ & $-0.6843$ & $-0.6905 \pm 0.0511$ & $-0.6685 \pm 0.0524$ \\
$P$ & $\phantom{-}0.0056$ & $\phantom{-}0.0111 \pm 0.0552$ & $\phantom{-}0.0892 \pm 0.0555$ \\
$C_x$ & $-0.4808$ & $-0.4809 \pm 0.0515$ & $-0.4728 \pm 0.0588$ \\
$O_x$ & $-0.5989$ & $-0.5971 \pm 0.0519$ & $-0.6149 \pm 0.0567$ \\
$E$ & $\phantom{-}0.5672$ & $\phantom{-}0.5655 \pm 0.0505$ & $\phantom{-}0.5695 \pm 0.0564$ \\
$F$ & $-0.4525$ & $-0.4411 \pm 0.0519$ & $-0.4486 \pm 0.0503$ \\
\hline
\noalign{\smallskip}\hline
\end{tabular}
\end{table}

\begin{table}[!b]
\caption{The model's values for the $\widehat{r}_i$ at $W = 1700$ MeV and $\cos\thcm = -0.5$, and the corresponding estimates from dataset A and dataset B.}\label{tab:compared_moduli}
\centering
\renewcommand{\arraystretch}{1.3}
\begin{tabular}{C{12.5pt}|C{45pt}|C{85pt}|C{85pt}}
\hline
\noalign{\smallskip}\hline
& {\bf Model} & {\bf Simulation A} & {\bf Simulation B}\\
\hline
$\widehat{r}_1$ & $0.304$ & $0.306 \pm 0.037$ & $0.311 \pm 0.037$ \\
$\widehat{r}_2$ & $0.657$ & $0.658 \pm 0.017$ & $0.653 \pm 0.018$ \\
$\widehat{r}_3$ & $0.640$ & $0.642 \pm 0.018$ & $0.639 \pm 0.018$ \\
$\widehat{r}_4$ & $0.256$ & $0.248 \pm 0.046$ & $0.263 \pm 0.044$ \\
\hline
\noalign{\smallskip}\hline
\end{tabular}
\end{table}

\begin{table}[!t]
\caption{The constraint of Eq.~\eq{eq:phases_constraint} for the four $\{\delta_1, \delta_2, \Delta_{13}, \Delta_{23}\}$ solutions extracted from the simulated data listed in Table~\ref{tab:compared_measurements}. The number of standard deviations, by which the constraint differs from zero, is denoted by $n_\sigma$. The \textsc{ci} are the confidence intervals corresponding with the $n_\sigma$.}\label{tab:compared_constraints}
\centering
\renewcommand{\arraystretch}{1.3}
\begin{tabular}{c|C{75pt}C{30pt}C{35pt}}
\hline
\noalign{\smallskip}\hline
\multirow{2}{*}{\bf Solution} & \multicolumn{3}{c}{\bf Simulation A}\\
& Constraint & $n_\sigma$ & \textsc{ci} (\%) \\
\hline
$1$ & $\phantom{-}0.029 \pm 3.866$ & $0.007$ & $99.4$ \\
$2$ & $-0.200 \pm 3.316$ & $0.060$ & $95.2$ \\
$3$ & $-2.513 \pm 3.316$ & $0.758$ & $44.9$ \\
$4$ & $-2.741 \pm 3.865$ & $0.709$ & $47.8$ \\
\hline
\noalign{\smallskip}\hline
\end{tabular}
\\ $ $ \\ $ $ \\
\begin{tabular}{c|C{75pt}C{30pt}C{35pt}}
\hline
\noalign{\smallskip}\hline
\multirow{2}{*}{\bf Solution} & \multicolumn{3}{c}{\bf Simulation B}\\
& Constraint & $n_\sigma$ & \textsc{ci} (\%) \\
\hline
$1$ & $\phantom{-}0.592 \pm 1.261$ & $0.467$ & $63.8$ \\
$2$ & $-0.229 \pm 0.898$ & $0.255$ & $79.9$ \\
$3$ & $-2.393 \pm 0.903$ & $2.651$ & $\phantom{0}0.8$ \\
$4$ & $\phantom{-}3.069 \pm 1.264$ & $2.428$ & $\phantom{0}1.5$ \\
\hline
\noalign{\smallskip}\hline
\end{tabular}
\end{table}

Estimating the $\delta_i$ is far less straightforward than estimating the $r_i$. Table~\ref{tab:compared_measurements} lists two simulated measurements of the complete set $\{C_x,O_x,E,F\}$ at $W = 1700$ MeV and $\cos\thcm = -0.5$, performed at an input resolution of $\sigma_\text{exp} = 0.05$. The mean value and standard deviation of the asymmetries were calculated from a sample of 50 events. Upon inspecting Table~\ref{tab:compared_measurements}, both simulations are seen to be qualitatively equivalent. This is also reflected by the accurate moduli estimates from both simulations, which are listed in Table~\ref{tab:compared_moduli}. In Sec.\ \ref{subsubsec:complete_sets} it was mentioned that four distinct solutions for $\{\delta_1, \delta_2, \Delta_{13}, \Delta_{23}\}$ exist for the concerned complete set. Only one satisfies the constraint \eq{eq:phases_constraint} in the exact case, though. When experimental error is involved, however, none of the four solutions satisfies the constraint. In Table~\ref{tab:compared_constraints}, the outcome of constraint \eq{eq:phases_constraint} is listed for each of the four solutions to both datasets. As the constraint is a sum of phases and its evaluation in the exact case yields a zero value, the mean values listed in Table~\ref{tab:compared_constraints} were rotated to $]-\pi,+\pi]$. The table also lists by how many standard deviations (denoted by $n_\sigma$) the mean deviates from zero. The solution with the lowest $n_\sigma$ has the highest likelihood. In principle a solution can be excluded if it is statistically insignificant, commonly quantified by a small confidence interval $\textsc{ci} = 1 - \text{erf} \left(n_\sigma/\sqrt{2}\right)$. From the $\textsc{ci}$ values of dataset B it follows that solution 3 and solution 4 can be excluded with significance. For dataset A, these solutions are also the least significant ones. Of the two remaining solutions, dataset A predicts that solution 1 is the real solution ($99.4\%$ \textsc{ci}). According to dataset B, 
however, solution 2 is the most likely one, though with a smaller confidence interval ($79.9\%$ \textsc{ci}). This example illustrates that ambiguities remain for (theoretically) complete sets when experimental uncertainties are taken into account.

\begin{table}[!b]
\caption{The model's values for the $\widetilde{\delta}_i$ at $W = 1700$ MeV and $\cos\thcm = -0.5$, along with the estimates resulting from the most significant solution of dataset A (solution 1) and dataset B (solution 2).}\label{tab:compared_phases}
\centering
\renewcommand{\arraystretch}{1.3}
\begin{tabular}{C{12.5pt}|C{45pt}|C{85pt}|C{85pt}}
\hline
\noalign{\smallskip}\hline
& {\bf Model} & {\bf Simulation A} & {\bf Simulation B}\\
\hline
$\widetilde{\delta}_1$ & $0.487$ & $0.424 \pm 0.801$ & $0.704 \pm 0.478$ \\
$\widetilde{\delta}_2$ & $2.458$ & $2.424 \pm 0.992$ & $2.637 \pm 0.348$ \\
$\widetilde{\delta}_3$ & $6.106$ & $6.090 \pm 0.235$ & $3.046 \pm 0.234$ \\
\hline
\noalign{\smallskip}\hline
\end{tabular}
\end{table}

\begin{figure*}[!t]
\begin{center}
\footnotesize
\begingroup
  \makeatletter
  \providecommand\color[2][]{%
    \GenericError{(gnuplot) \space\space\space\@spaces}{%
      Package color not loaded in conjunction with
      terminal option `colourtext'%
    }{See the gnuplot documentation for explanation.%
    }{Either use 'blacktext' in gnuplot or load the package
      color.sty in LaTeX.}%
    \renewcommand\color[2][]{}%
  }%
  \providecommand\includegraphics[2][]{%
    \GenericError{(gnuplot) \space\space\space\@spaces}{%
      Package graphicx or graphics not loaded%
    }{See the gnuplot documentation for explanation.%
    }{The gnuplot epslatex terminal needs graphicx.sty or graphics.sty.}%
    \renewcommand\includegraphics[2][]{}%
  }%
  \providecommand\rotatebox[2]{#2}%
  \@ifundefined{ifGPcolor}{%
    \newif\ifGPcolor
    \GPcolortrue
  }{}%
  \@ifundefined{ifGPblacktext}{%
    \newif\ifGPblacktext
    \GPblacktexttrue
  }{}%
  \let\gplgaddtomacro\g@addto@macro
  \gdef\gplbacktext{}%
  \gdef\gplfronttext{}%
  \makeatother
  \ifGPblacktext
    \def\colorrgb#1{}%
    \def\colorgray#1{}%
  \else
    \ifGPcolor
      \def\colorrgb#1{\color[rgb]{#1}}%
      \def\colorgray#1{\color[gray]{#1}}%
      \expandafter\def\csname LTw\endcsname{\color{white}}%
      \expandafter\def\csname LTb\endcsname{\color{black}}%
      \expandafter\def\csname LTa\endcsname{\color{black}}%
      \expandafter\def\csname LT0\endcsname{\color[rgb]{1,0,0}}%
      \expandafter\def\csname LT1\endcsname{\color[rgb]{0,1,0}}%
      \expandafter\def\csname LT2\endcsname{\color[rgb]{0,0,1}}%
      \expandafter\def\csname LT3\endcsname{\color[rgb]{1,0,1}}%
      \expandafter\def\csname LT4\endcsname{\color[rgb]{0,1,1}}%
      \expandafter\def\csname LT5\endcsname{\color[rgb]{1,1,0}}%
      \expandafter\def\csname LT6\endcsname{\color[rgb]{0,0,0}}%
      \expandafter\def\csname LT7\endcsname{\color[rgb]{1,0.3,0}}%
      \expandafter\def\csname LT8\endcsname{\color[rgb]{0.5,0.5,0.5}}%
    \else
      \def\colorrgb#1{\color{black}}%
      \def\colorgray#1{\color[gray]{#1}}%
      \expandafter\def\csname LTw\endcsname{\color{white}}%
      \expandafter\def\csname LTb\endcsname{\color{black}}%
      \expandafter\def\csname LTa\endcsname{\color{black}}%
      \expandafter\def\csname LT0\endcsname{\color{black}}%
      \expandafter\def\csname LT1\endcsname{\color{black}}%
      \expandafter\def\csname LT2\endcsname{\color{black}}%
      \expandafter\def\csname LT3\endcsname{\color{black}}%
      \expandafter\def\csname LT4\endcsname{\color{black}}%
      \expandafter\def\csname LT5\endcsname{\color{black}}%
      \expandafter\def\csname LT6\endcsname{\color{black}}%
      \expandafter\def\csname LT7\endcsname{\color{black}}%
      \expandafter\def\csname LT8\endcsname{\color{black}}%
    \fi
  \fi
  \setlength{\unitlength}{0.0500bp}%
  \begin{picture}(3542.00,2360.00)%
    \gplgaddtomacro\gplbacktext{%
      \csname LTb\endcsname%
      \put(-60,60){\makebox(0,0)[r]{\strut{} 0}}%
      \put(-60,515){\makebox(0,0)[r]{\strut{} 0.2}}%
      \put(-60,970){\makebox(0,0)[r]{\strut{} 0.4}}%
      \put(-60,1425){\makebox(0,0)[r]{\strut{} 0.6}}%
      \put(-60,1880){\makebox(0,0)[r]{\strut{} 0.8}}%
      \put(-60,2335){\makebox(0,0)[r]{\strut{} 1}}%
      \put(2321,-120){\makebox(0,0){\strut{} 0.01}}%
      \put(12,-120){\makebox(0,0){\strut{} 0.1}}%
      \put(-550,1197){\rotatebox{-270}{\makebox(0,0){\normalsize $\eta$}}}%
      \put(3364,826){\makebox(0,0)[r]{\strut{}$W = 1700$ MeV}}%
      \put(3364,590){\makebox(0,0)[r]{\strut{}$\cos\thcm = 0.9$}}%
    }%
    \gplgaddtomacro\gplfronttext{%
      \csname LTb\endcsname%
      \put(3049,2076){\makebox(0,0)[r]{\strut{}$\eta$ $\{C_x, O_x, E, F\}$\;}}%
      \csname LTb\endcsname%
      \put(3049,1840){\makebox(0,0)[r]{\strut{}$\eta_\text{imaginary}$ $\{C_x, O_x, E, F\}$\;}}%
      \csname LTb\endcsname%
      \put(3049,1604){\makebox(0,0)[r]{\strut{}$\eta$ $\{C_x, O_x, G, H\}$\;}}%
      \csname LTb\endcsname%
      \put(3049,1368){\makebox(0,0)[r]{\strut{}$\eta_\text{imaginary}$ $\{C_x, O_x, G, H\}$\;}}%
    }%
    \gplbacktext
    \put(0,0){\includegraphics{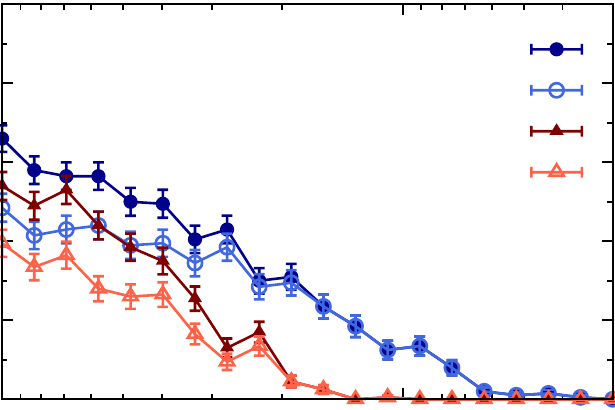}}%
    \gplfronttext
  \end{picture}%
\endgroup
\hspace{20pt}
\begingroup
  \makeatletter
  \providecommand\color[2][]{%
    \GenericError{(gnuplot) \space\space\space\@spaces}{%
      Package color not loaded in conjunction with
      terminal option `colourtext'%
    }{See the gnuplot documentation for explanation.%
    }{Either use 'blacktext' in gnuplot or load the package
      color.sty in LaTeX.}%
    \renewcommand\color[2][]{}%
  }%
  \providecommand\includegraphics[2][]{%
    \GenericError{(gnuplot) \space\space\space\@spaces}{%
      Package graphicx or graphics not loaded%
    }{See the gnuplot documentation for explanation.%
    }{The gnuplot epslatex terminal needs graphicx.sty or graphics.sty.}%
    \renewcommand\includegraphics[2][]{}%
  }%
  \providecommand\rotatebox[2]{#2}%
  \@ifundefined{ifGPcolor}{%
    \newif\ifGPcolor
    \GPcolortrue
  }{}%
  \@ifundefined{ifGPblacktext}{%
    \newif\ifGPblacktext
    \GPblacktexttrue
  }{}%
  \let\gplgaddtomacro\g@addto@macro
  \gdef\gplbacktext{}%
  \gdef\gplfronttext{}%
  \makeatother
  \ifGPblacktext
    \def\colorrgb#1{}%
    \def\colorgray#1{}%
  \else
    \ifGPcolor
      \def\colorrgb#1{\color[rgb]{#1}}%
      \def\colorgray#1{\color[gray]{#1}}%
      \expandafter\def\csname LTw\endcsname{\color{white}}%
      \expandafter\def\csname LTb\endcsname{\color{black}}%
      \expandafter\def\csname LTa\endcsname{\color{black}}%
      \expandafter\def\csname LT0\endcsname{\color[rgb]{1,0,0}}%
      \expandafter\def\csname LT1\endcsname{\color[rgb]{0,1,0}}%
      \expandafter\def\csname LT2\endcsname{\color[rgb]{0,0,1}}%
      \expandafter\def\csname LT3\endcsname{\color[rgb]{1,0,1}}%
      \expandafter\def\csname LT4\endcsname{\color[rgb]{0,1,1}}%
      \expandafter\def\csname LT5\endcsname{\color[rgb]{1,1,0}}%
      \expandafter\def\csname LT6\endcsname{\color[rgb]{0,0,0}}%
      \expandafter\def\csname LT7\endcsname{\color[rgb]{1,0.3,0}}%
      \expandafter\def\csname LT8\endcsname{\color[rgb]{0.5,0.5,0.5}}%
    \else
      \def\colorrgb#1{\color{black}}%
      \def\colorgray#1{\color[gray]{#1}}%
      \expandafter\def\csname LTw\endcsname{\color{white}}%
      \expandafter\def\csname LTb\endcsname{\color{black}}%
      \expandafter\def\csname LTa\endcsname{\color{black}}%
      \expandafter\def\csname LT0\endcsname{\color{black}}%
      \expandafter\def\csname LT1\endcsname{\color{black}}%
      \expandafter\def\csname LT2\endcsname{\color{black}}%
      \expandafter\def\csname LT3\endcsname{\color{black}}%
      \expandafter\def\csname LT4\endcsname{\color{black}}%
      \expandafter\def\csname LT5\endcsname{\color{black}}%
      \expandafter\def\csname LT6\endcsname{\color{black}}%
      \expandafter\def\csname LT7\endcsname{\color{black}}%
      \expandafter\def\csname LT8\endcsname{\color{black}}%
    \fi
  \fi
  \setlength{\unitlength}{0.0500bp}%
  \begin{picture}(3542.00,2360.00)%
    \gplgaddtomacro\gplbacktext{%
      \csname LTb\endcsname%
      \put(-60,60){\makebox(0,0)[r]{\strut{}}}%
      \put(-60,515){\makebox(0,0)[r]{\strut{}}}%
      \put(-60,970){\makebox(0,0)[r]{\strut{}}}%
      \put(-60,1425){\makebox(0,0)[r]{\strut{}}}%
      \put(-60,1880){\makebox(0,0)[r]{\strut{}}}%
      \put(-60,2335){\makebox(0,0)[r]{\strut{}}}%
      \put(2715,-120){\makebox(0,0){\strut{} 0.01}}%
      \put(12,-120){\makebox(0,0){\strut{} 0.1}}%
      \put(3364,2064){\makebox(0,0)[r]{\strut{}$W = 1800$ MeV}}%
      \put(3364,1828){\makebox(0,0)[r]{\strut{}$\cos\thcm = 0$}}%
    }%
    \gplgaddtomacro\gplfronttext{%
    }%
    \gplbacktext
    \put(0,0){\includegraphics{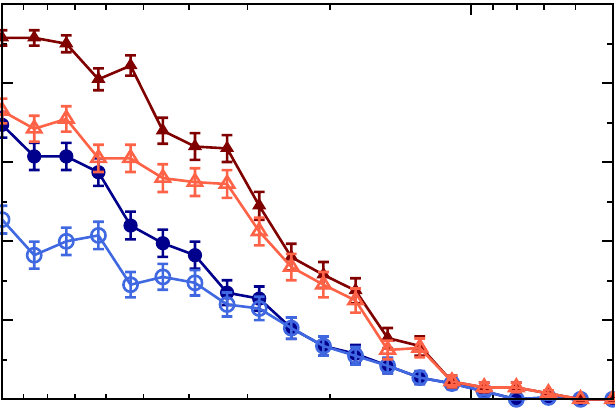}}%
    \gplfronttext
  \end{picture}%
\endgroup
\\ $ $ \\ $ $ \\
\begingroup
  \makeatletter
  \providecommand\color[2][]{%
    \GenericError{(gnuplot) \space\space\space\@spaces}{%
      Package color not loaded in conjunction with
      terminal option `colourtext'%
    }{See the gnuplot documentation for explanation.%
    }{Either use 'blacktext' in gnuplot or load the package
      color.sty in LaTeX.}%
    \renewcommand\color[2][]{}%
  }%
  \providecommand\includegraphics[2][]{%
    \GenericError{(gnuplot) \space\space\space\@spaces}{%
      Package graphicx or graphics not loaded%
    }{See the gnuplot documentation for explanation.%
    }{The gnuplot epslatex terminal needs graphicx.sty or graphics.sty.}%
    \renewcommand\includegraphics[2][]{}%
  }%
  \providecommand\rotatebox[2]{#2}%
  \@ifundefined{ifGPcolor}{%
    \newif\ifGPcolor
    \GPcolortrue
  }{}%
  \@ifundefined{ifGPblacktext}{%
    \newif\ifGPblacktext
    \GPblacktexttrue
  }{}%
  \let\gplgaddtomacro\g@addto@macro
  \gdef\gplbacktext{}%
  \gdef\gplfronttext{}%
  \makeatother
  \ifGPblacktext
    \def\colorrgb#1{}%
    \def\colorgray#1{}%
  \else
    \ifGPcolor
      \def\colorrgb#1{\color[rgb]{#1}}%
      \def\colorgray#1{\color[gray]{#1}}%
      \expandafter\def\csname LTw\endcsname{\color{white}}%
      \expandafter\def\csname LTb\endcsname{\color{black}}%
      \expandafter\def\csname LTa\endcsname{\color{black}}%
      \expandafter\def\csname LT0\endcsname{\color[rgb]{1,0,0}}%
      \expandafter\def\csname LT1\endcsname{\color[rgb]{0,1,0}}%
      \expandafter\def\csname LT2\endcsname{\color[rgb]{0,0,1}}%
      \expandafter\def\csname LT3\endcsname{\color[rgb]{1,0,1}}%
      \expandafter\def\csname LT4\endcsname{\color[rgb]{0,1,1}}%
      \expandafter\def\csname LT5\endcsname{\color[rgb]{1,1,0}}%
      \expandafter\def\csname LT6\endcsname{\color[rgb]{0,0,0}}%
      \expandafter\def\csname LT7\endcsname{\color[rgb]{1,0.3,0}}%
      \expandafter\def\csname LT8\endcsname{\color[rgb]{0.5,0.5,0.5}}%
    \else
      \def\colorrgb#1{\color{black}}%
      \def\colorgray#1{\color[gray]{#1}}%
      \expandafter\def\csname LTw\endcsname{\color{white}}%
      \expandafter\def\csname LTb\endcsname{\color{black}}%
      \expandafter\def\csname LTa\endcsname{\color{black}}%
      \expandafter\def\csname LT0\endcsname{\color{black}}%
      \expandafter\def\csname LT1\endcsname{\color{black}}%
      \expandafter\def\csname LT2\endcsname{\color{black}}%
      \expandafter\def\csname LT3\endcsname{\color{black}}%
      \expandafter\def\csname LT4\endcsname{\color{black}}%
      \expandafter\def\csname LT5\endcsname{\color{black}}%
      \expandafter\def\csname LT6\endcsname{\color{black}}%
      \expandafter\def\csname LT7\endcsname{\color{black}}%
      \expandafter\def\csname LT8\endcsname{\color{black}}%
    \fi
  \fi
  \setlength{\unitlength}{0.0500bp}%
  \begin{picture}(3542.00,2360.00)%
    \gplgaddtomacro\gplbacktext{%
      \csname LTb\endcsname%
      \put(-60,60){\makebox(0,0)[r]{\strut{} 0}}%
      \put(-60,515){\makebox(0,0)[r]{\strut{} 0.2}}%
      \put(-60,970){\makebox(0,0)[r]{\strut{} 0.4}}%
      \put(-60,1425){\makebox(0,0)[r]{\strut{} 0.6}}%
      \put(-60,1880){\makebox(0,0)[r]{\strut{} 0.8}}%
      \put(-60,2335){\makebox(0,0)[r]{\strut{} 1}}%
      \put(3529,-120){\makebox(0,0){\strut{} 0.0001}}%
      \put(2357,-120){\makebox(0,0){\strut{} 0.001}}%
      \put(1184,-120){\makebox(0,0){\strut{} 0.01}}%
      \put(12,-120){\makebox(0,0){\strut{} 0.1}}%
      \put(-550,1197){\rotatebox{-270}{\makebox(0,0){\normalsize $\eta$}}}%
      \put(1770,-490){\makebox(0,0){\normalsize $\sigma_\text{exp}$}}%
      \put(3364,2064){\makebox(0,0)[r]{\strut{}$W = 1800$ MeV}}%
      \put(3364,1828){\makebox(0,0)[r]{\strut{}$\cos\thcm = -0.3$}}%
    }%
    \gplgaddtomacro\gplfronttext{%
    }%
    \gplbacktext
    \put(0,0){\includegraphics{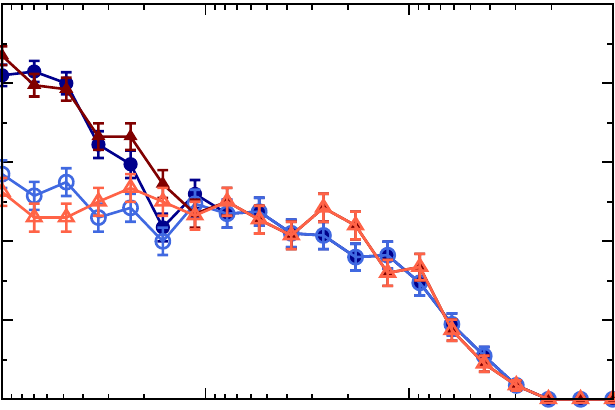}}%
    \gplfronttext
  \end{picture}%
\endgroup
\hspace{20pt}
\begingroup
  \makeatletter
  \providecommand\color[2][]{%
    \GenericError{(gnuplot) \space\space\space\@spaces}{%
      Package color not loaded in conjunction with
      terminal option `colourtext'%
    }{See the gnuplot documentation for explanation.%
    }{Either use 'blacktext' in gnuplot or load the package
      color.sty in LaTeX.}%
    \renewcommand\color[2][]{}%
  }%
  \providecommand\includegraphics[2][]{%
    \GenericError{(gnuplot) \space\space\space\@spaces}{%
      Package graphicx or graphics not loaded%
    }{See the gnuplot documentation for explanation.%
    }{The gnuplot epslatex terminal needs graphicx.sty or graphics.sty.}%
    \renewcommand\includegraphics[2][]{}%
  }%
  \providecommand\rotatebox[2]{#2}%
  \@ifundefined{ifGPcolor}{%
    \newif\ifGPcolor
    \GPcolortrue
  }{}%
  \@ifundefined{ifGPblacktext}{%
    \newif\ifGPblacktext
    \GPblacktexttrue
  }{}%
  \let\gplgaddtomacro\g@addto@macro
  \gdef\gplbacktext{}%
  \gdef\gplfronttext{}%
  \makeatother
  \ifGPblacktext
    \def\colorrgb#1{}%
    \def\colorgray#1{}%
  \else
    \ifGPcolor
      \def\colorrgb#1{\color[rgb]{#1}}%
      \def\colorgray#1{\color[gray]{#1}}%
      \expandafter\def\csname LTw\endcsname{\color{white}}%
      \expandafter\def\csname LTb\endcsname{\color{black}}%
      \expandafter\def\csname LTa\endcsname{\color{black}}%
      \expandafter\def\csname LT0\endcsname{\color[rgb]{1,0,0}}%
      \expandafter\def\csname LT1\endcsname{\color[rgb]{0,1,0}}%
      \expandafter\def\csname LT2\endcsname{\color[rgb]{0,0,1}}%
      \expandafter\def\csname LT3\endcsname{\color[rgb]{1,0,1}}%
      \expandafter\def\csname LT4\endcsname{\color[rgb]{0,1,1}}%
      \expandafter\def\csname LT5\endcsname{\color[rgb]{1,1,0}}%
      \expandafter\def\csname LT6\endcsname{\color[rgb]{0,0,0}}%
      \expandafter\def\csname LT7\endcsname{\color[rgb]{1,0.3,0}}%
      \expandafter\def\csname LT8\endcsname{\color[rgb]{0.5,0.5,0.5}}%
    \else
      \def\colorrgb#1{\color{black}}%
      \def\colorgray#1{\color[gray]{#1}}%
      \expandafter\def\csname LTw\endcsname{\color{white}}%
      \expandafter\def\csname LTb\endcsname{\color{black}}%
      \expandafter\def\csname LTa\endcsname{\color{black}}%
      \expandafter\def\csname LT0\endcsname{\color{black}}%
      \expandafter\def\csname LT1\endcsname{\color{black}}%
      \expandafter\def\csname LT2\endcsname{\color{black}}%
      \expandafter\def\csname LT3\endcsname{\color{black}}%
      \expandafter\def\csname LT4\endcsname{\color{black}}%
      \expandafter\def\csname LT5\endcsname{\color{black}}%
      \expandafter\def\csname LT6\endcsname{\color{black}}%
      \expandafter\def\csname LT7\endcsname{\color{black}}%
      \expandafter\def\csname LT8\endcsname{\color{black}}%
    \fi
  \fi
  \setlength{\unitlength}{0.0500bp}%
  \begin{picture}(3542.00,2360.00)%
    \gplgaddtomacro\gplbacktext{%
      \csname LTb\endcsname%
      \put(-60,60){\makebox(0,0)[r]{\strut{}}}%
      \put(-60,515){\makebox(0,0)[r]{\strut{}}}%
      \put(-60,970){\makebox(0,0)[r]{\strut{}}}%
      \put(-60,1425){\makebox(0,0)[r]{\strut{}}}%
      \put(-60,1880){\makebox(0,0)[r]{\strut{}}}%
      \put(-60,2335){\makebox(0,0)[r]{\strut{}}}%
      \put(3529,-120){\makebox(0,0){\strut{} 0.0001}}%
      \put(2357,-120){\makebox(0,0){\strut{} 0.001}}%
      \put(1184,-120){\makebox(0,0){\strut{} 0.01}}%
      \put(12,-120){\makebox(0,0){\strut{} 0.1}}%
      \put(1770,-490){\makebox(0,0){\normalsize $\sigma_\text{exp}$}}%
      \put(3364,2064){\makebox(0,0)[r]{\strut{}$W = 2200$ MeV}}%
      \put(3364,1828){\makebox(0,0)[r]{\strut{}$\cos\thcm = -0.9$}}%
    }%
    \gplgaddtomacro\gplfronttext{%
    }%
    \gplbacktext
    \put(0,0){\includegraphics{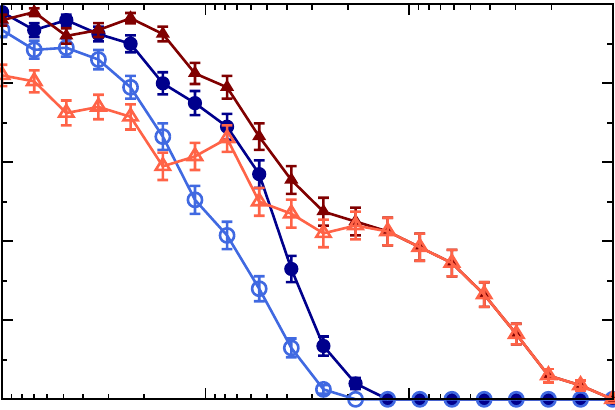}}%
    \gplfronttext
  \end{picture}%
\endgroup
\\ $ $ \\ $ $ \\
\end{center}
\caption{(Color online) The insolvability of the transversity amplitudes at four different kinematical points. The circles and triangles are simulated data generated from the complete sets $\{C_x,O_x,E,F\}$ and $\{C_x,O_x,G,H\}$. The filled and unfilled symbols correspond with $\eta = \eta_\text{imaginary} + \eta_\text{incorrect}$ and $\eta_\text{imaginary}$.}
\label{fig:insolvabilities}
\end{figure*}

\begin{figure*}[!t]
\begin{center}
\footnotesize
\begingroup
  \makeatletter
  \providecommand\color[2][]{%
    \GenericError{(gnuplot) \space\space\space\@spaces}{%
      Package color not loaded in conjunction with
      terminal option `colourtext'%
    }{See the gnuplot documentation for explanation.%
    }{Either use 'blacktext' in gnuplot or load the package
      color.sty in LaTeX.}%
    \renewcommand\color[2][]{}%
  }%
  \providecommand\includegraphics[2][]{%
    \GenericError{(gnuplot) \space\space\space\@spaces}{%
      Package graphicx or graphics not loaded%
    }{See the gnuplot documentation for explanation.%
    }{The gnuplot epslatex terminal needs graphicx.sty or graphics.sty.}%
    \renewcommand\includegraphics[2][]{}%
  }%
  \providecommand\rotatebox[2]{#2}%
  \@ifundefined{ifGPcolor}{%
    \newif\ifGPcolor
    \GPcolortrue
  }{}%
  \@ifundefined{ifGPblacktext}{%
    \newif\ifGPblacktext
    \GPblacktexttrue
  }{}%
  \let\gplgaddtomacro\g@addto@macro
  \gdef\gplbacktext{}%
  \gdef\gplfronttext{}%
  \makeatother
  \ifGPblacktext
    \def\colorrgb#1{}%
    \def\colorgray#1{}%
  \else
    \ifGPcolor
      \def\colorrgb#1{\color[rgb]{#1}}%
      \def\colorgray#1{\color[gray]{#1}}%
      \expandafter\def\csname LTw\endcsname{\color{white}}%
      \expandafter\def\csname LTb\endcsname{\color{black}}%
      \expandafter\def\csname LTa\endcsname{\color{black}}%
      \expandafter\def\csname LT0\endcsname{\color[rgb]{1,0,0}}%
      \expandafter\def\csname LT1\endcsname{\color[rgb]{0,1,0}}%
      \expandafter\def\csname LT2\endcsname{\color[rgb]{0,0,1}}%
      \expandafter\def\csname LT3\endcsname{\color[rgb]{1,0,1}}%
      \expandafter\def\csname LT4\endcsname{\color[rgb]{0,1,1}}%
      \expandafter\def\csname LT5\endcsname{\color[rgb]{1,1,0}}%
      \expandafter\def\csname LT6\endcsname{\color[rgb]{0,0,0}}%
      \expandafter\def\csname LT7\endcsname{\color[rgb]{1,0.3,0}}%
      \expandafter\def\csname LT8\endcsname{\color[rgb]{0.5,0.5,0.5}}%
    \else
      \def\colorrgb#1{\color{black}}%
      \def\colorgray#1{\color[gray]{#1}}%
      \expandafter\def\csname LTw\endcsname{\color{white}}%
      \expandafter\def\csname LTb\endcsname{\color{black}}%
      \expandafter\def\csname LTa\endcsname{\color{black}}%
      \expandafter\def\csname LT0\endcsname{\color{black}}%
      \expandafter\def\csname LT1\endcsname{\color{black}}%
      \expandafter\def\csname LT2\endcsname{\color{black}}%
      \expandafter\def\csname LT3\endcsname{\color{black}}%
      \expandafter\def\csname LT4\endcsname{\color{black}}%
      \expandafter\def\csname LT5\endcsname{\color{black}}%
      \expandafter\def\csname LT6\endcsname{\color{black}}%
      \expandafter\def\csname LT7\endcsname{\color{black}}%
      \expandafter\def\csname LT8\endcsname{\color{black}}%
    \fi
  \fi
  \setlength{\unitlength}{0.0500bp}%
  \begin{picture}(4626.00,3148.00)%
    \gplgaddtomacro\gplbacktext{%
      \csname LTb\endcsname%
      \put(2274,2881){\makebox(0,0){\normalsize $\eta \;(\sigma_\text{exp} = 0.1)$}}%
    }%
    \gplgaddtomacro\gplfronttext{%
      \csname LTb\endcsname%
      \put(937,211){\makebox(0,0){\strut{}}}%
      \put(1314,211){\makebox(0,0){\strut{}}}%
      \put(1692,211){\makebox(0,0){\strut{}}}%
      \put(2069,211){\makebox(0,0){\strut{}}}%
      \put(2446,211){\makebox(0,0){\strut{}}}%
      \put(2823,211){\makebox(0,0){\strut{}}}%
      \put(3200,211){\makebox(0,0){\strut{}}}%
      \put(3578,211){\makebox(0,0){\strut{}}}%
      \put(3955,211){\makebox(0,0){\strut{}}}%
      \put(438,438){\makebox(0,0)[r]{\strut{}-1}}%
      \put(438,983){\makebox(0,0)[r]{\strut{}-0.5}}%
      \put(438,1528){\makebox(0,0)[r]{\strut{} 0}}%
      \put(438,2073){\makebox(0,0)[r]{\strut{} 0.5}}%
      \put(438,2618){\makebox(0,0)[r]{\strut{} 1}}%
      \put(-96,1528){\rotatebox{-270}{\makebox(0,0){\normalsize $\cos\thcm$}}}%
    }%
    \gplbacktext
    \put(500,353){\includegraphics[scale=0.66]{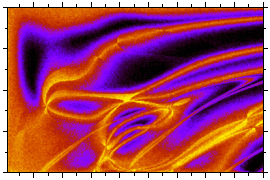}}%
    \gplfronttext
  \end{picture}%
\endgroup
\hspace{-41pt}\vspace{-15pt}
\begingroup
  \makeatletter
  \providecommand\color[2][]{%
    \GenericError{(gnuplot) \space\space\space\@spaces}{%
      Package color not loaded in conjunction with
      terminal option `colourtext'%
    }{See the gnuplot documentation for explanation.%
    }{Either use 'blacktext' in gnuplot or load the package
      color.sty in LaTeX.}%
    \renewcommand\color[2][]{}%
  }%
  \providecommand\includegraphics[2][]{%
    \GenericError{(gnuplot) \space\space\space\@spaces}{%
      Package graphicx or graphics not loaded%
    }{See the gnuplot documentation for explanation.%
    }{The gnuplot epslatex terminal needs graphicx.sty or graphics.sty.}%
    \renewcommand\includegraphics[2][]{}%
  }%
  \providecommand\rotatebox[2]{#2}%
  \@ifundefined{ifGPcolor}{%
    \newif\ifGPcolor
    \GPcolortrue
  }{}%
  \@ifundefined{ifGPblacktext}{%
    \newif\ifGPblacktext
    \GPblacktexttrue
  }{}%
  \let\gplgaddtomacro\g@addto@macro
  \gdef\gplbacktext{}%
  \gdef\gplfronttext{}%
  \makeatother
  \ifGPblacktext
    \def\colorrgb#1{}%
    \def\colorgray#1{}%
  \else
    \ifGPcolor
      \def\colorrgb#1{\color[rgb]{#1}}%
      \def\colorgray#1{\color[gray]{#1}}%
      \expandafter\def\csname LTw\endcsname{\color{white}}%
      \expandafter\def\csname LTb\endcsname{\color{black}}%
      \expandafter\def\csname LTa\endcsname{\color{black}}%
      \expandafter\def\csname LT0\endcsname{\color[rgb]{1,0,0}}%
      \expandafter\def\csname LT1\endcsname{\color[rgb]{0,1,0}}%
      \expandafter\def\csname LT2\endcsname{\color[rgb]{0,0,1}}%
      \expandafter\def\csname LT3\endcsname{\color[rgb]{1,0,1}}%
      \expandafter\def\csname LT4\endcsname{\color[rgb]{0,1,1}}%
      \expandafter\def\csname LT5\endcsname{\color[rgb]{1,1,0}}%
      \expandafter\def\csname LT6\endcsname{\color[rgb]{0,0,0}}%
      \expandafter\def\csname LT7\endcsname{\color[rgb]{1,0.3,0}}%
      \expandafter\def\csname LT8\endcsname{\color[rgb]{0.5,0.5,0.5}}%
    \else
      \def\colorrgb#1{\color{black}}%
      \def\colorgray#1{\color[gray]{#1}}%
      \expandafter\def\csname LTw\endcsname{\color{white}}%
      \expandafter\def\csname LTb\endcsname{\color{black}}%
      \expandafter\def\csname LTa\endcsname{\color{black}}%
      \expandafter\def\csname LT0\endcsname{\color{black}}%
      \expandafter\def\csname LT1\endcsname{\color{black}}%
      \expandafter\def\csname LT2\endcsname{\color{black}}%
      \expandafter\def\csname LT3\endcsname{\color{black}}%
      \expandafter\def\csname LT4\endcsname{\color{black}}%
      \expandafter\def\csname LT5\endcsname{\color{black}}%
      \expandafter\def\csname LT6\endcsname{\color{black}}%
      \expandafter\def\csname LT7\endcsname{\color{black}}%
      \expandafter\def\csname LT8\endcsname{\color{black}}%
    \fi
  \fi
  \setlength{\unitlength}{0.0500bp}%
  \begin{picture}(4626.00,3148.00)%
    \gplgaddtomacro\gplbacktext{%
      \csname LTb\endcsname%
      \put(2274,2881){\makebox(0,0){\normalsize $\eta_\text{incorrect}$ ($\sigma_\text{exp}$ = 0.1)}}%
    }%
    \gplgaddtomacro\gplfronttext{%
      \csname LTb\endcsname%
      \put(937,211){\makebox(0,0){\strut{}}}%
      \put(1314,211){\makebox(0,0){\strut{}}}%
      \put(1692,211){\makebox(0,0){\strut{}}}%
      \put(2069,211){\makebox(0,0){\strut{}}}%
      \put(2446,211){\makebox(0,0){\strut{}}}%
      \put(2823,211){\makebox(0,0){\strut{}}}%
      \put(3200,211){\makebox(0,0){\strut{}}}%
      \put(3578,211){\makebox(0,0){\strut{}}}%
      \put(3955,211){\makebox(0,0){\strut{}}}%
      \put(438,438){\makebox(0,0)[r]{\strut{}}}%
      \put(438,983){\makebox(0,0)[r]{\strut{}}}%
      \put(438,1528){\makebox(0,0)[r]{\strut{}}}%
      \put(438,2073){\makebox(0,0)[r]{\strut{}}}%
      \put(438,2618){\makebox(0,0)[r]{\strut{}}}%
      \put(4272,437){\makebox(0,0)[l]{\strut{} 0}}%
      \put(4272,873){\makebox(0,0)[l]{\strut{} 0.2}}%
      \put(4272,1309){\makebox(0,0)[l]{\strut{} 0.4}}%
      \put(4272,1745){\makebox(0,0)[l]{\strut{} 0.6}}%
      \put(4272,2181){\makebox(0,0)[l]{\strut{} 0.8}}%
      \put(4272,2618){\makebox(0,0)[l]{\strut{} 1}}%
    }%
    \gplbacktext
    \put(500,353){\includegraphics[scale=0.66]{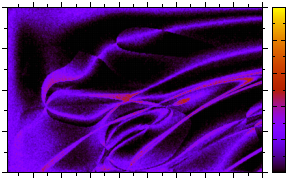}}%
    \gplfronttext
  \end{picture}%
\endgroup

\begingroup
  \makeatletter
  \providecommand\color[2][]{%
    \GenericError{(gnuplot) \space\space\space\@spaces}{%
      Package color not loaded in conjunction with
      terminal option `colourtext'%
    }{See the gnuplot documentation for explanation.%
    }{Either use 'blacktext' in gnuplot or load the package
      color.sty in LaTeX.}%
    \renewcommand\color[2][]{}%
  }%
  \providecommand\includegraphics[2][]{%
    \GenericError{(gnuplot) \space\space\space\@spaces}{%
      Package graphicx or graphics not loaded%
    }{See the gnuplot documentation for explanation.%
    }{The gnuplot epslatex terminal needs graphicx.sty or graphics.sty.}%
    \renewcommand\includegraphics[2][]{}%
  }%
  \providecommand\rotatebox[2]{#2}%
  \@ifundefined{ifGPcolor}{%
    \newif\ifGPcolor
    \GPcolortrue
  }{}%
  \@ifundefined{ifGPblacktext}{%
    \newif\ifGPblacktext
    \GPblacktexttrue
  }{}%
  \let\gplgaddtomacro\g@addto@macro
  \gdef\gplbacktext{}%
  \gdef\gplfronttext{}%
  \makeatother
  \ifGPblacktext
    \def\colorrgb#1{}%
    \def\colorgray#1{}%
  \else
    \ifGPcolor
      \def\colorrgb#1{\color[rgb]{#1}}%
      \def\colorgray#1{\color[gray]{#1}}%
      \expandafter\def\csname LTw\endcsname{\color{white}}%
      \expandafter\def\csname LTb\endcsname{\color{black}}%
      \expandafter\def\csname LTa\endcsname{\color{black}}%
      \expandafter\def\csname LT0\endcsname{\color[rgb]{1,0,0}}%
      \expandafter\def\csname LT1\endcsname{\color[rgb]{0,1,0}}%
      \expandafter\def\csname LT2\endcsname{\color[rgb]{0,0,1}}%
      \expandafter\def\csname LT3\endcsname{\color[rgb]{1,0,1}}%
      \expandafter\def\csname LT4\endcsname{\color[rgb]{0,1,1}}%
      \expandafter\def\csname LT5\endcsname{\color[rgb]{1,1,0}}%
      \expandafter\def\csname LT6\endcsname{\color[rgb]{0,0,0}}%
      \expandafter\def\csname LT7\endcsname{\color[rgb]{1,0.3,0}}%
      \expandafter\def\csname LT8\endcsname{\color[rgb]{0.5,0.5,0.5}}%
    \else
      \def\colorrgb#1{\color{black}}%
      \def\colorgray#1{\color[gray]{#1}}%
      \expandafter\def\csname LTw\endcsname{\color{white}}%
      \expandafter\def\csname LTb\endcsname{\color{black}}%
      \expandafter\def\csname LTa\endcsname{\color{black}}%
      \expandafter\def\csname LT0\endcsname{\color{black}}%
      \expandafter\def\csname LT1\endcsname{\color{black}}%
      \expandafter\def\csname LT2\endcsname{\color{black}}%
      \expandafter\def\csname LT3\endcsname{\color{black}}%
      \expandafter\def\csname LT4\endcsname{\color{black}}%
      \expandafter\def\csname LT5\endcsname{\color{black}}%
      \expandafter\def\csname LT6\endcsname{\color{black}}%
      \expandafter\def\csname LT7\endcsname{\color{black}}%
      \expandafter\def\csname LT8\endcsname{\color{black}}%
    \fi
  \fi
  \setlength{\unitlength}{0.0500bp}%
  \begin{picture}(4626.00,3148.00)%
    \gplgaddtomacro\gplbacktext{%
      \csname LTb\endcsname%
      \put(2274,2881){\makebox(0,0){\normalsize $\eta \;(\sigma_\text{exp} = 0.01)$}}%
    }%
    \gplgaddtomacro\gplfronttext{%
      \csname LTb\endcsname%
      \put(937,211){\makebox(0,0){\strut{}1.7}}%
      \put(1314,211){\makebox(0,0){\strut{}1.8}}%
      \put(1692,211){\makebox(0,0){\strut{}1.9}}%
      \put(2069,211){\makebox(0,0){\strut{}2.0}}%
      \put(2446,211){\makebox(0,0){\strut{}2.1}}%
      \put(2823,211){\makebox(0,0){\strut{}2.2}}%
      \put(3200,211){\makebox(0,0){\strut{}2.3}}%
      \put(3578,211){\makebox(0,0){\strut{}2.4}}%
      \put(3955,211){\makebox(0,0){\strut{}2.5}}%
      \put(2275,-159){\makebox(0,0){\normalsize $W$ (GeV)}}%
      \put(438,438){\makebox(0,0)[r]{\strut{}-1}}%
      \put(438,983){\makebox(0,0)[r]{\strut{}-0.5}}%
      \put(438,1528){\makebox(0,0)[r]{\strut{} 0}}%
      \put(438,2073){\makebox(0,0)[r]{\strut{} 0.5}}%
      \put(438,2618){\makebox(0,0)[r]{\strut{} 1}}%
      \put(-96,1528){\rotatebox{-270}{\makebox(0,0){\normalsize $\cos\thcm$}}}%
    }%
    \gplbacktext
    \put(500,353){\includegraphics[scale=0.66]{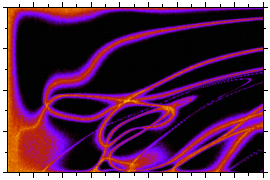}}%
    \gplfronttext
  \end{picture}%
\endgroup
\hspace{-41pt}
\begingroup
  \makeatletter
  \providecommand\color[2][]{%
    \GenericError{(gnuplot) \space\space\space\@spaces}{%
      Package color not loaded in conjunction with
      terminal option `colourtext'%
    }{See the gnuplot documentation for explanation.%
    }{Either use 'blacktext' in gnuplot or load the package
      color.sty in LaTeX.}%
    \renewcommand\color[2][]{}%
  }%
  \providecommand\includegraphics[2][]{%
    \GenericError{(gnuplot) \space\space\space\@spaces}{%
      Package graphicx or graphics not loaded%
    }{See the gnuplot documentation for explanation.%
    }{The gnuplot epslatex terminal needs graphicx.sty or graphics.sty.}%
    \renewcommand\includegraphics[2][]{}%
  }%
  \providecommand\rotatebox[2]{#2}%
  \@ifundefined{ifGPcolor}{%
    \newif\ifGPcolor
    \GPcolortrue
  }{}%
  \@ifundefined{ifGPblacktext}{%
    \newif\ifGPblacktext
    \GPblacktexttrue
  }{}%
  \let\gplgaddtomacro\g@addto@macro
  \gdef\gplbacktext{}%
  \gdef\gplfronttext{}%
  \makeatother
  \ifGPblacktext
    \def\colorrgb#1{}%
    \def\colorgray#1{}%
  \else
    \ifGPcolor
      \def\colorrgb#1{\color[rgb]{#1}}%
      \def\colorgray#1{\color[gray]{#1}}%
      \expandafter\def\csname LTw\endcsname{\color{white}}%
      \expandafter\def\csname LTb\endcsname{\color{black}}%
      \expandafter\def\csname LTa\endcsname{\color{black}}%
      \expandafter\def\csname LT0\endcsname{\color[rgb]{1,0,0}}%
      \expandafter\def\csname LT1\endcsname{\color[rgb]{0,1,0}}%
      \expandafter\def\csname LT2\endcsname{\color[rgb]{0,0,1}}%
      \expandafter\def\csname LT3\endcsname{\color[rgb]{1,0,1}}%
      \expandafter\def\csname LT4\endcsname{\color[rgb]{0,1,1}}%
      \expandafter\def\csname LT5\endcsname{\color[rgb]{1,1,0}}%
      \expandafter\def\csname LT6\endcsname{\color[rgb]{0,0,0}}%
      \expandafter\def\csname LT7\endcsname{\color[rgb]{1,0.3,0}}%
      \expandafter\def\csname LT8\endcsname{\color[rgb]{0.5,0.5,0.5}}%
    \else
      \def\colorrgb#1{\color{black}}%
      \def\colorgray#1{\color[gray]{#1}}%
      \expandafter\def\csname LTw\endcsname{\color{white}}%
      \expandafter\def\csname LTb\endcsname{\color{black}}%
      \expandafter\def\csname LTa\endcsname{\color{black}}%
      \expandafter\def\csname LT0\endcsname{\color{black}}%
      \expandafter\def\csname LT1\endcsname{\color{black}}%
      \expandafter\def\csname LT2\endcsname{\color{black}}%
      \expandafter\def\csname LT3\endcsname{\color{black}}%
      \expandafter\def\csname LT4\endcsname{\color{black}}%
      \expandafter\def\csname LT5\endcsname{\color{black}}%
      \expandafter\def\csname LT6\endcsname{\color{black}}%
      \expandafter\def\csname LT7\endcsname{\color{black}}%
      \expandafter\def\csname LT8\endcsname{\color{black}}%
    \fi
  \fi
  \setlength{\unitlength}{0.0500bp}%
  \begin{picture}(4626.00,3148.00)%
    \gplgaddtomacro\gplbacktext{%
      \csname LTb\endcsname%
      \put(2274,2881){\makebox(0,0){\normalsize $\eta_\text{incorrect}$ ($\sigma_\text{exp}$ = 0.01)}}%
    }%
    \gplgaddtomacro\gplfronttext{%
      \csname LTb\endcsname%
      \put(937,211){\makebox(0,0){\strut{}1.7}}%
      \put(1314,211){\makebox(0,0){\strut{}1.8}}%
      \put(1692,211){\makebox(0,0){\strut{}1.9}}%
      \put(2069,211){\makebox(0,0){\strut{}2.0}}%
      \put(2446,211){\makebox(0,0){\strut{}2.1}}%
      \put(2823,211){\makebox(0,0){\strut{}2.2}}%
      \put(3200,211){\makebox(0,0){\strut{}2.3}}%
      \put(3578,211){\makebox(0,0){\strut{}2.4}}%
      \put(3955,211){\makebox(0,0){\strut{}2.5}}%
      \put(2275,-159){\makebox(0,0){\normalsize $W$ (GeV)}}%
      \put(438,438){\makebox(0,0)[r]{\strut{}}}%
      \put(438,983){\makebox(0,0)[r]{\strut{}}}%
      \put(438,1528){\makebox(0,0)[r]{\strut{}}}%
      \put(438,2073){\makebox(0,0)[r]{\strut{}}}%
      \put(438,2618){\makebox(0,0)[r]{\strut{}}}%
      \put(4272,437){\makebox(0,0)[l]{\strut{} 0}}%
      \put(4272,873){\makebox(0,0)[l]{\strut{} 0.2}}%
      \put(4272,1309){\makebox(0,0)[l]{\strut{} 0.4}}%
      \put(4272,1745){\makebox(0,0)[l]{\strut{} 0.6}}%
      \put(4272,2181){\makebox(0,0)[l]{\strut{} 0.8}}%
      \put(4272,2618){\makebox(0,0)[l]{\strut{} 1}}%
    }%
    \gplbacktext
    \put(500,353){\includegraphics[scale=0.66]{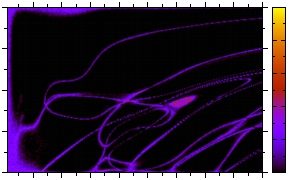}}%
    \gplfronttext
  \end{picture}%
\endgroup
\end{center}
\caption{(Color online) The $\{C_x,O_x,E,F\}$ insolvabilities $\eta = \eta_\text{imaginary} + \eta_\text{incorrect}$ and $\eta_\text{incorrect}$ as a function of $W$ and $\cos\thcm$ for two values of the input experimental resolution ($\sigma_\text{exp} = 0.1$ and $\sigma_\text{exp} = 0.01$).}
\label{fig:insolvability_maps}
\end{figure*}

In Table~\ref{tab:compared_phases}, the model's phases are compared with the corresponding estimates from solution 1 of dataset A and solution 2 of dataset B. From this table, it is seen that solution 1 is the actual solution. This finding could have been anticipated by comparing \textsc{ci} values  ($99.4\%$ versus $79.9\%$). However, a real experiment only yields a single dataset at a certain kinematical point. If an experiment were to yield dataset A, one could not exclude solutions 3 and 4 with significance and one could only conclude that solution 1 is most likely the real solution with $99.4\%$ \textsc{ci}. If dataset B were the result of a real experiment, then solutions 3 and 4 could be excluded with significance and one could state that solution 2 is most likely the actual solution with a \textsc{ci} value of only $79.9\%$, though. The confidence interval on the constraint \eq{eq:phases_constraint} is the only parameter able to distinguish between the four solutions. In some cases, however, the 
solution with the highest \textsc{ci} does not deliver the correct solution for the phases. 

In the beginning of this section, it was argued that an imaginary estimate for the $\widehat{r}_i$ leads to imaginary $\widetilde{\delta}_i$.  Even with real $\widehat{r}_i$, however, it cannot be excluded that imaginary $\widetilde{\delta}_i$ are retrieved. Indeed, from Eqs.\ \eq{eq:Cx-Ox} and \eq{eq:E-F} the phases are determined by solving a quadratic equation in the sine or cosine of the phase. For certain combinations of the measured double asymetries the discriminant of this quadratic equation becomes negative. In such case, the particular dataset can solely be used to estimate the moduli. In what follows the authors wish to quantify the insolvability of a complete set, i.e.\ how frequently it occurs that imaginary or incorrect solutions are obtained for the transversity amplitudes.

Figure \ref{fig:insolvabilities} shows the insolvability of the transversity amplitudes as a function of the input experimental resolution $\sigma_\text{exp}$. Simulations for the complete sets $\{C_x,O_x,E,F\}$ and $\{C_x,O_x,G,H\}$ are analyzed at four different kinematics. Each point in Fig.~\ref{fig:insolvabilities} results from an ensemble of 200 simulated datasets constructed from samples of 50 events for each of the asymmetries. The insolvability $\eta$ is the fraction of unsuccessful simulated datasets.  The $\eta$ is defined as the sum of the fraction of (two or four) imaginary solutions ($\eta_\text{imaginary}$) and the fraction of incorrect solutions ($\eta_\text{incorrect}$). An incorrect solution is a solution with the highest \textsc{ci} value that does not correspond with the model's value in the limit $\sigma_\text{exp} \to 0$. The insolvability $\eta$ is binomially distributed and hence the corresponding error is calculated as
\begin{align}
\sigma(\eta) = \sqrt{\frac{\eta(1-\eta)}{N}},
\end{align}
where $N$ represents the number of simulated datasets. It is seen that the insolvability decreases with decreasing $\sigma_\text{exp}$, as is expected. Further, the $\eta_\text{imaginary}$ has the largest contribution and $\eta_\text{incorrect}$ vanishes more rapidly with $\sigma_\text{exp}$ than $\eta_\text{imaginary}$. In some situations, the insolvability only vanishes for very challenging experimental resolutions, and it is seen that this behavior depends on the complete set in question.

Figure \ref{fig:insolvability_maps} shows the $\{C_x,O_x,E,F\}$ insolvabilities $\eta$ and $\eta_\text{incorrect}$ as a function of $W$ and $\cos\thcm$ for $\sigma_\text{exp} = 0.1$ and $\sigma_\text{exp} = 0.01$. For each of the $375\times241 = 90\,375$ kinematical points, $1000$ simulated datasets were generated (from samples of 50 events for all asymmetries) from which the average $\eta_\text{imaginary}$ and $\eta_\text{incorrect}$ are calculated. It is seen that for $\sigma_\text{exp} = 0.1$ the insolvability can become quite substantial ($0.7 \lesssim \eta \lesssim 0.9$) for some regions in the $(W, \cos\thcm)$ space. In other regions of the phase space the insolvability vanishes. For $\sigma_\text{exp} = 0.1$, $\eta_\text{imaginary}$ has the largest contribution to $\eta$. In Fig.~\ref{fig:insolvability_maps}, the effect of improving the experimental resolution by an order of magnitude is clearly visible. Indeed, both the $\eta_\text{imaginary}$ and $\eta_\text{incorrect}$ surfaces are significantly 
reduced.
The overall insolvability, however, still occupies a fair fraction of the phase space. It is also striking that the regions for $|\cos\thcm| \approx 1$ and/or $W$ approaching the threshold are highly insolvable, both for $\sigma_\text{exp} = 0.1$ and $\sigma_\text{exp} = 0.01$.

\begin{figure*}[!t]
\begin{center}
\begingroup
  \makeatletter
  \providecommand\color[2][]{%
    \GenericError{(gnuplot) \space\space\space\@spaces}{%
      Package color not loaded in conjunction with
      terminal option `colourtext'%
    }{See the gnuplot documentation for explanation.%
    }{Either use 'blacktext' in gnuplot or load the package
      color.sty in LaTeX.}%
    \renewcommand\color[2][]{}%
  }%
  \providecommand\includegraphics[2][]{%
    \GenericError{(gnuplot) \space\space\space\@spaces}{%
      Package graphicx or graphics not loaded%
    }{See the gnuplot documentation for explanation.%
    }{The gnuplot epslatex terminal needs graphicx.sty or graphics.sty.}%
    \renewcommand\includegraphics[2][]{}%
  }%
  \providecommand\rotatebox[2]{#2}%
  \@ifundefined{ifGPcolor}{%
    \newif\ifGPcolor
    \GPcolortrue
  }{}%
  \@ifundefined{ifGPblacktext}{%
    \newif\ifGPblacktext
    \GPblacktexttrue
  }{}%
  \let\gplgaddtomacro\g@addto@macro
  \gdef\gplbacktext{}%
  \gdef\gplfronttext{}%
  \makeatother
  \ifGPblacktext
    \def\colorrgb#1{}%
    \def\colorgray#1{}%
  \else
    \ifGPcolor
      \def\colorrgb#1{\color[rgb]{#1}}%
      \def\colorgray#1{\color[gray]{#1}}%
      \expandafter\def\csname LTw\endcsname{\color{white}}%
      \expandafter\def\csname LTb\endcsname{\color{black}}%
      \expandafter\def\csname LTa\endcsname{\color{black}}%
      \expandafter\def\csname LT0\endcsname{\color[rgb]{1,0,0}}%
      \expandafter\def\csname LT1\endcsname{\color[rgb]{0,1,0}}%
      \expandafter\def\csname LT2\endcsname{\color[rgb]{0,0,1}}%
      \expandafter\def\csname LT3\endcsname{\color[rgb]{1,0,1}}%
      \expandafter\def\csname LT4\endcsname{\color[rgb]{0,1,1}}%
      \expandafter\def\csname LT5\endcsname{\color[rgb]{1,1,0}}%
      \expandafter\def\csname LT6\endcsname{\color[rgb]{0,0,0}}%
      \expandafter\def\csname LT7\endcsname{\color[rgb]{1,0.3,0}}%
      \expandafter\def\csname LT8\endcsname{\color[rgb]{0.5,0.5,0.5}}%
    \else
      \def\colorrgb#1{\color{black}}%
      \def\colorgray#1{\color[gray]{#1}}%
      \expandafter\def\csname LTw\endcsname{\color{white}}%
      \expandafter\def\csname LTb\endcsname{\color{black}}%
      \expandafter\def\csname LTa\endcsname{\color{black}}%
      \expandafter\def\csname LT0\endcsname{\color{black}}%
      \expandafter\def\csname LT1\endcsname{\color{black}}%
      \expandafter\def\csname LT2\endcsname{\color{black}}%
      \expandafter\def\csname LT3\endcsname{\color{black}}%
      \expandafter\def\csname LT4\endcsname{\color{black}}%
      \expandafter\def\csname LT5\endcsname{\color{black}}%
      \expandafter\def\csname LT6\endcsname{\color{black}}%
      \expandafter\def\csname LT7\endcsname{\color{black}}%
      \expandafter\def\csname LT8\endcsname{\color{black}}%
    \fi
  \fi
  \setlength{\unitlength}{0.0500bp}%
  \begin{picture}(4626.00,3148.00)%
    \gplgaddtomacro\gplbacktext{%
      \csname LTb\endcsname%
      \put(2274,2881){\makebox(0,0){\normalsize $\text{Sample} \;(\sigma_\text{exp} = 0.1)$}}%
    }%
    \gplgaddtomacro\gplfronttext{%
      \csname LTb\endcsname%
      \put(937,211){\makebox(0,0){\strut{}1.7}}%
      \put(1314,211){\makebox(0,0){\strut{}1.8}}%
      \put(1692,211){\makebox(0,0){\strut{}1.9}}%
      \put(2069,211){\makebox(0,0){\strut{}2.0}}%
      \put(2446,211){\makebox(0,0){\strut{}2.1}}%
      \put(2823,211){\makebox(0,0){\strut{}2.2}}%
      \put(3200,211){\makebox(0,0){\strut{}2.3}}%
      \put(3578,211){\makebox(0,0){\strut{}2.4}}%
      \put(3955,211){\makebox(0,0){\strut{}2.5}}%
      \put(2275,-159){\makebox(0,0){\normalsize $W$ (GeV)}}%
      \put(438,438){\makebox(0,0)[r]{\strut{}-1}}%
      \put(438,983){\makebox(0,0)[r]{\strut{}-0.5}}%
      \put(438,1528){\makebox(0,0)[r]{\strut{} 0}}%
      \put(438,2073){\makebox(0,0)[r]{\strut{} 0.5}}%
      \put(438,2618){\makebox(0,0)[r]{\strut{} 1}}%
      \put(-96,1528){\rotatebox{-270}{\makebox(0,0){\normalsize $\cos\thcm$}}}%
    }%
    \gplbacktext
    \put(490,353){\includegraphics[scale=0.66]{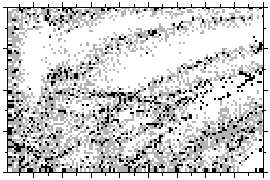}}%
    \gplfronttext
  \end{picture}%
\endgroup
\hspace{-41pt}
\begingroup
  \makeatletter
  \providecommand\color[2][]{%
    \GenericError{(gnuplot) \space\space\space\@spaces}{%
      Package color not loaded in conjunction with
      terminal option `colourtext'%
    }{See the gnuplot documentation for explanation.%
    }{Either use 'blacktext' in gnuplot or load the package
      color.sty in LaTeX.}%
    \renewcommand\color[2][]{}%
  }%
  \providecommand\includegraphics[2][]{%
    \GenericError{(gnuplot) \space\space\space\@spaces}{%
      Package graphicx or graphics not loaded%
    }{See the gnuplot documentation for explanation.%
    }{The gnuplot epslatex terminal needs graphicx.sty or graphics.sty.}%
    \renewcommand\includegraphics[2][]{}%
  }%
  \providecommand\rotatebox[2]{#2}%
  \@ifundefined{ifGPcolor}{%
    \newif\ifGPcolor
    \GPcolortrue
  }{}%
  \@ifundefined{ifGPblacktext}{%
    \newif\ifGPblacktext
    \GPblacktexttrue
  }{}%
  \let\gplgaddtomacro\g@addto@macro
  \gdef\gplbacktext{}%
  \gdef\gplfronttext{}%
  \makeatother
  \ifGPblacktext
    \def\colorrgb#1{}%
    \def\colorgray#1{}%
  \else
    \ifGPcolor
      \def\colorrgb#1{\color[rgb]{#1}}%
      \def\colorgray#1{\color[gray]{#1}}%
      \expandafter\def\csname LTw\endcsname{\color{white}}%
      \expandafter\def\csname LTb\endcsname{\color{black}}%
      \expandafter\def\csname LTa\endcsname{\color{black}}%
      \expandafter\def\csname LT0\endcsname{\color[rgb]{1,0,0}}%
      \expandafter\def\csname LT1\endcsname{\color[rgb]{0,1,0}}%
      \expandafter\def\csname LT2\endcsname{\color[rgb]{0,0,1}}%
      \expandafter\def\csname LT3\endcsname{\color[rgb]{1,0,1}}%
      \expandafter\def\csname LT4\endcsname{\color[rgb]{0,1,1}}%
      \expandafter\def\csname LT5\endcsname{\color[rgb]{1,1,0}}%
      \expandafter\def\csname LT6\endcsname{\color[rgb]{0,0,0}}%
      \expandafter\def\csname LT7\endcsname{\color[rgb]{1,0.3,0}}%
      \expandafter\def\csname LT8\endcsname{\color[rgb]{0.5,0.5,0.5}}%
    \else
      \def\colorrgb#1{\color{black}}%
      \def\colorgray#1{\color[gray]{#1}}%
      \expandafter\def\csname LTw\endcsname{\color{white}}%
      \expandafter\def\csname LTb\endcsname{\color{black}}%
      \expandafter\def\csname LTa\endcsname{\color{black}}%
      \expandafter\def\csname LT0\endcsname{\color{black}}%
      \expandafter\def\csname LT1\endcsname{\color{black}}%
      \expandafter\def\csname LT2\endcsname{\color{black}}%
      \expandafter\def\csname LT3\endcsname{\color{black}}%
      \expandafter\def\csname LT4\endcsname{\color{black}}%
      \expandafter\def\csname LT5\endcsname{\color{black}}%
      \expandafter\def\csname LT6\endcsname{\color{black}}%
      \expandafter\def\csname LT7\endcsname{\color{black}}%
      \expandafter\def\csname LT8\endcsname{\color{black}}%
    \fi
  \fi
  \setlength{\unitlength}{0.0500bp}%
  \begin{picture}(4626.00,3148.00)%
    \gplgaddtomacro\gplbacktext{%
      \csname LTb\endcsname%
      \put(2274,2881){\makebox(0,0){\normalsize $\text{Sample} \;(\sigma_\text{exp} = 0.01)$}}%
    }%
    \gplgaddtomacro\gplfronttext{%
      \csname LTb\endcsname%
      \put(937,211){\makebox(0,0){\strut{}1.7}}%
      \put(1314,211){\makebox(0,0){\strut{}1.8}}%
      \put(1692,211){\makebox(0,0){\strut{}1.9}}%
      \put(2069,211){\makebox(0,0){\strut{}2.0}}%
      \put(2446,211){\makebox(0,0){\strut{}2.1}}%
      \put(2823,211){\makebox(0,0){\strut{}2.2}}%
      \put(3200,211){\makebox(0,0){\strut{}2.3}}%
      \put(3578,211){\makebox(0,0){\strut{}2.4}}%
      \put(3955,211){\makebox(0,0){\strut{}2.5}}%
      \put(2275,-159){\makebox(0,0){\normalsize $W$ (GeV)}}%
      \put(438,438){\makebox(0,0)[r]{\strut{}}}%
      \put(438,983){\makebox(0,0)[r]{\strut{}}}%
      \put(438,1528){\makebox(0,0)[r]{\strut{}}}%
      \put(438,2073){\makebox(0,0)[r]{\strut{}}}%
      \put(438,2618){\makebox(0,0)[r]{\strut{}}}%
    }%
    \gplbacktext
    \put(490,353){\includegraphics[scale=0.66]{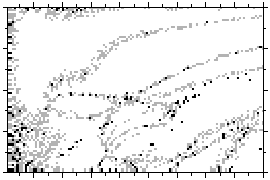}}%
    \gplfronttext
  \end{picture}%
\endgroup
\end{center}
\caption{Two random samples of the individual events that give rise to the left plots shown in Fig.\ \ref{fig:insolvability_maps}. The gray, black, and white dots correspond with imaginary moduli and/or phases, incorrect phases, and correct phases.}
\label{fig:insolvability_maps_single_sim}
\end{figure*}

Figure \ref{fig:insolvability_maps_single_sim} presents a sample extraction analysis of the transversity amplitudes for both $\sigma_\text{exp} = 0.1$ and $\sigma_\text{exp} = 0.01$. At each of the $125\times81 = 10\,125$ kinematical points a single simulated dataset was generated and subsequently analyzed. The samples represent how the result of a measurement of the $(W, \cos\thcm)$ dependence of the asymmetries might look like. For $\sigma_\text{exp} = 0.1$, a large fraction of the kinematical phase space suffers from imaginary or incorrect solutions. Increasing the resolution to $\sigma_\text{exp} = 0.01$ substantially improves the situation. For $\sigma_\text{exp} = 0.1$, the fraction of incorrect solutions is rather small and for $\sigma_\text{exp} = 0.01$ their contribution is almost negligible.

Though having the smaller contribution to the insolvability, incorrect solutions are spurious and should therefore be eliminated, if possible. Incorrect solutions arise from simply marking the solution with the highest \textsc{ci} value as the correct solution. Such a procedure, however, does not stand on statistically solid grounds. A more conservative approach would consist of imposing a tolerance level on the \textsc{ci} values. This would mean that the solution with the highest \textsc{ci} value is only accepted as the correct solution if its confidence interval is greater than or equals the tolerance level. Imposing a tolerance level would not come without a cost, however. The downside of this `filtering' procedure would be a substantial increase in the overall insolvability. Indeed, all of the correct solutions that do not pass the tolerance level would also be rejected and therefore lead to an increase in the insolvability. Nevertheless, the question remains whether such a tolerance requirement can 
really suppress the fraction of incorrect solutions entirely?

\begin{table}[!t]
\caption{The mean confidence intervals $\overline{\textsc{ci}}$ for 1000 correct and 1000 incorrect solutions at random kinematical points ($W < 2500$ MeV and $\cos\thcm \in [-1,1]$).}\label{tab:CI_versus_sigma_exp}
\centering
\renewcommand{\arraystretch}{1.3}
\begin{tabular}{c|C{50pt}C{50pt}}
\hline
\noalign{\smallskip}\hline
\multirow{2}{*}{$\bm{\sigma_{\text{\bf exp}}}$} & \multicolumn{2}{c}{$\overline{\text{\scriptsize\bf CI}}$ $\bm{(\%)}$}\\
& Correct & Incorrect \\
\hline
$0.1\phantom{00}$ & $93.0^{+4.4}_{-8.3}$ & $91.8^{+5.4}_{-11.0}$ \\
$0.01\phantom{0}$ & $91.5^{+5.3}_{-9.1}$ & $88.1^{+7.9}_{-18.2}$ \\
$0.001$ & $91.7^{+5.0}_{-8.5}$ & $84.9^{+10.0}_{-22.6}$\\
\hline
\noalign{\smallskip}\hline
\end{tabular}
\end{table}

In Table~\ref{tab:CI_versus_sigma_exp}, the mean confidence intervals are listed for both 1000 correct and 1000 incorrect solutions at random kinematical points. While the ``correct'' $\overline{\textsc{ci}}$ values are higher than the ``incorrect'' ones, for each $\sigma_\text{exp}$ value listed, the difference between both is not at all substantial. Moreover, it is seen that a fair fraction of the incorrect solutions have a confidence interval of at least $95\%$. As there is a significant overlap between the ``correct'' and ``incorrect'' $\overline{\textsc{ci}}$ values, the entire elimination of the incorrect solutions would require a confidence level so high that nearly all of the correct solutions would get rejected as well, thereby resulting in a practically $100\%$ insolvability. Even lowering the confidence level so as to remove the bulk of the incorrect solutions, would equally wipe out the majority of the correct solutions and hence substantially increase the insolvability.

So, the spurious incorrect solutions can never really be identified, at least not at achievable experimental resolutions. It was shown in the above analysis that even when the confidence interval of a solution is significantly high, it is simply not possible to state whether the solution in question is a correct or an incorrect one. This is in stark contrast with the observation that for a considerable experimental resolution the fraction of incorrect solutions is nearly negligible.

\section{Conclusions}
\label{sec:conclusion}
In this paper, the issue of extracting complete information about reaction amplitudes from pseudoscalar-meson photoproduction data is addressed. The merits of employing the transversity basis (for finite experimental resolution) have been highlighted. Indeed, linear equations connect the moduli of the amplitudes to single-polarization observables. Nonlinear equations connect the relative phases of the amplitudes to double-polarization observables, which are less readily available.

The observables of pseudoscalar-meson photoproduction have been expanded in the transversity basis. An inconsistency with the existing literature is discovered. An independent test of the derived expressions is presented. By rotating the obtained transversity expressions to both the helicity and the CGLN basis, expressions from literature are retrieved. Therefore, convincing evidence has been provided that the derived transversity expansion for the observables is correct.

The extraction of the moduli $r_i$ of the normalized transversity amplitudes is a rather straightforward procedure. The authors have performed this analysis for the $\gamma p \to K^+\Lambda$ reaction with sets of $\{\Sigma, T, P\}$ data from the GRAAL collaboration. These data cover the range $1.65 \lesssim W \lesssim 1.91$~GeV. For the moduli, imaginary solutions can be obtained at some isolated kinematics due to finite experimental resolution. In the performed analysis of the GRAAL data, less than $6.5\%$ of the extracted $\widehat{r}_i$ are complex. Upon improving the experimental resolution, the fraction of complex $\widehat{r}_i$ can be reduced. It is found that the $(W, \cos\thcm)$ dependence of the extracted $\widehat{r}_i$ can be nicely reproduced by the RPR-2011 model.

The formalism of Wen-Tai Chiang and Tabakin for solving complete sets has been extended so as to provide consistent estimates for the independent relative phases for data with finite error bars. As a check of this formalism, it was applied to Monte Carlo simulations of the complete set $\{C_x,O_x,E,F\}$ for the reaction $\gamma p \to K^+ \Lambda$. The simulations are generated with the RPR-2011 model.

Estimating the independent phases is far more challenging than determining the moduli. In order to quantify the phase-related issues, the insolvability of the transversity amplitudes has been introduced. The insolvability is a measure for the fraction of complete measurements which does not result in a successful determination of the transversity amplitudes. The insolvability receives contributions from both imaginary and incorrect solutions. It was observed that the fraction of imaginary solutions is much larger for the phases than for the moduli. However, the amount of imaginary solutions can be reduced by increasing the experimental resolution. The `incorrect' component of the insolvability is much more troublesome. It originates from a discrete phase ambiguity that cannot be resolved for finite uncertainties of the asymmetries. Although this component is not the dominant one and decreases for increasing experimental resolution, it was found that at achievable experimental conditions it is impossible to 
discriminate between correct and incorrect solutions for the phases with statistical significance.

It remains to be investigated whether the measurement of an additional double asymmetry, or multiple ones, could help in resolving the phase ambiguities for realistic experimental resolutions.

\acknowledgments
This work is supported by the Research Council of Ghent University and the Flemish Research Foundation (FWO Vlaanderen).

\appendix*
\section{Other representations of the asymmetries}

\subsection{Helicity representation}
\begin{table}[!t]
\caption{The expressions for the single and double asymmetries in the normalized helicity basis.}\label{tab:helicity_representation}
\centering
\vspace{10pt}
\renewcommand{\arraystretch}{1.3}
\begin{tabular}{c|c}
\hline
\noalign{\smallskip}\hline
& \textbf{Helicity representation}  \\
\hline
$\Sigma$& $-\Re(h_1h_4^* - h_2h_3^*)\phantom{-}$\\
$T$ & $-\Im(h_1h_2^* + h_3h_4^*)\phantom{-}$\\
$P$ & $-\Im(h_1h_3^* + h_2h_4^*)\phantom{-}$\\
\hline
$C_{x'}$ & $-\Re(h_1h_3^* + h_2h_4^*)\phantom{-}$\\
$C_{z'}$ & $-\tfrac{1}{2}\left(|h_1|^2 + |h_2|^2 - |h_3|^2 - |h_4|^2\right)$\\
$O_{x'}$ & $+\Im(h_1h_2^* - h_3h_4^*)\phantom{-}$ \\
$O_{z'}$ & $-\Im(h_1h_4^* - h_2h_3^*)\phantom{-}$\\
\hline
$E$ & $-\tfrac{1}{2}\left(|h_1|^2 - |h_2|^2 + |h_3|^2 - |h_4|^2\right)$\\
$F$ & $-\Re(h_1h_2^* + h_3h_4^*)\phantom{-}$\\
$G$ & $+\Im(h_1h_4^* + h_2h_3^*)\phantom{-}$\\
$H$ & $+\Im(h_1h_3^* - h_2h_4^*)\phantom{-}$\\
\hline
$T_{x'}$ & $+\Re(h_1h_4^* + h_2h_3^*)\phantom{-}$\\
$T_{z'}$ & $+\Re(h_1h_2^* - h_3h_4^*)\phantom{-}$\\
$L_{x'}$ & $-\Re(h_1h_3^* - h_2h_4^*)\phantom{-}$\\
$L_{z'}$ & $-\tfrac{1}{2}\left(|h_1|^2 - |h_2|^2 - |h_3|^2 + |h_4|^2\right)$\\
\hline
\noalign{\smallskip}\hline
\end{tabular}
\end{table}

\label{app:helicity_representation}
The helicity amplitudes $H_1, H_2, H_3,$ and $H_4$ are defined as
\begin{align}
H_1 &= {}_R\bra{+}J_+\ket{-}_{T},\nonumber\\
H_2 &= {}_R\bra{+}J_+\ket{+}_{T},\nonumber\\
H_3 &= {}_R\bra{-}J_+\ket{-}_{T},\nonumber\\
H_4 &= {}_R\bra{-}J_+\ket{+}_{T}.\label{helicityamplitudes}
\end{align}
Here, $\ket{\pm}_T$ and $\ket{\pm}_R$ represent the target and recoil helicity eigenstates, respectively, and $J_+$ is defined in Eq.\ \eq{eq:Jpm}. As the target (recoil) momentum $\vec{p}_T$ ($\vec{p}_R$) is directed along the negative $z$--axis ($z'$--axis), it follows that
\begin{align}
\ket{\pm}_T &= \ket{\pm}_{-z} = u_\pm(p_T,\pi,0),\nonumber\\
\ket{\pm}_R &= \ket{\pm}_{-z'} = u_\pm(p_R,\pi-\thcm,\pi).\label{eq:TRkets}
\end{align}
From Eqs.\ \eq{eq:spinor}, \eq{eq:rotationmatrices}, \eq{eq:xyzkets}, and \eq{eq:TRkets} one deduces the following relations
\begin{align}
\ket{\pm}_T &= \mp\frac{1}{\sqrt{2}}\Bigl(i\ket{\pm}_y - \ket{\mp}_y\Bigr),\nonumber\\
\ket{\pm}_R &= \pm\frac{1}{\sqrt{2}}\Bigl(ie^{\mp i\thcm/2}\ket{\pm}_y - e^{\pm i\thcm/2}\ket{\mp}_y\Bigr),\label{TR-helicity_to_transversity}
\end{align}
and in combination with the definition of $J_-$ in Eq.~\eq{eq:Jpm}, one can readily show that the following properties hold
\begin{align}
H_1 &= {}_R\bra{+}J_+\ket{-}_{T} = +{}_R\bra{-}J_-\ket{+}_{T},\nonumber\\
H_2 &= {}_R\bra{+}J_+\ket{+}_{T} = -{}_R\bra{-}J_-\ket{-}_{T},\nonumber\\
H_3 &= {}_R\bra{-}J_+\ket{-}_{T} = -{}_R\bra{+}J_-\ket{+}_{T},\nonumber\\
H_4 &= {}_R\bra{-}J_+\ket{+}_{T} = +{}_R\bra{+}J_-\ket{-}_{T}.\label{helicityamplitudes2}
\end{align}
From Eqs.\ \eq{eq:transversityamplitudes}, \eq{eq:Jpm}, \eq{helicityamplitudes}, and \eq{TR-helicity_to_transversity}, one then obtains
\begin{align}
H_i = \frac{1}{\sqrt{2}}U_{ij}b_j,
\end{align}
with $U$ a unitary matrix $ \left( U^\dagger U = UU^\dagger = 1 \right)$
\begin{align}
U = \frac{e^{i\thcm/2}}{2}
\begin{pmatrix}
1 & -e^{-i\thcm} & -1 & -e^{-i\thcm} \\
i & ie^{-i\thcm} & i & -ie^{-i\thcm} \\
i & ie^{-i\thcm} & -i & ie^{-i\thcm} \\
-1 & e^{-i\thcm} & -1 & -e^{-i\thcm}
\end{pmatrix}.\label{eq:U}
\end{align}
The `normalized' helicity amplitudes $h_i$
\begin{align}
h_i = \frac{\sqrt{2}H_i}{\sqrt{|H_1|^2 + |H_2|^2 + |H_3|^2 + |H_4|^2}},
\end{align}
can be written in terms of the normalized transversity amplitudes $a_i$ of Eq.~\eq{eq:normalizedtransversityamplitudes}
\begin{align}
a_i = \frac{1}{\sqrt{2}}(U^\dagger)_{ij}h_j = \frac{1}{\sqrt{2}}U^*_{ji}h_j.
\end{align}
Hereby, use is made of the unitarity of $U$. From the above relations and the expressions of Table~\ref{tab:transversity_representation}, one obtains the helicity representation (Table~\ref{tab:helicity_representation}) of the polarization observables. The $\{A_{x'},A_{z'}\}$ ($A \in \{C, O, T, L\}$) are related to the $\{A_x,A_z\}$ through
\begin{align}
\begin{pmatrix}
A_{x'} \\
A_{z'}
\end{pmatrix}
=
\begin{pmatrix}
\cos\thcm & -\sin\thcm \\
\sin\thcm & \cos\thcm
\end{pmatrix}
\begin{pmatrix}
A_x \\
A_z
\end{pmatrix}.\label{eq:primed_doubles}
\end{align}
The expressions of Table~\ref{tab:helicity_representation} coincide with those in Ref.\ \cite{Fasano:1992es}.

\subsection{CGLN representation}
\label{app:CGLN_representation}
The Chew-Goldberger-Low-Nambu (CGLN) amplitudes $F_i$ are, for example, defined in Eq.~(8) of Ref.~\cite{Sandorfi:2011nv}. In what follows a connection is established between the the CGLN and the transversity amplitudes, as this might be useful for future analyses. From the definition of the Dirac spinor \eq{eq:spinor}, the reduced CGLN amplitudes $f_i = \sqrt{\rho_0}F_i$ can be related to the $a_i$ of Eq.~\eq{eq:normalizedtransversityamplitudes}. Here, $\rho_0$ is the density-of-states factor, defined in Eq.~(5) of Ref.~\cite{Sandorfi:2011nv}. The following relation holds
\begin{align}
f_i = V_{ij}a_j,
\end{align}
with $V$ a non-unitary matrix, given by
\begin{align}
V = -\frac{i}{\sin^2\thcm}
\begin{pmatrix}
i e^{i\thcm} & i e^{-i\thcm} & 0 & 0 \\
i \sin\thcm & i \sin\thcm & 0 & 0 \\
-e^{i\thcm} & e^{-i\thcm} & e^{i\thcm} & e^{-i\thcm} \\
1 & -1 & -1 & -1
\end{pmatrix}.
\end{align}
This means that the CGLN basis is not orthogonal. By substituting $a_i = (V^{-1})_{ij}f_j$ in Table~\ref{tab:transversity_representation}, with
\begin{align}
V^{-1} = \frac{i}{2}
\begin{pmatrix}
-1 & e^{-i\thcm} & 0 & 0 \\
1 & -e^{i\thcm} & 0 & 0 \\
-1 & e^{-i\thcm} & -i\sin\thcm & -ie^{-i\thcm}\sin\thcm \\
-1 & e^{i\thcm} & i\sin\thcm & ie^{i\thcm}\sin\thcm
\end{pmatrix},
\end{align}
one retrieves the CGLN expansion of the asymmetries as is listed in Eqs.~(58b--p) of Ref.~\cite{Sandorfi:2011nv}.

As $V \propto \sin^{-2}\thcm$, it is divergent at $\cos\thcm = \pm 1$. In these two cases, the CGLN amplitudes cannot be expressed in terms of the transversity ones. Conversely, for $\cos\thcm = \pm 1$ the $a_i$ can be expanded in the $f_i$ basis, though. By substituting $\cos\thcm = \pm 1$ in $a_i = (V^{-1})_{ij}f_j$, it is found that
\begin{align}
a_1 = -a_2 = a_3 = a_4 = -\frac{i}{2}(f_1 \mp f_2).
\end{align}
The normalization condition \eq{eq:normalization} then leads to $|a_i| = \frac{1}{2}$ for $\cos\thcm = \pm 1$. As a consequence, by invoking Table~\ref{tab:transversity_representation}, all the single and double asymmetries can be quantified at these two extreme angles. At $\cos\thcm = \pm 1$ one has that $C_z = E = 1$ and $L_z = -1$, while all other asymmetries vanish. This result holds for any possible value of $W$. It is quite remarkable that this interesting general result can be derived by altering the representation of the amplitudes.


\begin{thebibliography}{99}
\bibitem{CLAS2012a}
E.~Pasyuk,
\href{http://dx.doi.org/10.1051/epjconf/20123706013}{EPJ Web Conf.\ {\bf 37}, 06013 (2012)}.

\bibitem{CLAS2012b}
F.~J.~Klein (CLAS Collaboration),
\href{http://dx.doi.org/10.1063/1.3701188}{AIP Conf.\ Proc.\ 1432, 51 (2012)}.

\bibitem{LEGS-GRAAL2012}
A.~D'Angelo \textit{et al}.,
\href{http://dx.doi.org/10.1063/1.3701189}{AIP Conf.\ Proc.\ 1432, 56 (2012)}.

\bibitem{MAMI2012}
H.-J.~Arends,
\href{http://dx.doi.org/10.1063/1.3701203}{AIP Conf.\ Proc.\ 1432, 142 (2012)}.

\bibitem{ELSA2012}
F.~Klein,
\href{http://dx.doi.org/10.1063/1.3701204}{AIP Conf.\ Proc.\ 1432, 150 (2012)}.

\bibitem{Barker:1975bp}
I.~S.~Barker, A.~Donnachie, and J.~K.~Storrow,
\href{http://dx.doi.org/10.1016/0550-3213(75)90049-8}{Nucl.\ Phys.\ B \textbf{95}, 347 (1975)}.

\bibitem{Keaton1996}
G.~Keaton and R.~Workman,
\href{http://dx.doi.org/10.1103/PhysRevC.53.1434}{Phys.\ Rev.\ C \textbf{53}, 1434 (1996)}.

\bibitem{Chiang:1996em} 
Wen-Tai~Chiang and F.~Tabakin,
\href{http://dx.doi.org/10.1103/PhysRevC.55.2054}{Phys.\ Rev.\ C \textbf{55}, 2054 (1997)}.

\bibitem{Ireland2010}
D.~G.~Ireland,
\href{http://dx.doi.org/10.1103/PhysRevC.82.025204}{Phys.\ Rev.\ C \textbf{82}, 025204 (2010)}.

\bibitem{Hoblit2009}
S.~Hoblit \textit{et al.},
\href{http://dx.doi.org/10.1103/PhysRevLett.102.172002}{Phys.\ Rev.\ Lett.\ \textbf{102}, 172002 (2009)}.

\bibitem{BjorkenDrellBook}
J.~D.~Bjorken and S.~D.~Drell, Relativistic Quantum Mechanics, McGraw-Hill, New York (1964).

\bibitem{Sandorfi:2011nv}
A.~M.~Sandorfi, S.~Hoblit, H.~Kamano, and T.-S.~H.~Lee,
\href{http://dx.doi.org/10.1088/0954-3899/38/5/053001}{J.\ Phys.\ \textbf{38}, 053001 (2011)}.

\bibitem{Adelseck1990}
R.~A.~Adelseck and B.~Saghai,
\href{http://dx.doi.org/10.1103/PhysRevC.42.108}{Phys.\ Rev.\ C \textbf{42}, 108 (1990)}.

\bibitem{Fasano:1992es}
C.~G.~Fasano, F.~Tabakin, and B.~Saghai,
\href{http://dx.doi.org/10.1103/PhysRevC.46.2430}{Phys.\ Rev.\ C \textbf{46}, 2430 (1992)}. 

\bibitem{tamaraprc}
T.~Corthals, J.~Ryckebusch, and T.~Van~Cauteren,
\href{http://dx.doi.org/10.1103/PhysRevC.73.045207}{Phys.\ Rev.\ C \textbf{73}, 045207 (2006)}.

\bibitem{lesleyprl}
L.~De~Cruz, T.~Vrancx, P.~Vancraeyveld, and J.~Ryckebusch,
\href{http://dx.doi.org/10.1103/PhysRevLett.108.182002}{Phys.\ Rev.\ Lett.\ \textbf{108}, 182002 (2012)}.

\bibitem{lesleyprc}
L.~De~Cruz, J.~Ryckebusch, T.~Vrancx, and P.~Vancraeyveld,
\href{http://dx.doi.org/10.1103/PhysRevC.86.015212}{Phys.\ Rev.\ C \textbf{86}, 015212 (2012)}.

\bibitem{Lleres2007}
A.~Lleres \textit{et al}.\ (GRAAL Collaboration),
\href{http://dx.doi.org/10.1140/epja/i2006-10167-8}{Eur.\ Phys.\ J.\ A \textbf{31}, 79 (2007)}.

\bibitem{Lleres2009}
A.~Lleres \textit{et al}.\ (GRAAL Collaboration),
\href{http://dx.doi.org/10.1140/epja/i2008-10713-4}{Eur.\ Phys.\ J.\ A \textbf{39}, 149 (2009)}.

\bibitem{cracken2010}
M.~E.~McCracken \textit{et al}.\ (CLAS Collaboration),
\href{http://dx.doi.org/10.1103/PhysRevC.81.025201}{Phys.\ Rev.\ C \textbf{81}, 025201 (2010)}.

\bibitem{Workman:2011hi}
\href{http://dx.doi.org/10.1140/epja/i2011-11143-y}{R.~L.~Workman, M.~W.~Paris, W.~J.~Briscoe, L.~Tiator, S.~Schumann, M.~Ostrick, and S.~S.~Kamalov,
Eur.\ Phys.\ J.\ A {\bf 47} 143 (2011)}.

\bibitem{Tiator:2011tu}
L.~Tiator,
\href{http://dx.doi.org/10.1063/1.3701206}{AIP Conf.\ Proc.\ {\bf 1432}, 162 (2012)}.

\bibitem{Tiator:2012ah}
\href{http://arxiv.org/abs/arXiv:1211.3927}{L.~Tiator (Bled Workshops in Physics. Vol.~13 No.~1), arXiv:1211.3927 (2012)}.

\end{thebibliography}
\end{document}